\documentclass[lettersize,onecolumn]{IEEEtran}

\usepackage[dvipsnames]{xcolor}
\usepackage{changepage}
\usepackage{tikz}
\usetikzlibrary{calc}
\usepackage{color,hyperref,listings}
\usepackage[shortlabels]{enumitem}
\usepackage{algorithm}
\usepackage{algorithmic}
\usepackage{amssymb}
\usepackage{tkz-euclide}
\usepackage{siunitx}
\usepackage{physics}
\usepackage{multicol}
\usepackage{pgfplots}
\usepackage{adjustbox}
\usepackage{cancel}
\usepackage{commath}
\usepackage{graphicx}
\usepackage{mathdots}
\usepackage{tabularx}
\usepackage{mathtools}
\usepackage{booktabs}
\usepackage{environ}
\usepackage{cleveref}
\usepackage{fancyhdr}
\usepackage{centernot}
\usepackage[margin=0.75in]{geometry}
\usepackage[american,siunitx,arrowmos]{circuitikz}
\usepackage[final]{pdfpages}
\pgfplotsset{compat=1.18}
\usepackage[english]{babel}
\usepackage{fontawesome}
\usepackage{listings}
\usepackage{fixmath}
\usepackage{xspace}
\usepackage{listings}
\usepackage{amsthm}
\usepackage{upgreek}
\usetikzlibrary{trees}
\usepackage{amsmath,amsfonts}
\usepackage{array}
\usepackage[caption=false,font=normalsize,labelfont=sf,textfont=sf]{subfig}
\usepackage{textcomp}
\usepackage{stfloats}
\usepackage{url}
\usepackage{verbatim}
\usepackage{cite}
\usepackage{graphicx}
\usepackage{prettyref}
\usepackage{wrapfig}
\usepackage{textcomp}% http://ctan.org/pkg/textcomp
\newcommand{\Ebb}{\mathbb{E}}
\newcommand{\Hbb}{\mathbb{H}}
\newcommand{\Nbb}{\mathbb{N}}
\newcommand{\Pbb}{\mathbb{P}}

\newcommand{\uv}{\mathbf{u}}

\newcommand{\xv}{\mathbf{x}}
\newcommand{\yv}{\mathbf{y}}

\newcommand{\Xv}{\mathbf{X}}

\newcommand{\Acal}{\mathcal{A}}
\newcommand{\Bcal}{\mathcal{B}}
\newcommand{\Lcal}{\mathcal{L}}
\newcommand{\Fcal}{\mathcal{F}}
\newcommand{\Mcal}{\mathcal{M}}
\newcommand{\Zcal}{\mathcal{Z}}

\renewcommand{\Gamma}{\Upgamma}
\renewcommand{\Theta}{\Uptheta}
\renewcommand{\Omega}{\Upomega}

\newcommand{\indic}[1]{\mathbf{1}\left\{#1\right\}}

\DeclareMathOperator*{\supp}{supp}

\renewcommand{\norm}[1]{\left\lVert#1\right\rVert}

\newcommand{\st}{\text{\,s.t.\,}}

\newcommand{\bigO}{\mathrm{O}}
\renewcommand{\limsup}{\varlimsup}
\renewcommand{\liminf}{\varliminf}

\def\ie{\textit{i.e.}\@\xspace}
\def\eg{\textit{e.g.}\@\xspace}
\def\etc{\textit{etc.}\@\xspace}

\newtheorem{theorem}{Theorem}[section]

\newtheorem{lemma}[theorem]{Lemma}
\newtheorem{corollary}[theorem]{Corollary}
\newtheorem{fact}[theorem]{Fact}

\theoremstyle{definition}
\newtheorem{definition}[theorem]{Definition}
\newtheorem{construction}[theorem]{Construction}

\newtheorem{exercise-easy}[theorem]{Exercise}
\newtheorem{exercise-med}[theorem]{Exercise}
\newtheorem{exercise-hard}[theorem]{Exercise$^\star$}
\newtheorem{claim}[theorem]{Claim}
\newtheorem*{claim*}{Claim}

\newtheorem{remark}[theorem]{Remark}
\newtheorem*{remark*}{Remark}

\newtheorem*{observation*}{Observation}

\newcommand{\savehyperref}[2]{\texorpdfstring{\hyperref[#1]{#2}}{#2}}

\newrefformat{eq}{\savehyperref{#1}{\textup{(\ref*{#1})}}}
\newrefformat{ineq}{\savehyperref{#1}{\textup{(\ref*{#1})}}}
\newrefformat{eqn}{\savehyperref{#1}{\textup{(\ref*{#1})}}}
\newrefformat{lem}{\savehyperref{#1}{Lemma~\ref*{#1}}}
\newrefformat{con}{\savehyperref{#1}{Construction~\ref*{#1}}}
\newrefformat{def}{\savehyperref{#1}{Definition~\ref*{#1}}}
\newrefformat{thm}{\savehyperref{#1}{Theorem~\ref*{#1}}}
\newrefformat{cor}{\savehyperref{#1}{Corollary~\ref*{#1}}}
\newrefformat{cha}{\savehyperref{#1}{Chapter~\ref*{#1}}}
\newrefformat{sec}{\savehyperref{#1}{Section~\ref*{#1}}}
\newrefformat{app}{\savehyperref{#1}{Appendix~\ref*{#1}}}
\newrefformat{tab}{\savehyperref{#1}{Table~\ref*{#1}}}
\newrefformat{fig}{\savehyperref{#1}{Figure~\ref*{#1}}}
\newrefformat{hyp}{\savehyperref{#1}{Hypothesis~\ref*{#1}}}
\newrefformat{alg}{\savehyperref{#1}{Algorithm~\ref*{#1}}}
\newrefformat{rem}{\savehyperref{#1}{Remark~\ref*{#1}}}
\newrefformat{item}{\savehyperref{#1}{Item~\ref*{#1}}}
\newrefformat{step}{\savehyperref{#1}{step~\ref*{#1}}}
\newrefformat{conj}{\savehyperref{#1}{Conjecture~\ref*{#1}}}
\newrefformat{fact}{\savehyperref{#1}{Fact~\ref*{#1}}}
\newrefformat{prop}{\savehyperref{#1}{Proposition~\ref*{#1}}}
\newrefformat{prob}{\savehyperref{#1}{Problem~\ref*{#1}}}
\newrefformat{claim}{\savehyperref{#1}{Claim~\ref*{#1}}}
\newrefformat{relax}{\savehyperref{#1}{Relaxation~\ref*{#1}}}
\newrefformat{rem}{\savehyperref{#1}{Remark~\ref*{#1}}}
\newrefformat{red}{\savehyperref{#1}{Reduction~\ref*{#1}}}
\newrefformat{part}{\savehyperref{#1}{Part~\ref*{#1}}}
\newrefformat{ex}{\savehyperref{#1}{Exercise~\ref*{#1}}}
\newrefformat{property}{\savehyperref{#1}{Property~\ref*{#1}}}
\newrefformat{type}{\savehyperref{#1}{Type~\ref*{#1}}}
\newrefformat{eg}{\savehyperref{#1}{Example~\ref*{#1}}}
\newrefformat{obs}{\savehyperref{#1}{Observation~\ref*{#1}}}

\definecolor{deepblue}{rgb}{0,0,0.5}
\definecolor{deepred}{rgb}{0.6,0,0}
\definecolor{deepgreen}{rgb}{0,0.5,0}

\begin{document}

\title{A Family of LZ78-based Universal Sequential Probability Assignments}

\author{Naomi Sagan and Tsachy Weissman}

        % <-this % stops a space
% \thanks{Manuscript received October 26, 2023; revised December 8, 2023.}}

\markboth{IEEE Transactions on Information Theory, Submitted for Review}%
{}
\IEEEpubid{0000--0000~\copyright~2023 IEEE}
\newcommand{\tablewidth}{400pt}

\maketitle

\begin{abstract}
We propose and study a family of universal sequential probability assignments on individual sequences, based on the incremental parsing procedure of the Lempel-Ziv (LZ78) compression algorithm.
We show that the normalized log loss under any of these models converges to the normalized LZ78 codelength, uniformly over all individual sequences.  To establish the universality of these models, 
we consolidate a set of results from the literature relating finite-state compressibility to optimal log-loss under Markovian and finite-state models. We also consider some theoretical and computational properties of these models when viewed as probabilistic sources. Finally, 
we present experimental results showcasing the potential benefit of using this family---as models and as sources---for compression, generation, and classification.  
\end{abstract}


\section{Introduction}
\IEEEPARstart{I}{n} the celebrated \cite{originalLZ77paper,originalLZ78paper}, Lempel and Ziv introduced two compression schemes that are universal; \ie,  achieving, among other things, the fundamental limits of compression in an individual sequence setting.
In particular, LZ78 \cite{originalLZ78paper} incrementally parses a sequence into phrases based on an efficiently-computable prefix tree.
Since then, LZ78 (and, to a lesser extent, LZ77) have been used in numerous universal sequential schemes in probability modeling, decision making, filtering, \etc\,\!
% In the process of developing these universal schemes, connections have been drawn between fundamental limits of individual sequence compression and different classes of models for individual sequences.

Intuition behind the universality of LZ78 via a prefix-tree-based probability model and arithmetic coding \cite{rissanenLangdon1979} is described in \cite{langdon1983LZNote}, hinting at the use of LZ78 for universal sequence modeling in the process.
This probability model has been used in a broad range of fields, including but not limited to: universal gambling \cite{feder1991gambling}, universal sequence prediction \cite{federMerhavGutman1992}, universal decoding of finite-state channels \cite{ziv1984FSchannels}, universal denoising \cite{weissman2007Filtering}, and universal guessing of individual sequences \cite{merhav2020guessing}.
For more uses of LZ78 incremental parsing, see \cite{merhav2024jacobzivsindividualsequenceapproach}.
\cite{begleiter2004predictionVMM} uses several prefix tree schemes, including one based on LZ78, for sequence prediction in a setting with training and test data.
Interestingly, though the LZ78-based predictor did not perform the best overall, it outperformed all other schemes on a protein classification task.
Significant among the other predictors mentioned is context tree weighting \cite{willems1995CTW}, which achieves universality over the class of finite-depth tree sources.

Correspondence between finite-state compressibility, finite-state predictability, and Markov model predictability have been explored in several works.
\cite{federMerhavGutman1992} also establishes a correspondence between finite-state compressibility and finite-state predictability under Hamming prediction loss, and \cite{weinberger1994optimalSPA} establishes part of a similar result under log loss.
\cite{merhavFeder1993SequentialDecision} explores optimality in different classes of sequential modeling for individual sequences, concluding that, for any bounded loss function, Markovian schemes asymptotically achieve the same performance as finite state machines.
In this paper, we build off of such results and concretely establish the equivalence of these three quantities (under log loss).

Also of interest in universal modeling of individual sequences is mixture distributions under Dirichlet priors.
Gilbert \cite{gilbert1971codes} was among the first to describe such a model for encoding i.i.d. sources with unknown distribution via an additive perturbation of the empirical distribution, which is  Bayesian scheme under a Dirichlet prior \cite{cover1972admissibility}.
Similarly, a Bayesian sequential probability assignment under the Jeffreys prior in \cite{krichevskiyTrofimov1981} yielded the Krichevskiy-Trofimov estimator for universal encoding.
\cite{merhavFeder1998Universal} contains a broader discussion of these probability models, in the context of both individual sequence and probability source modeling.
As an alternative to the Krichevskiy-Trofimov mixture, \cite{ingber2006FSM} presents an explicit finite-state machine formulation.

More recently, the LZ78-based universal lossy compression algorithms have arisen.
A universal rate-distortion code was proposed in \cite{yang1996lossy}, where the reconstruction codebook is ordered by LZ78 codelength.
A possibly more computationally-efficient variation was described in \cite{merhav2022universalrandomcodingensemble}, which uses a randomly generated codebook, with the codewords drawn with log likelihood equal to the LZ78 codelength (up to a constant).
In \cite{merhav2024LossyCompressionLimits}, fundamental limits of such universal lossy compressors are established.
Recent efforts have also been made to understand the finite-state redundancy of LZ78, \eg, \cite{jacquet2014limiting,jacquet:hal-03139593}.
In addition, the use of compressors for tasks such as classification.
The first LZ-based universal classification scheme was Ziv-Merhav cross parsing \cite{zivMerhav1993CrossParsing}, which uses how many phrases a sequence can be parsed given a reference sequence as a measure of universal relative entropy.
Ziv-Merhav cross parsing has been successfully used in several domains, including biometrics identification \cite{coutinho2010ECG}, document similarity \cite{Helmer2007MeasuringTS}, and genomics \cite{pratas2018primate}.
An LZ77-based scheme is proposed in \cite{benedetto2002Language}, which concatenates test sequences to the training sequence and then compresses.
LZ78 has also been used for similar tasks, for instance determining cellular phone location \cite{bhattacharya2002leziUpdate} and smart home device usage \cite{gopalratnam2007activeLeZi}.

In the past several years, compression and information theory techniques have been incorporated into deep learning methods, improving generalization and accuracy.
% Particularly significant are the principles of minimum description length (MDL) \cite{RISSANEN1978mdl}\footnote{Minimizing the size of finite-state models leads to generalization.} and information bottleneck \cite{tishby2000informationbottleneckmethod}\footnote{Proposes minimizing the mutual information between the model input and latent variable while maximizing the mutual information between the latent variable and desired output to achieve generalization.}.
Particularly significant are the principles of minimum description length (MDL) \cite{RISSANEN1978mdl} and information bottleneck \cite{tishby2000informationbottleneckmethod}.
\cite{lan2022mdlrnn,abudy2025minimumdescriptionlengthapproach}, \eg, use a recurrent neural network trained with an MDL-based loss function for formal language modeling.
In classification, \cite{pmlr-v202-kawaguchi23a,sefidgaran2024minimumdescriptionlengthgeneralization} use MDL and information bottleneck, respectively, to prove bounds on the generalization gap of neural networks.
Compression can also directly improve neural network architectures, \eg, \cite{rae2019compressivetransformerslongrangesequence} applies neural lossy compression to long contexts for transformer language models for improved accuracy and \cite{pmlr-v139-jaegle21a,jaegle2022perceiveriogeneralarchitecture} use cross-attention-based lossy compression to map input byte streams from any domain to a fixed length.
These works demonstrate that compression-based methods are successful in improving neural network performance.
In this paper, we return to the basics of compression-based learning, directly using LZ78 to produce an efficient, universal model.

In a similar direction, \cite{deletang2024languagemodelingcompression} explores connections between language modeling and compression, using language models for compression and the GZIP compressor for sequence generation.
Language models performed quite well as compressors, but at a large computational cost.
GZIP outperformed language models for audio generation but fell behind in text generation, a task for which we hope to close the gap between neural networks and compression-based models.

In this paper, we concretely define a family of LZ78-based sequential probability assignments, of which formulations from \cite{langdon1983LZNote,feder1991gambling,federMerhavGutman1992,weissman2007Filtering} are a special case.
Loosely, models in this family are mixture distributions over an arbitrary prior, conditioned on the LZ78 context of each symbol.
Each SPA in this family, to first order, incurs a log loss that is a scaled version of the LZ78 codelength.
This correspondence, though intuitive to expect, has not been formally established in the literature outside of limited special cases.
This family of sequential probability assignments can also induce a rich family of LZ78-based compressors (\ie, via arithmetic coding \cite{rissanenLangdon1979}) that have the same asymptotic universality guarantees as LZ78 but whose performance may differ in a finite-sequence environment.
An empirical exploration of these compression properties will be included in future work.

We prove that the LZ78 family of models is universal, in the sense that its log-loss asymptotically matches or outperforms any finite-state sequential probability assignment.
In the process, we consolidate a set of results from throughout the literature \cite{merhavFeder1993SequentialDecision,federMerhavGutman1992,weissman2007Filtering,originalLZ78paper}: the optimal asymptotic performance of finite-state probability models (under log loss) is no better than the optimal performance of Markovian models, and the corresponding log loss is a scaled version of the finite-state compressibility, as defined in \cite{originalLZ78paper}.

Additionally, we define a family of probability sources based on the LZ78 sequential probability assignments.
A thorough theoretical investigation of these sources is beyond the scope of this work (cf. \cite{sagan2025lz78source} for a theoretical analysis), but we show initial results pertaining to the compression framework of \cite{merhav2022universalrandomcodingensemble}.
Unlike the LZ78 probability model, which has a limited number of simple-to-compute formulations, this probability source generates realizations from any prior with roughly the same computational cost.
Using this probability source, we generate a family of sequences for which our family of sequential probability assignments achieves substantially better asymptotic performance than any finite-state machine.

The rest of the paper is organized as follows.
In \prettyref{sec:prelim}, we introduce notation that will be used throughout the remainder of the paper.
\prettyref{sec:framework-spa-construction} defines the LZ78 family of sequential probability assignments, with special cases highlighted in \prettyref{sec:special-case-spas}, and the uniform convergence of the self-entropy log loss (for any model in the family) to the LZ78 codelength proven in \prettyref{sec:lz78-log-loss-eq-codelength}.
Then, \prettyref{sec:markov-and-fs-spa} defines two notions of universality for sequential probability assignments, and the equivalence of both of those notions and finite-state compressibility is in \prettyref{sec:lambda-eq-mu-eq-rho}.
\prettyref{sec:lz78-spa-optimality} contains the universality result of the LZ78 family of models with respect to individual sequences, with extensions to stationary and ergodic probability sources in \prettyref{sec:stationary-ergodic-results}.
In \prettyref{sec:computational-analysis}, we provide a brief analysis of the LZ78 SPA's computational complexity.
In \prettyref{sec:lz78-probability-source}, we define the LZ78-based probability source, and elaborate on a special case in \prettyref{sec:bernoulli-lz78-source}.
Finally, \prettyref{sec:experiments} is dedicated to the potential use of the LZ78 sequential probability assignments for text generation and classification, and the LZ78 probability source for compression.
We conclude in \prettyref{sec:conclusion}.

\section{General Notation and Conventions}\label{sec:prelim}
\textbf{Individual sequences}. We refer to a deterministic (albeit arbitrary) infinite sequence of symbols as an \textit{individual sequence}.
An individual sequence is denoted $\xv = \begin{pmatrix}
    x_1 & x_2 & \cdots & x_n & \cdots
\end{pmatrix}$, where the symbols $x_i$ take on values in a fixed alphabet $\Acal$.
For instance, a sequence that takes on values from $1$ to $m$ will have $\Acal = \{1, 2, \dots, m\}$, denoted in shorthand as $[m]$.
A \textit{binary sequence} has alphabet $\Acal = \{0, 1\}$.
We assume the size of the alphabet is fixed as $|\Acal| = A < \infty$.

For sequence $\xv$, $x^n$ denotes the first $n$ symbols and $x_k^\ell$ is the window $\begin{pmatrix}
    x_k & x_{k+1} & \cdots & x_\ell
\end{pmatrix}$.
If $k > \ell$, we take $x_k^\ell$ to be the empty string.

$\Acal^* \triangleq \bigcup_{k \geq 0} \Acal^k$ is defined as the set of all finite sequences, of any length (including $0$) over the alphabet $\Acal$.
For a pair of sequences $x^n \in \Acal^n$ and $y^m \in \Acal^m$, $(x^n \,^\frown y^m) \in \Acal^{m+n}$ is their concatenation, with $x^n$ first and then $y^n$.

Given finite sequence $x^n$ and $a \in \Acal$, $N(a|x^n)$ is the number of times that the symbol $a$ appears: $N(a|x^n) = \sum_{i=1}^n \indic{x_i = a}$.

\textbf{Probability}.
A probability source is denoted $\Xv = \begin{pmatrix}
    X_1 & X_2 & \cdots X_n & \cdots
\end{pmatrix}$, where each symbol $X_i$ is a random variable with a distribution from $\Mcal(\Acal)$, \ie,  the simplex of probability mass functions (PMFs) over $\Acal$.
For PMF $\theta \in \Mcal(\Acal)$ and $a \in \Acal$, $\theta[a]$ represents the probability that $\theta$ assigns to $a$.

\textbf{Miscellanea}.
For this paper, $\log$ refers to the base-$2$ logarithm unless otherwise specified.

\textbf{Notation Summary}. See \prettyref{app:notation-table} for a table of the notation used throughout this paper (both those defined here and those defined later in the paper).

\section{The LZ78 Family of Sequential Probability Assignments}\label{sec:lz78-spa}
In this section, we discuss how the LZ78 compression algorithm \cite{originalLZ78paper} induces a sequential probability assignment (SPA) on individual sequences.
Before precisely defining this SPA, we review SPAs on individual sequences, the SPA log loss function, and the LZ78 compression algorithm.

\subsection{Sequential Probability Assginments}
\begin{definition}[Sequential Probability Assignment]
    For sequence $\xv$, a sequential probability assignment, $q$, maps each finite sequence $x^{t-1}$ to the simplex of probability assignments for the next symbol, $x_{t}$:
    \[q \triangleq \left\{q_t(x_t|x^{t-1})\right\}_{t \geq 1} \text{ where } q_t(\cdot|x^{t-1}) \in \Mcal(\Acal).\]
    In other words, given the prefix, $x^{t-1}$, of a sequence, $q_t$ produces an ``estimated probability distribution'' for the next symbol.
    It is understood that $q(x_t|x^{t-1})$ refers to $q_t(x_t|x^{t-1})$, so we will omit the subscript in $q_t$ when possible.
\end{definition}

We evaluate the accuracy of a SPA via log loss:
\begin{definition}[SPA Log Loss]
    The asymptotic log loss incurred by an SPA, $q$, on infinite sequence $\xv$, is
    \[\limsup_{n\to\infty} \frac{1}{n} \log \frac{1}{q(x^n)} = \limsup_{n\to\infty} \frac{1}{n} \sum_{t=1}^n \log \frac{1}{q(x_t|x^{t-1})}.\]
    The objective of minimizing log loss provides a framework for discussing the universality of the LZ78 family of SPAs in \prettyref{sec:lz78-spa-optimality}.
    In particular, we will show that the asymptotic log loss of the LZ78 family of SPAs is at most that of the best finite-state SPA.
\end{definition}

\subsection{Review: LZ78 Compression Algorithm}\label{sec:lz78-review}
LZ78 encoding \cite{originalLZ78paper} forms a prefix tree by parsing an individual sequence into a list of consecutive subsequences called \textit{phrases}.
The family of SPAs we present in this paper is heavily based on the LZ78 incremental parsing procedure and resulting prefix tree (as is described in \prettyref{con:lz78-spa} and visualized in \prettyref{app:lz78-spa-examples}).

We demonstrate the formation of the LZ78 prefix tree via an example on a binary alphabet.

    Consider the sequence
    \[x^n = \texttt{01100110011}.\]
    The LZ78 prefix tree begins as a singular root node.
    We start at the beginning of the sequence and add the first symbol, \texttt{0}, as a branch to the root node.
    \texttt{0} is then the first phrase.
    We then do the same thing with the next symbol, \texttt{1}, which becomes the second phrase.
    After encoding the first two phrases, the tree is as follows:
    \begin{center}
        \begin{tikzpicture}[
        level distance=1cm,
          level 1/.style={sibling distance=6cm},
          level 2/.style={sibling distance=2cm},
          level 3/.style={sibling distance=1.5cm},
          level 4/.style={sibling distance=1cm}
          ]
          \node[rectangle, fill=yellow!25, rounded corners, draw]{\parbox[t]{4em}{\centering\small root}}
            child {
                node[rectangle, fill=gray!5, rounded corners, draw]{\parbox[t]{4em}{\centering\small node \texttt{0} \\{\small(phrase 1)}}}
            }
            child {
                node[rectangle, fill=gray!5, rounded corners, draw]{\parbox[t]{4em}{\centering\small node \texttt{1} \\{\small(phrase 2)}}}
            };
        \end{tikzpicture}
    \end{center}
    
    So far, we have parsed $x^n$ as \texttt{0,1} and still have to encode \texttt{01100110011}.
    
    The next symbol is \texttt{1}, which is already present as a branch off of the root.
    So, we traverse from the root to the node \texttt{1} and move onto the next symbol, \texttt{0}, which can be added as a branch off of the node \texttt{1}.
    After adding the new leaf, we say that the third phrase is \texttt{10}.
    \begin{center}
        \begin{tikzpicture}[
        level distance=1.2cm,
          level 1/.style={sibling distance=6cm},
          level 2/.style={sibling distance=2cm},
          level 3/.style={sibling distance=1.5cm},
          level 4/.style={sibling distance=1cm}
          ]
          \node[rectangle, fill=yellow!25, rounded corners, draw]{\parbox[t]{4em}{\centering\small root}}
            child {
                node[rectangle, fill=gray!5, rounded corners, draw]{\parbox[t]{4em}{\centering\small node \texttt{0} \\{\small(phrase 1)}}}
            }
            child {
                node[rectangle, fill=gray!5, rounded corners, draw]{\parbox[t]{4em}{\centering\small node \texttt{1} \\{\small(phrase 2)}}}
                child {
                    node[rectangle, fill=gray!5, rounded corners, draw]{\parbox[t]{4em}{\centering\small node \texttt{10} \\{\small(phrase 3)}}}
                }
            };
        \end{tikzpicture}
    \end{center}

    By the process of LZ78 parsing, $x^n$ is divided into the phrases \texttt{0,1,10,01,100,11}, forming the tree:
    \begin{center}
        \begin{tikzpicture}[
        level distance=1.2cm,
          level 1/.style={sibling distance=6cm},
          level 2/.style={sibling distance=2cm},
          level 3/.style={sibling distance=1.5cm},
          level 4/.style={sibling distance=1cm}
          ]
          \node[rectangle, fill=yellow!25, rounded corners, draw]{\parbox[t]{4em}{\centering\small root}}
            child {
                node[rectangle, fill=gray!5, rounded corners, draw]{\parbox[t]{4em}{\centering\small node \texttt{0} \\{\small(phrase 1)}}}
                child {
                    node[rectangle, fill=gray!5, rounded corners, draw]{\parbox[t]{4em}{\centering\small node \texttt{01} \\{\small(phrase 4)}}}
                }
            }
            child {
                node[rectangle, fill=gray!5, rounded corners, draw]{\parbox[t]{4em}{\centering\small node \texttt{1} \\{\small(phrase 2)}}}
                child {
                    node[rectangle, fill=gray!5, rounded corners, draw]{\parbox[t]{4em}{\centering\small node \texttt{10} \\{\small(phrase 3)}}}
                    child {
                        node[rectangle, fill=gray!5, rounded corners, draw]{\parbox[t]{4em}{\centering\small node \texttt{100} \\{\small(phrase 5)}}}
                    }
                }
                child {
                    node[rectangle, fill=gray!5, rounded corners, draw]{\parbox[t]{4em}{\centering\small node \texttt{11} \\{\small(phrase 6)}}}
                }
            };
        \end{tikzpicture}
    \end{center}
    
This algorithm can be described in words as follows (refer to \prettyref{app:lz78-algorithm} for a full algorithmic description):
\begin{construction}[LZ78 Tree]
    In general, a sequence is parsed into phrases as follows:
    \begin{enumerate}
        \itemsep-0.5em 
        \item The LZ78 tree starts off as a singular root node.
        \item Repeat the following until we reach the end of the sequence:
        
        \vspace{-0.5em}
        \begin{enumerate}
            \itemsep-0.5em
            \item Starting at the root, traverse the prefix tree according to the next symbols in the sequence until we reach a leaf.
            \item Add a new node branching off of the leaf, corresponding to the next symbol in the input sequence.
        \item The LZ78 phrase is defined as the slice of the input sequence used in traversing the tree, including the symbol corresponding to the new branch.
        \end{enumerate}
        
    \end{enumerate}
    Each node of the prefix tree corresponds to an LZ78 phrase, so the number of nodes is equivalent to the number of phrases that have been parsed.
    
    For sequence compression, we assign an index to each node of the tree in the order that the nodes were created, and encode each phrase by the index of the leaf node found in step (1), plus the new symbol used to create the branch in step (2).
    For sequential probability assignment, we keep track of the number of times we traverse each node of the tree, as we will describe in detail later.
\end{construction}

From the LZ78 parsing algorithm, we define some useful quantities:
\begin{center}
    \small
    \renewcommand{\arraystretch}{1.3}
    \begin{tabular}{ll}\toprule
        \textbf{Notation} & \textbf{Description}\\
        \midrule
        $\Zcal(x^t)$ & List of \textbf{all LZ78 phrases} in the parsing of $x^t$, including the empty phrase. \\
        $C(x^n)$ & \textbf{Number of LZ78 phrases}; $|\Zcal(x^n)|$. \\
        $z_c(x^{t-1})$ & \parbox[t][][t]{\tablewidth}{
            The \textbf{LZ78 context} associated  $x_t$, \ie, the beginning of the phrase to which $x_{t}$ belongs, not including the symbol $x_t$ itself.\footnote{The notation $z_c(x^{t-1})$ is used to highlight the dependence only on $x^{t-1}$, rather than $x^t$.}
            If $x_{t}$ is the beginning of a new phrase, then $z_c(x^{t-1})$ will be the empty sequence.
            This is the current node of the prefix tree being traversed when parsing $x_t$.
        } \\
         $\mathcal{Y}\{x^n, z\}$ & \parbox[t][][t]{\tablewidth}{
            For $z \in \mathcal{Z}(x^n)$, this is the \textbf{ordered subsequence} of $x^n$ that has LZ78 context $z$, \ie, the subsequence of symbols parsed while at node $z$ of the LZ78 tree. \textit{E.g.}, in the example above, $\mathcal{Y}\{x^n, \texttt{1}\} = (0, 0, 1)$: from node \texttt{0}, we first parse a \texttt{0} (from phrase 3), then another \texttt{0} (from phrase 5), and finally a \texttt{1} (from phrase 6).
         } \\
         $N_\text{LZ}(a|x^t, z)$ & \parbox[t][][t]{\tablewidth}{
            For $a \in \Acal$ and $z \in \Zcal(x^t)$, this is the number of phrases in $\Zcal(x^t)$ that start with $z^\frown a$, \ie, the number of times that $a$ appears in $\mathcal{Y}\{x^t, z\}$.
            $z$ is an LZ78 context, and $a$ is a potential ``next symbol,'' which may or may not equal $x_{t+1}$.
         } \\
        $N_\text{LZ}(x^t, z)$ & The number of phrases in $\Zcal(x^t)$ that start with $z$, \ie, the length of $\mathcal{Y}\{x^t, z\}$. \\
        \bottomrule
    \end{tabular}
\end{center}

The LZ78 family of SPAs, loosely speaking attains universality by conditioning on the LZ78 context of each symbol.
As we prove in \prettyref{lem:phrase-length-infty} of the Appendix, for any individual sequence, the phrases in the corresponding LZ78 parsing will grow infinitely long as $n \to \infty$.
This allows us to capture a context length that grows as the sequence length grows, forming a natural hyperparameter-free method of growing the context.
This fact manifests in the redundancy analysis via a closely-related property: that the number of phrases in an LZ78 parsing of an individual sequence is sub-linear in the sequence length; by equation (9) of \cite{originalLZ78paper}, the number of phrases in the LZ78 parsing of any individual sequence is $\bigO\!\left(\frac{n \log A}{\log n}\right)$, uniformly over all input sequences.

\subsection{Defining the LZ78 Family of Sequential Probability Assignments}\label{sec:framework-spa-construction}
In this section, we develop a family of SPAs that is universal (as per the discussion in \prettyref{sec:markov-and-fs-spa}) and computable in $\bigO(n)$ time with a fairly small implicit constant (as discussed in \prettyref{sec:computational-analysis}).
This family of SPAs generalizes formulations from \cite{langdon1983LZNote,feder1991gambling,federMerhavGutman1992,weissman2007Filtering}.
Recognizing that such SPAs are implicitly Bayesian mixtures with a specific prior, we make this aspect explicit and extend it to a general prior.
This additional flexibility can be essential to empirical accuracy in a finite-sequence setting, which we explore for a genomics classification application in \cite{omri2025genomic}.
In addition, this generalization presents the opportunity for novel theoretical analysis, as discussed in the remainder of the paper, especially in Sections \ref{sec:lz78-log-loss-eq-codelength}, \ref{sec:lz78-spa-optimality}, and \ref{sec:lz78-probability-source}.

\textbf{Towards the LZ78 SPA family}.
To motivate the specific form of the LZ78 SPA family, we consider one of the simplest possible SPAs, the one that defines $q(a|x^{t-1})$ based on the empirical distribution of $x^{t-1}$:
\[q^\text{naive}(a|x^{t-1}) = \frac{N(a|x^{t-1})}{t-1}.\]

This SPA, however, incurs infinite loss if $\exists t$ s.t. $N(x_t | x^{t-1}) = 0$.
This can be amended via a Bayesian mixture approach, \ie, by placing a prior distribution on the frequencies of each symbol.
If the prior distribution is not degenerate, then the issue of unbounded loss is alleviated.

\begin{definition}[Bayesian Mixture SPA]\label{con:bayesian-mixture}
     Define $q^\Uppi$ as the probability mass function of the following  mixture distribution:\\[-2em]
    \begin{enumerate}
        \itemsep-0.5em 
        \item Let $\Uppi$ be a prior distribution on the simplex $\Mcal(\Acal)$.
        We first sample $\Theta \sim \Uppi$, which is a PMF over $\Acal$.
        \item We then sample $X_1, \dots, X_n \stackrel{\text{i.i.d.}}{\sim} \Theta$.
        Define the SPA $q^\Uppi(x^n)$ as the joint PMF for $X_1, \dots, X_n$.
    \end{enumerate}
    By Bayes' theorem, $q^\Uppi(x_t|x^{t-1})$ is
    \begin{equation}
        q^\Uppi(x_t|x^{t-1}) = \frac{q^\Uppi(x^t)}{q^\Uppi(x^{t-1})} = \frac{\int_{\Mcal(\Acal)} \left(\prod_{i=1}^t \Theta[x_i]\right) d\Uppi(\Theta)}{\int_{\Mcal(\Acal)} \left(\prod_{i=1}^{t-1} \Theta[x_i]\right) d\Uppi(\Theta)} = \frac{\int_{\Mcal(\Acal)} \prod_{a \in \Acal} \Theta[a]^{N(a|x^t)} d\Uppi(\Theta)}{\int_{\Mcal(\Acal)} \prod_{a \in \Acal} \Theta[a]^{N(a|x^{t-1})} d\Uppi(\Theta)}. \label{eqn:q-with-prior-integral}
    \end{equation}
\end{definition}
For certain choices of prior, this integral expression becomes simple to compute, as is discussed in \prettyref{sec:special-case-spas}.

Such a mixture, however, only uses zero-order information about the sequence; it does not take into consideration the different contexts that can precede a symbol.
For instance, if in a binary sequence, the sequence $(0, 0, 0)$ is always followed by a $1$, a reasonable SPA should eventually be able to predict that pattern.
To mitigate this, we could consider a fixed-length context preceding each symbol, as a $k$-order Markov SPA (\prettyref{def:markov-spa}) does.

However, any fixed context length of $k$ will not capture patterns that depend on contexts longer than $k$, so it is desirable to have a context length that is allowed to grow unbounded.
By \prettyref{lem:phrase-length-infty}, the LZ78 context associated with a symbol is guaranteed to grow unbounded as the length of the input sequence tends to infinity.
So, we modify the SPA from \eqref{eqn:q-with-prior-integral} by conditioning on the LZ78 context associated with the current symbol.

\begin{construction}[LZ78 Sequential Probability Assignment]\label{con:lz78-spa}
    Let $\Uppi$ be a prior distribution on $\Mcal(\Acal)$.
    We define the LZ78 SPA for prior $\Uppi$ as
    \[q^{LZ78, \Uppi}(a|x^{t-1}) = q^\Uppi\left(a\,\big|\, \mathcal{Y}\{x^{t-1}, z_c(x^{t-1})\}\right),\]
    where $q^\Uppi$ is as defined in \eqref{eqn:q-with-prior-integral}, and $\mathcal{Y}\{x^{t-1}, z_c(x^{t-1})\}$ is the subsequence of $x^{t-1}$ that has the same LZ78 context as $x_t$ (as per \prettyref{sec:lz78-review}).
\end{construction}

This forms the LZ78 family of SPAs.
In the subsequent sections, we will define the form of the LZ78 SPA for specific priors that result in a simple form of \eqref{eqn:q-with-prior-integral}, and show that the log loss of this SPA approaches a scaled version of the LZ78 codelength.
Then, we discuss the performance of this SPA with respect to log loss, proving that its loss is upper-bounded by the optimal log loss of a broad class of SPAs.
\begin{remark}
    For the definition of the LZ78 family of SPAs, we do not place any restrictions on the prior distribution.
    For subsequent theoretical results, however, we stipulate that the prior has full support on $\Mcal(\Acal)$.
\end{remark}

A step-by-step example of evaluating the LZ78 SPA can be found in \prettyref{app:lz78-spa-examples}.

\subsection{Special Cases of the LZ78 Sequential Probability Assignment}\label{sec:special-case-spas}
A canonical example of when the LZ78 SPA becomes tractable is when the prior $\Uppi$ is Dirichlet$(\gamma, \dots, \gamma$), for $0 < \gamma \leq 1$, which reduces $q^\Uppi(x_t|x^{t-1})$ to a simple perturbation of the empirical distribution:
\begin{construction}[Dirichlet SPA]\label{con:jeffreys-prior-spa}
    If the prior defining $q^\Uppi$ in \eqref{eqn:q-with-prior-integral} is $\text{Dirichlet}\left(\gamma, \dots, \gamma\right)$, for positive $\gamma$, then, due to \cite{cover1972admissibility},
    \begin{align*}
        q^\Uppi(x_t|x^{t-1}) = \frac{N(x_t|x^{t-1}) + \gamma}{(t-1) + \gamma A}.
    \end{align*}

    The corresponding LZ78 SPA evaluates to
    \[q^{LZ78, \Uppi}(a|x^{t-1}) = \frac{N_\text{LZ}(a|x^{t-1}, z_c(x^{t-1})) + \gamma}{\sum_{b \in \Acal} N_\text{LZ}(b | x^{t-1}, z_c(x^{t-1})) + \gamma A}.\]
\end{construction}

\begin{remark}\label{rem:jeffreys-prior}
    As per, \eg, \cite{merhavFeder1998Universal}, the choice $\gamma = \frac{1}{2}$ in \prettyref{con:jeffreys-prior-spa} is essentially (to the leading term) minimax optimal with respect to the log loss incurred by the SPA on any individual sequence.
\end{remark}

The structure of the LZ78-based universal predictor from \cite{langdon1983LZNote,feder1991gambling,federMerhavGutman1992,weissman2007Filtering} is a special case the case of \prettyref{con:jeffreys-prior-spa} with $\gamma = \frac{1}{A-1}$.
The structure of this SPA makes it possible to directly show that the log loss incurred on any individual sequence approaches the LZ78 codelength, scaled by $\frac{1}{n}$.
\begin{construction}\label{con:tree-spa}
    Let $\Uppi_0$ be the Dirichlet prior on $\Mcal(\Acal)$ with parameter $\gamma = \frac{1}{A-1}$.
    
    The SPA $q^{\text{LZ78}, \Uppi_0}(x^n)$ is then
    \[q^{\text{LZ78},\Uppi_0}(x_t|x^{t-1}) = \frac{(A-1) N_\text{LZ}(x_t|x^{t-1}, z_c(x^{t-1})) + 1}{(A-1) \left(\sum_{a \in \Acal} N_\text{LZ}(a|x^{t-1}, z_c(x^{t-1}))\right) + A}.\]
    This SPA can be alternatively understood using the following variant of the LZ78 prefix tree:\\[-2em]
    \begin{enumerate}
        \itemsep-0.5em
        \item The root node starts out with $A$ branches, one for each possible first symbol.
        \item When parsing a phrase, we traverse the tree until we reach a leaf.
        Upon reaching a leaf, we add $A$ branches to the leaf (one for each symbol in $\Acal$).
        \item We label each node with the number of leaves that are descendents of the node (including, when relevant, the node itself).
        Let us call this number $\Lcal(z)$, where $z \in \Zcal$ is the phrase corresponding to the node of interest.
        \item $q^{\text{LZ78}, \Uppi_0}(x_t|x^{t-1})$ is equal to the ratio of the label, $\Lcal$, of the current node in the LZ78 tree (after traversing the tree according to the current symbol) to that of the parent node.

        This is because every new phrase in the LZ78 parsing of a sequence removes one leaf from the tree and adds $A$ leaves (by adding $A$ branches to an additional leaf).
        So, for each phrase that a node is a part of, its label is incremented by $A-1$ .
        Also, every node except for the root (which can never be the current node), starts off with a $\Lcal = 1$, making the label of the current node equal to $(A-1) N(x_t|x^{t-1},z_c(x^{t-1})) + 1$, \ie, the numerator of $q^{\text{LZ78}, \Uppi_0}(x_t|x^{t-1})$.
        The label of the parent node is the sum of the labels of its children, \ie, the denominator of $q^{\text{LZ78}, \Uppi_0}(x_t|x^{t-1})$.
    \end{enumerate}

    For the example in \prettyref{sec:lz78-review} where $x^n$ is parsed into \texttt{0,1,10,01,100,11}, the prefix tree would be:
    \vspace{-1em}
    
    \begin{center}
        \begin{tikzpicture}[level distance=1.4cm,
          level 1/.style={sibling distance=10cm},
          level 2/.style={sibling distance=4.5cm},
          level 3/.style={sibling distance=2cm},
          level 4/.style={sibling distance=2cm}]
          \node[rectangle, fill=green!15, rounded corners, draw]{\parbox[t]{4em}{\centering\small \texttt{root} \\$\Lcal=8$}}
            child {
                node[rectangle, fill=green!15, rounded corners, draw]{\parbox[t]{4em}{\centering\small node \texttt{0} \\ (phrase 1) \\ $\Lcal=3$}}
                child {
                    node[rectangle, fill=blue!10, rounded corners, draw]{\parbox[t]{4em}{\centering\small node \texttt{00} \\ $\Lcal=1$}}
                }
                child {
                    node[rectangle, fill=green!15, rounded corners, draw]{\parbox[t]{4em}{\centering\small node \texttt{01} \\ (phrase 4) \\ $\Lcal=2$}}
                    child {
                        node[rectangle, fill=blue!10, rounded corners, draw]{\parbox[t]{4em}{\centering\small node \texttt{010} \\ $\Lcal=1$}}
                    }
                    child {
                        node[rectangle, fill=blue!10, rounded corners, draw]{\parbox[t]{4em}{\centering\small node \texttt{011} \\ $\Lcal=1$}}
                    }
                }
            }
            child {
                node[rectangle, fill=green!15, rounded corners, draw]{\parbox[t]{4em}{\centering\small node \texttt{1} \\ (phrase 2) \\ $\Lcal=5$}}
                child {
                    node[rectangle, fill=green!15, rounded corners, draw]{\parbox[t]{4em}{\centering\small node \texttt{10} \\ (phrase 3) \\ $\Lcal=3$}}
                    child {
                        node[rectangle, fill=green!15, rounded corners, draw]{\parbox[t]{4em}{\centering\small node \texttt{100} \\ (phrase 5) \\ $\Lcal=2$}}
                        child {
                            node[rectangle, fill=blue!10, rounded corners, draw]{\parbox[t]{4.3em}{\centering\small node \texttt{1000} \\ $\Lcal=1$}}
                        }
                        child {
                            node[rectangle, fill=blue!10, rounded corners, draw]{\parbox[t]{4.3em}{\centering\small node \texttt{1001} \\ $\Lcal=1$}}
                        }
                    }
                    child {
                        node[rectangle, fill=blue!10, rounded corners, draw]{\parbox[t]{4em}{\centering\small node \texttt{101} \\ $\Lcal=1$}}
                    }
                }
                child {
                    node[rectangle, fill=green!15, rounded corners, draw]{\parbox[t]{4em}{\centering\small node \texttt{11} \\ (phrase 6) \\ $\Lcal=2$}}
                    child {
                        node[rectangle, fill=blue!10, rounded corners, draw]{\parbox[t]{4em}{\centering\small node \texttt{110} \\ $\Lcal=1$}}
                    }
                    child {
                        node[rectangle, fill=blue!10, rounded corners, draw]{\parbox[t]{4em}{\centering\small node \texttt{111} \\ $\Lcal=1$}}
                    }
                }
            };
        \end{tikzpicture}
    \end{center}
    \vspace{-1em}
    Internal nodes, which correspond to phrases in the LZ78 parsing, are colored {\color{deepgreen} green}, and leaves, which do not yet correspond to phrases, are colored {\color{deepblue} blue}.
\end{construction}

 For this SPA, there is a direct and exact connection between the log loss incurred on each phrase and the number of phrases in the LZ78 parsing thus far.
 In each phrase, the log loss incurred is, up to constant terms, the logarithm of the number of phrases that have been parsed thus far, as proven in \prettyref{lem:tree-spa-phrase-loss}.

Using this, we can directly show that the log loss incurred by $q^{\text{LZ78}, \Uppi_0}$ is asymptotically equivalent to the $\frac{1}{n}$-scaled LZ78 codelength.
This is a crucial result for \prettyref{sec:lz78-log-loss-eq-codelength}, where we show that the same holds for any SPA in the LZ78 family.

\begin{lemma}[Log loss of \prettyref{con:tree-spa}]\label{lem:tree-spa-log-loss}
    For any individual sequence and $q^{\text{LZ78}, \Uppi_0}$ from \prettyref{con:tree-spa},
    \[\max_{x^n} \left|\frac{1}{n}\log \frac{1}{q^{\text{LZ78}, \Uppi_0}(x^n)} - \frac{C(x^n)\log C(x^n)}{n}\right| = \epsilon(A, n),\]
    where $\epsilon(A, n) = \bigO\!\left(\frac{(\log A)^2}{\log n}\right)$.
    
    Proof sketch (full proof in \prettyref{app:proofs-special-case-spas}):
        \textup{An upper bound, $\max_{x^n}\left(\cdots \right) = \bigO\!\left(\frac{(\log A)^2}{\log n}\right)$, can be achieved via direct computation.
        The lower bound, $\min_{x^n} \left(\cdots\right) = \bigO\!\left(\frac{\log A}{\log n}\right)$, relies on Stirling's approximation for $\log (C(x^n)!)$.
        Taking the maximum of the two bounds produces the final result.}
\end{lemma}
Note that, by Theorem 2 of \cite{originalLZ78paper}, \prettyref{lem:tree-spa-log-loss} implies that $-\frac{1}{n}\log q^{\text{LZ78}, \Uppi_0}(x^n)$ uniformly converges to $\frac{1}{n}$ times the LZ78 codelength (which we will refer to as the \textit{normalized LZ78 codelength}).

\subsection{Correspondence of LZ78 Sequential Probability Assignment Log Loss and LZ78 Codelength}\label{sec:lz78-log-loss-eq-codelength}
In this section, we prove one critical result of this paper: the asymptotic correspondence between the normalized LZ78 codelength and the log loss incurred by any SPA of the family \prettyref{con:lz78-spa}.
Specifically, we prove that the distance between the log loss and $\frac{C(x^n)\log C(x^n)}{n}$ approaches $0$ as $n\to\infty$, uniformly over individual sequences.
As, by Theorem 2 of \cite{originalLZ78paper}, the same holds for the normalized LZ78 codelength, the correspondence of the SPA log loss and the scaled codelength directly follows via the triangle inequality.

\begin{theorem}\label{thm:lz78-log-loss-codelength-correspondence}
    For any prior such that $\supp(\Uppi) = \Mcal(\Acal)$,
    \[ \lim_{n\to\infty} \max_{x^n} \left| \frac{1}{n} \log \frac{1}{q^{\text{LZ78},\Uppi}(x^n)} - \frac{C(x^n) \log C(x^n)}{n}\right| = 0.\]

    Proof sketch (full proof in \prettyref{app:lz78-log-loss-eq-codelength}):
        \textup{\prettyref{lem:tree-spa-log-loss} shows that this holds for the specific instance of the LZ78 SPA from \prettyref{con:tree-spa}, so this result reduces to showing that all SPAs in the LZ78 family have asymptotically-equivalent log losses.\footnote{\ie, that the absolute distance in log loss between any two LZ78 SPAs approaches $0$, uniformly over all individual sequences.}
        We then show that the log loss of any Bayesian mixture SPA with $\supp(\Uppi) = \Mcal(\Acal)$ approaches the empirical entropy of the input sequence, uniformly over all inputs.
        This is achieved by evaluating the integral expression in \eqref{eqn:q-with-prior-integral} over a shrinking section of the simplex, and then applying simple bounds on relative entropy.
        From there, we show that the absolute distance between the LZ78 SPA log loss and the empirical entropy of $x^n$, conditioned on LZ78 prefix, uniformly approaches $0$.
        Finally, the triangle inequality produces the desired result.}
\end{theorem}

\section{Universality of SPAs}\label{sec:universality}
\subsection{Classes of Sequential Probability Assignments and Associated Log Loss}\label{sec:markov-and-fs-spa}
In order to characterize the accuracy of the LZ78 family of SPAs, we will show that its log loss is upper-bounded by the optimal log loss of any finite-state SPA.
As a prerequisite, we must understand finite-state SPAs, their optimal log loss, and its relationship with other relevant quantities.

\begin{definition}[Finite-State SPA]
    A finite-state SPA is such that the probabilities assigned depend solely on the current state of an underlying finite-state mechanism.
    Concretely, $q$ is a $M$-state SPA if $\exists$ next-state function $g: [M] \times \Acal \to [M]$, prediction function $f: [M] \to \Mcal(\Acal)$, and initial state $s_1 \in [M]$ such that
    \[\forall t \geq 1,\, q(\cdot | x^{t-1}) = f(s_t),\quad \text{and}\quad s_{t+1} = g(s_t, x_t). \]
    The set of all $M$-state SPAs is denoted $\Fcal_M$.
\end{definition}

The optimal log loss of any finite-state SPA is defined as follows:
\begin{definition}[Optimal Finite-State Log Loss]
    The minimum log loss of a $M$-state SPA for sequence $\xv$ is defined as
    \[\lambda_M(\xv) \triangleq \limsup_{n \to \infty} \lambda_M(x^n)\quad \text{for}\quad \lambda_M(x^n) \triangleq \min_{q \in \Fcal_M} \frac{1}{n} \sum_{t=1}^n \log \frac{1}{q(x_t|x^{t-1})}.\]
    The optimal finite-state log loss takes the number of states to infinity,
    \[\lambda(\xv) \triangleq \lim_{M \to \infty} \limsup_{n \to \infty} \lambda_M(x^n).\]
     $\lambda_M(\xv)$ is monotonically non-increasing in $M$ and bounded below, so the outer limit is guaranteed to exist.
\end{definition}

As the set of finite-state SPAs is quite broad, it is often simpler to consider the family of Markov SPAs, where $q(x_t|x^{t-1})$ only depends only on a fixed-length context $x_{t-k}^{t-1}$.
In \prettyref{sec:lambda-eq-mu-eq-rho}, we will show that, although the set of finite-state SPAs is more broad than that of Markov SPAs, they are asymptotically equivalent for any individual sequence.

\begin{definition}[Markov SPA]\label{def:markov-spa}
    An SPA $q$ is Markov of order $k$ if $\exists g: \Acal^k \to \Mcal(\Acal)$ such that $q(\cdot|x^{t-1}) = g\left(x_{t-k}^{t-1}\right)$, $\forall \xv$ and $t \geq k+1$.
    The set of all $k$-order Markov SPAs is denoted $\Mcal_k$.
\end{definition}

\begin{remark}
    For a $k$-order Markov SPA, $q_{t}$, $t \leq k$ is fully arbitrary.
    For instance, if evaluating a loss function, those $q_{t}$ can be chosen to incur zero loss.
\end{remark}

\begin{remark}\label{fact:markov-finite-state-correspondence}
    If SPA $q$ is Markov of order $k$, it is an $A^k$-state SPA.
    This can be seen by defining the state as the $k$-tuple consisting of the previous $k$ symbols.
\end{remark}

The definition of the optimal Markov SPA log loss is analogous to that of the optimal finite-state SPA log loss:
\begin{definition}[Optimal Markov Log Loss]
    For any sequence $\xv$, the optimal log loss of a $k$-order Markov SPA is
    \[\mu_k(\xv) \triangleq \limsup_{n\to\infty} \mu_k(x^n) \quad\text{for}\quad \mu_k(x^n) \triangleq \min_{q \in \Mcal_k} \frac{1}{n} \sum_{t=1}^n \log \frac{1}{q(x_t | x^{t-1})}.\]
    The optimal Markov SPA log loss is defined by taking the context length to $\infty$:
    \[\mu(\xv) \triangleq \lim_{k\to\infty} \mu_k(\xv) = \lim_{k\to\infty} \limsup_{n\to\infty} \mu_k(x^n).\]
    As with $\lambda(\xv)$, the outer limit is guaranteed to exist.
\end{definition}
 
\subsubsection{Optimal Markov and Finite-State Log Loss in terms of Empirical Entropies}\label{sec:mu-lambda-entropies}

Essential to the proofs in \prettyref{sec:lambda-eq-mu-eq-rho} (and fundamental to the understanding of optimal SPAs) is the relationship between $\mu_k(x^n)$, $\lambda_m(x^n)$, and empirical entropies on individual sequence $x^n$.
The specifics of these relationships, as well as the corresponding proofs, are detailed in \prettyref{app:classes-of-spas} and summarized below.
\begin{itemize}
    \item $\mu_0(x^n)$ is equal to the zero-order empirical entropy of $x^n$.
    \item $\mu_k(x^n)$ is, up to $o(1)$ terms, equal to the empirical entropy of $x^n$, conditioned on the length-$k$ context of each symbol.
    \item $\lambda_M^{g, s_1}(x^n)$, where $g$ is a fixed state transition function and $s_1$ is a fixed initial state, is equal to the empirical entropy of $x^n$, conditioned on the current state.
\end{itemize}

\subsubsection{Finite-State Compressibility}
Also closely related to optimal SPAs is finite-state compressibility \cite{originalLZ78paper}, as, in many cases, limits of compressibility and probability assignment log loss coincide.
For instance, the entropy of an i.i.d. probability source is both the theoretical limit of compression and of the log loss for a probability assignment on that source.
In addition, in \prettyref{sec:lz78-log-loss-eq-codelength}, we showed that the log loss of any LZ78 SPA from \prettyref{con:lz78-spa} is asymptotically equivalent to the scaled LZ78 codelength.

In \prettyref{sec:lambda-eq-mu-eq-rho}, we will show that a similar result holds for finite state compressibility; \ie, that it is equal to the optimal finite-state SPA log loss (with a $\frac{1}{\log A}$ scaling factor).
For the sake of that result being self-contained, we provide definitions of finite-state compressibility and some prerequisite concepts.
For more detailed descriptions of these quantities, see \cite{originalLZ78paper}.
\begin{definition}[Encoder]
    For an individual sequence $\xv$ from alphabet $\Acal$, an encoder is a mapping of input symbols $x_t \in \Acal$ to output symbols $y_t \in \Bcal$, where $\Bcal$ is the output alphabet.
    The elements of $\Bcal$ have varying bitwidths, and $\Bcal$ can include the empty sequence.
\end{definition}

\begin{definition}[Information Lossless Finite-State Encoder]
     An $M$-state encoder consists of an initial state $s_1 \in [m]$, an encoding function $f: \Acal \times [M] \to \Bcal$, and state-transition function $g: \Acal \times [M] \to [M]$ such that
    \[y_t = f(x_t, s_t),\, s_{t+1} = g(x_t, s_t),\, \forall t \geq 1.\]

    A finite-state encoder is \textbf{information lossless} if the initial state, output signal $y^n$, and set of states $s^n$ uniquely determine the input signal $x^n$.
    Let the set of all information lossless $M$-state compressors be $\rho_M$.
\end{definition}

\begin{definition}[Compression Ratio]
    The compression ratio of an encoder on an individual sequence is $\frac{\ell(y^n)}{\ell(x^n)}$, where $\ell(\cdot)$ represents the number of bits required to directly represent a sequence.
    We take the number of bits required to represent the input sequence, $\ell(x^n)$, to be $n\log A$.
\end{definition}
\begin{definition}[Finite-State Compressibility]
   
    For any finite sequence $x^n$, the minimum $M$-state compression ratio is
    \[\rho_M(x^n) \triangleq \min_{E \in \rho_M} \frac{\ell(y^n)}{\ell(x^n)} = \min_{E \in \rho_M} \frac{\ell(y^n)}{n\log A},\]
    where $y^n$ is understood to be the output produced by applying encoder $E$ to sequence $x^n$.

    Analogous to $\lambda_M(\xv)$ and $\lambda(\xv)$, the $M$-state compressibility and finite-state compressibility of $\xv$ are, respectively,
    \begin{align*}
        \rho_M(\xv) &\triangleq \limsup_{n\to\infty} \rho_M(x^n), \text{ and }
        \rho(\xv) \triangleq \lim_{M \to\infty} \rho_M(\xv) = \lim_{M \to\infty} \limsup_{n\to\infty} \rho_M(x^n).
    \end{align*}
\end{definition}

\subsection{Equivalence of Optimal Finite-State Log Loss, Optimal Markov Log Loss, and Finite-State Compressibility}\label{sec:lambda-eq-mu-eq-rho}

One of our primary goals is to prove that the log loss of the LZ78 family of SPAs is asymptotically upper-bounded by the optimal finite-state log loss, for any individual sequence.
To do so, we would like to utilize the relationship between the LZ78 SPA log loss and $\rho(\xv)$ established by \prettyref{sec:lz78-log-loss-eq-codelength} and \cite{originalLZ78paper}.
To this extent, we consolidate a set of results that have been alluded to throughout the literature, as well as the viewpoints of individual sequence compressibility and optimal sequential probability assignment: for any individual sequence, $\lambda(\xv) = \mu(\xv) = \rho(\xv) \log A$.
Though not formally stated as a theorem, the bulk of the $\lambda(\xv) = \mu(\xv)$ result is present in the analysis of \cite{federMerhavGutman1992,merhavFeder1993SequentialDecision} (\hspace{0.1em}\cite{federMerhavGutman1992} presents this result in the context of ``probability of error'' loss, with \cite{merhavFeder1993SequentialDecision} extending it to other bounded loss functions).
In addition, much of the analysis that $\mu(\xv) = \rho(\xv) \log A$ is discussed in \cite{originalLZ78paper}, and is also alluded to in \cite{federMerhavGutman1992,weinberger1994optimalSPA}.

\begin{theorem}\label{thm:lambda-eq-mu-eq-rho}
    For any infinite individual sequence, the optimal finite-state log loss, optimal Markov SPA log loss, and finite-state compressibility are equivalent:
    \[\lambda(\xv) = \mu(\xv) = \rho(\xv) \log A.\]

    Proof sketch (full proof in \prettyref{app:proofs-lambda-eq-mu-eq-rho}):
    \textup{
        We first prove that $\lambda(\xv) = \mu(\xv)$ via a lower and upper bound \ie, $\lambda(\xv) \leq \mu(\xv)$ and $\lambda(\xv) \geq \mu(\xv)$.
        The upper bound, $\lambda(\xv) \leq \mu(\xv)$, follows directly from \prettyref{fact:markov-finite-state-correspondence}.
        To achieve the lower bound, we first replace $\mu_k(x^n)$ and $\lambda_M^{g,s_1}(x^n)$ by their corresponding empirical entropies.
        Using the fact that conditioning reduces entropy, as well as the chain rule of entropy, we can show that $\mu_k(x^n) - \lambda_M^{g, s_1}(x^n)$ is upper-bounded by $\frac{\log M}{k+1}$, regardless of the choice of state transition function or initial state.
        From here, taking $k\to\infty$, followed by $M\to\infty$, completes the proof that $\lambda(\xv) = \mu(\xv)$.
    }

    \textup{
        The result that $\rho(\xv) \log A = \mu(\xv)$ follows from Theorem 3 of \cite{originalLZ78paper}, along with an application of the chain rule of entropy and some minor further analysis.
    }
\end{theorem}

\section{Universality of the LZ78 Family of SPAs}\label{sec:lz78-spa-optimality}
Given the work thus far, it becomes simple to prove that, for any individual sequence, the limit supremum of the log loss of any LZ78 SPA is at most $\lambda(\xv)$.
\ie, in terms of log loss, the LZ78 family of SPAs either matches or outperforms any finite-state SPA.

\begin{theorem}[Universality of LZ78 SPA]\label{thm:universality}
    For prior distribution $\Uppi$ such that $\supp(\Uppi) = \Mcal(\Acal)$, $q^{\text{LZ78},\Uppi}$ from \prettyref{con:lz78-spa} satisfies,  for any individual sequence, 
    \[\limsup_{n\to\infty} \frac{1}{n}\log \frac{1}{q^{\text{LZ78},\Uppi}(x^n)} \leq \lambda(\xv).\]

    \begin{proof}
        \[\limsup_{n\to\infty} \frac{1}{n}\log \frac{1}{q^{\text{LZ78},\Uppi}(x^n)} \stackrel{(a)}{=} \limsup_{n\to\infty} \frac{1}{n}C(x^n) \log C(x^n) \stackrel{(b)}{\leq} \rho(\xv) \log A \stackrel{(c)}{=} \lambda(\xv),\]
        where (a) is implied by \prettyref{thm:lz78-log-loss-codelength-correspondence}, (b) is equation (7a) from \cite{originalLZ78paper}, and (c) is a result from \prettyref{thm:lambda-eq-mu-eq-rho}.
    \end{proof}
\end{theorem}
The inequality, in fact, can be strict (\ie, the LZ78 SPA strictly outperforms any sequence of finite-state SPAs).
In \prettyref{sec:bernoulli-lz78-source}, we define a class of sequences for which this is the case.
% \begin{remark}[Generalization of LZ78 SPA Family]
%     \todo{Once the Philosophy Transactions paper is posted on Arxiv, add a remark about the LZ transform.}
% \end{remark}
\begin{remark}\label{rem:lz-vs-ml}
    The universality result of \prettyref{thm:universality} is asymptotic and guaranteed to hold as $n\to\infty$.
    The finite-sample performance of the LZ78 SPA with respect to various finite-state SPAs may vary.
    In \cite{ding2025lzmidicompressionbasedsymbolicmusic}, the finite-sample performance is competitive for a music generation task.
    However, in other domains (\eg, natural language), LZ78 is not expected to outperform machine learning-based methods (which can be viewed as finite-state SPAs with a very large number of states).
\end{remark}

\subsection{Results for Stationary and Ergodic Probability Sources}\label{sec:stationary-ergodic-results}
If, instead of an individual sequence, we consider a stationary stochastic process $\Xv$, the expected log loss incurred by any SPA in the LZ78 family approaches the entropy rate of the source in the limit $n\to\infty$.
Specifically, the following results hold, and follow without much effort from the corresponding results in the individual sequence setting:
\begin{itemize}
    \item $\Ebb \mu_0(X^n) \leq H(X_1)$,
    \item $\Ebb \mu_k(X^n) \leq H(X_{k+1}|X_k)$,
    \item For any SPA such that the limit supremum of the log loss is at most $\mu(\xv)$, \eg, any SPA in the LZ78 family, the expected log loss approaches the entropy rate,
    \item If the process is also ergodic, then the above result holds almost surely rather than in expectation.
\end{itemize}
Details about these results can be found in \prettyref{app:stationary-and-ergodic-sources}.

\section{LZ78 SPA Implementation Details}\label{sec:implementation}
In this section, we consider the task of evaluating the LZ SPA for $x^n$ via evaluating $q^{\text{LZ78}, \Uppi}(x_1)$, $q^{\text{LZ78}, \Uppi}(x_2|x_1)$, \\$q^{\text{LZ78}, \Uppi}(x_3|x^2)$, \etc, in sequence.
In particular, when evaluating $q^{\text{LZ78}, \Uppi}(x_t|x^{t-1})$, we assume that the SPA has been evaluated for timesteps $1$ through $(t-1)$ and only those timesteps.
WLOG, we set the alphabet to be the range $\{0, \dots, A-1\}$.

For the implementation details, we take $\Uppi$ to be a Dirichlet prior with parameter $\gamma$, as per \prettyref{con:jeffreys-prior-spa}.
For the complexity analysis, we assume that the Bayesian mixture SPA $q^\Uppi(y_t|y^{t-1})$ can be computed in constant time with respect to $t$.
As $q^\Uppi$ purely depends on $N(a|y^{t-1})$, $\forall a \in \Acal$ (see \ref{eqn:q-with-prior-integral}), this is a reasonable assumption to make.

In this section, data structures are given in pseudocode and all lists are zero-indexed.
Elements of lists and maps are accessed using square brackets, and properties of objects are accessed using ``\texttt{.}'' notation

\subsection{Algorithms and Data Structures}
\textbf{Na\"ive implementation}.
To evaluate the LZ78 SPA, we need to keep track of an LZ78 prefix tree of the sequence that has been processed so far.
In addition, for each node $z$, we need to know $N_\text{LZ}(a|x^{t-1}, z)$, $\forall a \in \Acal$.
One simple way to do so is to store a tree with the following node data structure:
\begin{verbatim}
Node:
    num_times_traversed: integer
    children: HashMap[(integer between 0 and A-1)-> Node]
\end{verbatim}
\texttt{num\_times\_traversed} starts at $0$ when a node is added as a leaf, and is incremented by $1$ every time a node is traversed.
\texttt{children} represents the branches associated with the current node, and maps a symbol in the alphabet to the corresponding child node.

While traversing the tree, we keep track of a pointer to the current node of the tree, denoted $z$.
The tree starts out with a single root node, initialized with \texttt{num\_times\_traversed} $=0$ and \texttt{children} empty.
$z$ points to this node.

Computing $q^{\text{LZ78}, \Uppi}(x_t|x^{t-1})$ is as follows:
\begin{enumerate}
    \item If $x_t$ is present in the branches from the current node, $N_\text{LZ}(x_t|x^{t-1}, z)$ is equal to $z.$\texttt{children}[$x_t$]$+1$: from node $z$, we parsed the symbol $x_t$ once while adding $z.$\texttt{children}[$x_t$] as a leaf node, and $c$ times thereafter.
    Otherwise, $N_\text{LZ}(x_t|x^{t-1}, z) = 0$.

    $\sum_{b\in\Acal} N_\text{LZ}(b|x^{t-1}, z)$ is the number of times the current node has been traversed, \ie, the \texttt{num\_times\_traversed} field of $z$.
    So, for a Dirichlet prior, the SPA evaluates to
    \[q^{\text{LZ78}, \Uppi}(x_t|x^{t-1}) = \frac{N_\text{LZ}(x_t|x^{t-1}, z) + \gamma}{z.\texttt{num\_times\_traversed} + A\gamma}.\]
    For a prior that is not Dirichlet, we can calculate $N_\text{LZ}(a|x^{t-1}, z)= z.$\texttt{children}[$a$]$+1$, for all $a \in \Acal$ that are present in $z.$\texttt{children} ($N_\text{LZ}(a|x^{t-1}, z)=0$ otherwise).
    This gives us all of the information necessary to evaluate the form of the Bayesian mixture SPA in \eqref{eqn:q-with-prior-integral} using numerical integration.

    \item Increment $z.$\texttt{num\_times\_traversed} for the current node by $1$.
    \item If $z.$\texttt{children}[$x_t$] does not exist, create a new \texttt{Node} with \texttt{num\_times\_traversed} $=0$ and \texttt{children} empty. 
    Set $z.$\texttt{children}[$x_t$] to that new node, and then set $z$ to point to the root node.

    Otherwise, set $z$ to point to $z.$\texttt{children}[$z_t$].
\end{enumerate}

\textbf{Memory-efficient implementation}.
In practice, keeping a \texttt{Node} object with a small \texttt{HashMap} for each node of the LZ78 prefix tree imposes an unnecessarily-high memory overhead.\footnote{This overhead is constant per node, so it does not affect any asymptotic analysis, but it ends up being of practical concern.}
To alleviate this, we use a Lempel-Ziv-Welch \cite{lzwpaper} approach: instead keeping one object per node, we have two global data structures:
\begin{enumerate}
    \item \texttt{node\_traversed\_counts}: a list such that \texttt{node\_traversed\_counts}$[0]$ is the number of times the root has been traversed, and \texttt{node\_traversed\_counts}$[k]$ for $k \leq 1 \leq C(x^t)$ is the number of times that the node corresponding to phrase $k$ in the LZ78 parsing of $x^t$ has been traversed.

    We refer to the position of a node in the \texttt{node\_traversed\_counts} list as its ``index.'' 
    \item \texttt{branches}: a \texttt{HashMap} mapping the tuple $(\texttt{node index},\,\texttt{symbol in alphabet})$ to the index of the corresponding child node.
\end{enumerate}
When traversing the tree, we keep track of the index of current node, which we denote $k$.
To compute $q^{\text{LZ78}, \Uppi}(x_t|x^{t-1})$,
\begin{enumerate}
    \item If $(k, x_t)$ is present in \texttt{branches}, then the index of the child corresponding to $x_t$ is $k' = $\texttt{branches}[($k, x_t$)].
    Then, $N_\text{LZ}(x_t|x^{t-1}, z)=$\texttt{node\_traversed\_counts}[$k'$]$+1$, as in the na\"ive implementation.
    $\sum_{b\in\Acal} N_\text{LZ}(b|x^{t-1}, z)$ is \texttt{node\_traversed\_counts}[$k$].
    From these quantities, we can compute the SPA.
    \item Increment \texttt{node\_traversed\_counts}[$k$] by $1$.
    \item If $(k, x_t)$ is not present in \texttt{branches}, we are adding a new leaf to the LZ78 tree.
    This is achieved by setting \texttt{branches}[$(k, x_t)$] to $|$\texttt{node\_traversed\_counts}$|$, adding a $0$ to the end of \texttt{node\_traversed\_counts}, and then setting $k$ to $0$.

    Otherwise, set $k$ to $k'$ from step 1.
\end{enumerate}

\textbf{Code implementation}.
Code for the LZ78 SPA under a Dirichlet prior, implemented in \texttt{Rust}, can be found at \\\href{https://github.com/NSagan271/lz78_rust}{https://github.com/NSagan271/lz78\_rust}.

\subsection{Complexity Analysis} \label{sec:computational-analysis}
Let $\omega(A)$ be the complexity of computing the Bayesian mixture SPA under the chosen prior for alphabet size $A$, given the implementation of the LZ78 tree described above.
We assume that this computation is independent of the sequence length, which is reasonable to assume based on the form of \eqref{eqn:q-with-prior-integral}.
For a Dirichlet prior, $\omega(A) = \bigO(1)$, as the required quantities and $N_\text{LZ}(x_t|x^{t-1}, z)$ and $\sum_{b\in\Acal} N_\text{LZ}(b|x^{t-1}, z)$ can both be computed in constant time (with only a handful of operations).

\textbf{Time complexity}.
Then, computing $q^{\text{LZ78}, \Uppi}(x_t|x^{t-1})$ takes amortized $\omega(A)$ time: it requires computing the Bayesian mixture SPA for the current node, traversing the tree for one step, and possibly adding a new leaf.
All operations other than computing the Bayesian mixture SPA involve a few memory accesses, \texttt{HashMap} accesses, \texttt{if} statements, and additions, which are all amortized constant time operations.

So, the overall process of computing $q^{\text{LZ78}, \Uppi}(x^n)$ takes $\bigO(\omega(A) n)$ time, with a relatively small underlying constant factor.

\textbf{Memory complexity}.
Here, we analyze the memory-efficient implementation detailed above (though both implementations have asymptotically the same memory usage).
The \texttt{node\_traversed\_counts} list has $C(x^n)$ elements, and \texttt{branches} has $C(x^n) - 1$ elements (a new element is added for every new phrase), so the memory usage of the LZ78 SPA for $x^n$ is $\bigO\!\left(\frac{n\log A}{\log n}\right)$ (as per the asymptotic upper bound on $C(x^n)$ from \cite{originalLZ78paper}).

\textbf{Some comparisons}.
The computation (time and memory) required for the LZ78 compression algorithm is equivalent to the process of building and traversing the prefix tree in our family of SPAs.
Beyond the computation required for LZ78 compression, LZ78 family of SPAs has the additional overhead of storing the number of times each node has been traversed and of using those quantities to compute $q^{\text{LZ78}, \Uppi}(x_t|x^{t-1})$.
This results in a small constant memory overhead for each node of the LZ78 tree, and an $\omega(A)$ time overhead for each symbol in the input sequence.
LZ78 compression itself has $\bigO(n)$ time complexity and $\bigO\!\left(\frac{n\log A}{\log n}\right)$ memory complexity, whereas the LZ78 SPA family has the same memory complexity but $\bigO(\omega(A)n)$ time complexity.
If we take the alphabet size as a constant, or if $\omega(A)$ is constant, the LZ78 SPA family and LZ78 compression only differ by a constant factor.

Another significant tree-based SPA is context-tree weighting (CTW) \cite{willems1995CTW}, which is universal over the class of depth-$D$ tree probability sources (cf. \cite{willems1995CTW} for more details).
As stated in \cite{willems1995CTW}, the complexity (for a binary alphabet) is $\bigO(2^D)$ memory and $\bigO(nD)$ computation.
Unlike the LZ78 SPA family, the memory complexity is constant with respect to the sequence length; the improved memory complexity is traded off for a narrower definition of universality.
The time complexity of CTW and the LZ78 SPAs are both linear in terms of the sequence length.
For a Dirichlet prior, the LZ78 SPA is more efficient by a constant factor, as computation of a Dirichlet SPA is a subset of the computation required by each step of CTW.

Direct computation against machine-learning based SPAs is beyond the scope of this paper and are being considered in a subsequent work.
Some empirical comparisons for genomics compression and music generation can be found in \cite{omri2025genomic,ding2025lzmidicompressionbasedsymbolicmusic}.

\section{The LZ78 Sequential Probability Assignment as a Probability Source}\label{sec:lz78-probability-source}
\subsection{Motivation and Definition of LZ78 Probability Source}
While LZ78 has been studied extensively in the context of compression, sequence modeling, and other universal schemes, it has not been studied as a probability source.
Beyond intrinsic understanding, it is worthwhile to study this source due to its potential for lossy compression.
As described in \cite{merhav2022universalrandomcodingensemble}, a universal rate-distortion code can be achieved via a randomly generated codebook, where reconstruction codewords have log likelihood that scales with their LZ78 codelength.
As is further discussed in \prettyref{sec:prob-source-experiment}, the LZ78 probability source naturally generates sequences according to this distribution, so it can be a computationally-feasible way of realizing the theoretical results from \cite{merhav2022universalrandomcodingensemble}.

There are two general techniques for defining a probability source based on the LZ78 SPA, both of which are statistically equivalent but have different computational properties.

The first directly uses the perspective of the $q^\Uppi(x^n)$, the Bayesian mixture SPA of \prettyref{con:bayesian-mixture}, as the density of a process that draws $\Theta \in \Mcal(\Acal)$ according to the given prior, and then generates a sequence i.i.d. according to $\Theta$.
\begin{construction}[LZ78 Probability Source, Mixture Perspective]\label{con:lz78-source-1}
    Given prior distribution $\Uppi$, we can generate samples from the corresponding LZ78 probability source, $Q_t^{\text{LZ78},\Uppi}$, as follows:\\[-2em]
    \begin{enumerate}
        \itemsep-0.5em
        \item Generate an infinite series of random variables, $\Theta_1, \Theta_2, \cdots$, i.i.d. according to $\Uppi$.
        Computationally, this step should be performed lazily, \ie, only generating new values as they are needed for subsequent steps.
        \item Grow an LZ78 prefix tree, assigning a $\Theta$ value generated from step (1) to each node and using the $\Theta$ at the current node of the tree to generate the source.
        Concretely, this is done by repeating the following steps, starting at the root node of the prefix tree:

        \vspace{-0.5em}
        \begin{enumerate}
            \itemsep-0.5em
            \item If the current node of the LZ78 tree does not have an assigned $\Theta$, select the next value generated in step (1) and assign it to the current node.
            \item Generate the next output of $Q^{\text{LZ78}, \Uppi}$ as $X_t \sim \eta_t$, where $\eta_t$ is the value of $\Theta$ assigned to the current node.
            \item Traverse the LZ78 tree for the newly-drawn symbol $X_t$.
        \end{enumerate}
    \end{enumerate}
\end{construction}

The second formulation of an LZ78 probability source directly uses the value of the LZ78 SPA at each current timestep.
\begin{construction}[LZ78 Probability Source, SPA Perspective]\label{con:lz78-source-2}
    Given prior distribution $\Uppi$ with density $f$, we can also use the following procedure to generate samples from $Q_t^{\text{LZ78},\Uppi}$: \\[-2em]
    \begin{enumerate}
        \item Starting at the root of the LZ78 tree, repeat the following procedure:

        \vspace{-0.5em}
        \begin{enumerate}[a.]
            \itemsep-0.5em
            \item Draw $X_t$ according to the probability mass function $q^{\text{LZ78},\Uppi}(\cdot|X^{t-1})$, where $X^{t-1}$ is the realized sequence thus far.
            \item Traverse the LZ78 tree for the newly-drawn symbol $X_t$.
        \end{enumerate}
    \end{enumerate}
    This scheme also extends to any general strongly sequential SPA (\ie, one that only requires knowledge of $x^{t-1}$ to compute $q(x_t|x^{t-1})$), including those that are not based on mixture distributions over a prior.
\end{construction}

\begin{remark}
    If the prior distribution, $\Uppi$, is Dirichlet, then the formulation of the LZ78 probability source from \prettyref{con:lz78-source-2} is simple to compute \prettyref{sec:special-case-spas}.
    For a general prior, however, $q^{\text{LZ78},\Uppi}(\cdot|X^{t-1})$, the formulation of the source from \prettyref{con:lz78-source-1} can be easier evaluate and simulate.
\end{remark}

Detailed theoretical results about this source, including its entropy rate, are explored in \cite{sagan2025lz78source}.
For the purposes of this paper, we consider an extreme yet illustrative example of the source to  answer the question posed at a end of \prettyref{sec:lz78-spa-optimality} regarding existence of sequences for which the limit supremum of the LZ78 SPA log loss is strictly better (less) than $\mu(\xv)$.
\subsection{Example: Bernoulli LZ78 Probability Source}\label{sec:bernoulli-lz78-source}
We consider the binary source corresponding to $\Uppi=\mathrm{Ber}(1/2)$ (\ie, $\Theta$ is $0$ and $1$ with equal probability).
This means that each node of the LZ78 prefix tree only generates all ones or all zeros, and each new leaf has equal probability of having $\Theta=1$ or $\Theta=0$.

As a result, $Q^{\text{LZ78},\Uppi}_t$ is uniform if the context of $X_t$ is a leaf of the LZ78 tree; otherwise, each node of the LZ78 tree may only have one child, to which it must traverse each time.
It can easily be verified that the sequence realized by this probability source has the following properties: \\[-2em]
\begin{enumerate}
    \item Each phrase in the LZ78 parsing of the realized sequence is equal to the previous phrase, with one new symbol at the end that is equally likely to be $0$ or $1$.
    \item The $k$\textsuperscript{th} phrase, denoted $Z_k$, has length $\ell(C_k) = k$.
    \item As a result, the number of phrases in the realization $X^n$ is $C(X^n) = \bigO(\sqrt{n})$.
\end{enumerate}
It is worthwhile to note that these properties hold, deterministically, for any sequence that can be realized from this source.

Any LZ78 SPA of the class in \prettyref{con:lz78-spa} satisfies $\limsup_{n\to\infty}-\frac{1}{n}\log q^{\text{LZ78},\Uppi}(X^n) = 0$, for any possible realization of this source and $\supp(\Uppi) = \Mcal(\Acal)$.
However, $\mu(\Xv) = 1$ (a.s.), as we prove in \prettyref{app:bernoulli-lz78-source}.

\section{Experiments}\label{sec:experiments}
\subsection{LZ78 as a Sequential Probability Assignment}
In this section, we briefly explore the capabilities of LZ78 for sequence generation and classification.
For the generation experiment, we ``train'' an LZ78 SPA (\ie, building the prefix tree and storing the number of times each node was visited) on a provided set of training data.
Then, we generate values based on $q^{\text{LZ78}, \Uppi}$ (continuing to traverse the prefix tree for the newly-generated symbols), with some simple tricks for improving the quality of the generated text.
For the classification experiment, we train an LZ78 SPA for each distinct label, and classify test points based on the LZ78 SPA on which they achieve the smallest log loss.

In general, these methods will not produce better results than neural-network-based approaches, given the same amount of data.
However, they generally use much less compute than neural networks (\eg, they only perform a small number of mathematical operations per input datapoint, and do not require use of a GPU to run quickly).
Direct comparison of LZ78-SPA-based generation and classification to competing methods such as neural networks is beyond the scope of the paper, as are ablation studies on the choice of prior.
The experiments contained here provide preliminary examples that the LZ78 SPA can be used for such tasks with a degree of success.
Our follow-up work in \cite{omri2025genomic,ding2025lzmidicompressionbasedsymbolicmusic} shows favorable performance of the LZ78 SPA in genomics classification and music generation settings (against transformer and diffusion baselines), and \cite{omri2025genomic} additionally includes ablation studies over the family of Dirichlet priors.

\subsubsection{Text Generation}\label{sec:generation-experiment}
This experiment includes two phases: a ``training'' phase and a ``generation'' phase.
During the training phase, an LZ78 prefix tree is formed using the available training sequences (see \prettyref{sec:implementation}).
For simplicity, each character is a separate symbol, and the alphabet consists of all lowercase and uppercase English letters, digits, and common punctuation marks.
All other characters are omitted from the training data, for the sake of filtering out uncommon characters that unnecessarily increase the size of the alphabet.

Given the LZ78 prefix tree from the training phase, generation proceeds as follows:
\begin{enumerate}
    \item We start at the root of the prefix tree, and optionally traverse the tree using a provided sequence of ``seeding data.''
    For this step, and the entirety of the generation procedure, no new leaves are added to the prefix tree, nor are the counts, $N_\text{LZ}(x^n, z)$, updated.
    We keep track of the currently-traversed node, which we all the \texttt{state}.
    \item Loop:
    \begin{enumerate}
        \item Compute the sequential probability assignment (using the fixed counts, $N_\text{LZ}(x^n, z)$, from the training phase, and the current \texttt{state}), of the next symbol for all $a \in \mathcal{A}$.
        \item Denote the $k$ symbols with the highest probabilities by $\mathcal{K}$.
        Draw a new symbol for the output sequence using the probability distribution:
        \[\Pbb(X=a) = \begin{cases}
            \frac{2^{\log (q^\text{LZ78}(a|\texttt{state})) / T}}{\sum_{x\in\mathcal{K}} 2^{\log (q^\text{LZ78}(x|\texttt{state})) / T}},&a \in \mathcal{K} \\ 0,& a \not\in\mathcal{K}
        \end{cases},\]
        where $T$ is a temperature parameter.
        \item Traverse the LZ78 prefix tree using the newly-drawn symbol.
        \item If the \texttt{state} is the root of the tree or a leaf, then we don't have any useful information to be derived from the LZ78 prefix (for a leaf, the sequential probability assignment is uniform, and, for the root, the LZ78 prefix is $\emptyset$.
        In that case, we take the last $M$ characters of the generated sequence (where $M$ is a hyperparameter), and repeat step (1).
        If we are still at the root or a leaf, repeat with the last $M-1$ symbols, \etc\,
        This is a heuristic similar to the back-shift parsing of the \texttt{LZ-MS} algorithm in  \cite{begleiter2004predictionVMM}.
    \end{enumerate}
\end{enumerate}

For this experiment, we trained the LZ78 SPA, under the Jeffreys Prior (\ie, a Dirichlet prior with parameter $\frac{1}{2}$), on the \texttt{tiny-shakespeare} dataset \cite{tinyshakespeare}, which is 1 MB, the first eight partitions of the \texttt{realnewslike} segment of C4 \cite{dodge2021documentinglargewebtextcorpora}, which is approximately 500 MB, and the first 500 MB of the \texttt{tinystories} dataset \cite{eldan2023tinystoriessmalllanguagemodels}.
Training took 0.4 seconds for \texttt{tiny-shakespeare} and 6 minutes each for the C4 and \texttt{tinystories} subsets.

The results of the generation experiments are in \Cref{fig:shakespeare-gen,fig:c4-gen,fig:stories-gen}.
For each experiment, $M = 500$, $k = 5$, and $T = 0.1$.
Considering the LZ78 SPA is directly applied to text at a one-character-per-symbol level, without any pre-processing techniques like tokenization, the generated text is surprisingly high-quality.
Both examples capture the general structure and tone of their training data and consist of plausible phrases (and occasionally, sentences).
Higher-quality generated text will require more training data and perhaps more sophisticated techniques, which we will explore in future work.

\begin{figure}[htbp]
    \centering
    \begin{quote}
    \fbox{\begin{minipage}{46em}
This is the moon with the fairest charge thee stay;

Which is not so. \vskip1em

Provost:

Art thou not

That which withal; you go you to Baptista; or, but I am not Lucentio,

Red the beasts, that Warwick,

And those that runatice of his auture of our straight and will play the his profane eyes, came to see thy servant so die. \vskip1em

LUCIO:

But what lives in Signior Gremio: fools

His will I may have auture of his auture of our common good time

Unto the Tower,

Give me thy hand as come enough

And then he shall not rather with the Lord Stand be so longer see that are fond, as thou say the orace to speak brother, or oints? \vskip1em

BIONDELLO:

I may more straight and will play the his punish his convey much this leavine art thou that will not should be thus sir, there's face.

Go you to Baptista; or, lo, here all abroad in the 
    \end{minipage}}
    \end{quote}
    \caption{800 symbols generated via the LZ78 SPA, trained on the \texttt{tiny-shakespeare} dataset, seeded with the sequence ``This.''}
    \label{fig:shakespeare-gen}
\end{figure}

\begin{figure}[htbp]
    \centering
    \begin{quote}
    \fbox{\begin{minipage}{46em}
This is a story about the version of Macron said he was ``gration of the Tigers' hands of the department kept the world around us do well to start the second half.

The Wildcats last three or four days ago. In fact, I appreciate the opposition to END TO PUT I think the time to make the roster before being shot wide with the driver isn't a single speaking to the public.

There’s a lot more spared. The ranking community is just one of the many leaving these players came close to that point.

Last year, the Steelers travel to West Virginia and two more as a president who said the right to cross Russia investigation is ongoing. To put this in countries including the United States. The plant for a special permission for some Democrats say the Red Cross and the small company culture in which a bad rap, and the former president of the Virginia Tech shootings. The report found that millennials are also included.

Congo’s early results have been produced and directed by James Baker. She was the wife of 
 \end{minipage}}
    \end{quote}
    \caption{1000 symbols generated via the LZ78 SPA, trained on a subset of the \texttt{realnewslike} segment of C4, seeded with the sequence ``This.''}
    \label{fig:c4-gen}
\end{figure}

\begin{figure}
    \centering
    \begin{quote}
        \fbox{\begin{minipage}{46em}
This is my boat," he said.

Anna and Ben felt sorry for the stars"

"Can we play with the ducks. They are nice. They were happy. They forget about the card. She picked it up and showed it to her mom.

"Look, Mommy A zoo with their mom. They saw many animals, like lions, monkeys, and elephants. But they are not careful. You were just curious and asked her mom what it meant. They thought it was so cool that he wanted to be friends with them.

The moral of the story is that it's important to be kind to others. And he also learned that it's important to take care of things that are not yours. You have to ask first. And you have to be polite and ask nicely inside. They heard a loud noise.

Anna and Ben are scared. They drop the tree and the fox were playing in the park. They were both happy and brave. He thought it was fun to see the dentist was not fun. It was dangerous and silly. They said they were sorry. They said they wanted to go home.

Mom hugs them and says, "I love you, bikes and ran to the
        \end{minipage}}
    \end{quote}
    \caption{1000 symbols generated via the LZ78 SPA, trained on a subset of the \texttt{tinystories} dataset, seeded with the sequence ``This.''}\label{fig:stories-gen}
\end{figure}

\subsubsection{Image and Text Classification}
We use the LZ78 SPA for classification as follows: first, we divide the training data according to the label, or class, of each sample.
Then, considering each class independently, we concatenate the relevant samples and construct an LZ78 prefix tree.
As with the text generation experiment, we start at the root of the appropriate tree for each training sample.
This results in $c$ different prefix trees, where $c$ is the number of classes in the dataset.

To classify test samples, we compute the log loss of the sample on all $c$ of the prefix trees from the training phrase (without adding new leaves or incrementing the counts, $N_\text{LZ}(a|x^n, z)$, at each node).
The sample is classified according to the prefix tree with the smallest log loss over that sample.

Classification experiments were performed on the MNIST \cite{deng2012mnist}, Fashion-MNIST \cite{xiao2017fashionmnistnovelimagedataset}, IMDB \cite{maas2011imdb}, and Enron-Spam \cite{enronspam} datasets.
The first two datasets consist of binary and grayscale $28\times 28$ images,\footnote{MNIST consists of handwritten digits and Fashion-MNIST consists of basic articles of clothing.} respectively.
Both datasets have with 60,000 training examples and 10,000 test examples, and are divided into 10 classes.
To produce a sequence that could be fed into the LZ78 SPA, the raw pixels of the images were concatenated in row-major order.
For Fashion-MNIST, we uniformly quantized the 8-bit pixels to 2 bits, which resulted in a performance boost.
The second two datasets are text datasets, and are processed exactly like the training data in  \prettyref{sec:generation-experiment}.
The IMDB dataset consists of 50,000 ``highly polar'' movie reviews, and the Enron-Spam dataset consists of approximately 33,000 spam and non-spam emails.
Both datasets are evenly divided between training and test segments.

The results of these classification experiments are in \prettyref{tab:classification-experiments}.
As in \prettyref{sec:generation-experiment}, we used the LZ78 SPA under a Dirichlet prior with parameter $0.1$.
To achieve an accuracy boost at the cost of run time, we formed the prefix tree for each class by looping through the training set 20 times for the image datasets and 5 times for the text datasets.
Training is parallelized across the classes but otherwise unoptimized.

\begin{table}[htbp]
    \centering
    \caption{Results of classification experiments using the LZ78 SPA.}
    \begin{tabular}{c|cccc}
         Dataset & MNIST & Fashion-MNIST & IMBD & Spam Emails \\
         \midrule
         Accuracy (\%) & 75.36 & 72.16 & 75.62 & 98.12 \\
         Training Time (s) & 14 & 15 & 16 & 14 
    \end{tabular}
    \label{tab:classification-experiments}
\end{table}

\subsubsection{Discussion}
Given the same amount of data, deep learning-based approaches are generally  expected to be more accurate than the LZ78-based methods discussed here.
They, however, can be prohibitively expensive, requiring orders of magnitude more compute depending on the complexity of the network.
Though a thorough investigation comparing LZ-based generation and classification to neural networks is beyond the scope of this work, in \cite{omri2025genomic,ding2025lzmidicompressionbasedsymbolicmusic} we show that the LZ78 SPA presents a competitive trade-off between accuracy and efficiency in certain domains.

In ongoing work, we will perform broader comparisons, including against methods not based on deep learning. 
For instance, it is worthwhile to compare the classification results to Ziv-Merhav Cross Parsing \cite{zivMerhav1993CrossParsing}, which shares some similarities to  the classification setup here.
As per \cite{coutinho2010ECG,Helmer2007MeasuringTS,pratas2018primate}, Ziv-Merhav Cross Parsing has enjoyed success in several classification tasks.
For language modeling tasks, it is natural to compare against n-gram models, which use a fixed-length context window to make predictions.

\subsection{LZ78 as a Probability Source}\label{sec:prob-source-experiment}

\subsubsection{Compression via LZ78 Probability Source}
As per recent results from \cite{merhav2022universalrandomcodingensemble}, the LZ78 probability source from \prettyref{con:lz78-source-1} or \prettyref{con:lz78-source-2} can be practically useful in lossy and lossless compression.
\cite{merhav2022universalrandomcodingensemble} states that a universal lossy compressor can be constructed via a codebook of samples generated by a distribution proportional to $2^{-LZ(\hat{x}^n)}$, where $LZ(\hat{x}^n)$ is the LZ78 codelength of the reconstruction vector $\hat{x}^n$.

By \prettyref{thm:lz78-log-loss-codelength-correspondence}, for any prior bounded away from zero, the log loss incurred by $q^{\text{LZ78}, \Uppi}$ asymptotically approaches the LZ78 codelength, uniformly over all individual sequences.
In addition, by construction, the probability that sequence $x^n$ is drawn from $Q^{\text{LZ78}, \Uppi}$ from \prettyref{con:lz78-source-2} is equal to $q^{\text{LZ78}, \Uppi}(x^n)$.
Therefore, it is reasonable to expect a codebook generated via $Q^{\text{LZ78}, \Uppi}$ to have similar universality properties to the codebook from \cite{merhav2022universalrandomcodingensemble}.
Though a detailed examination of the compression properties of $Q^{\text{LZ78}, \Uppi}$ is beyond the scope of this work, we provide some promising empirical results.

For the purposes of this paper, we consider some simple yet illustrative examples: three short, highly-compressible sequences that illustrate the potential of $Q^\text{LZ}$ for sequence compression.
Though we only consider lossless compression, it is possible to extend this experiment to lossy compression, as discussed in \cite{merhav2022universalrandomcodingensemble}.
Specifically, we consider: an all-zero sequence, which has entropy of $0$, a sequence generated from the Bernoulli LZ78 probability source of \prettyref{sec:bernoulli-lz78-source}, which has an entropy rate of $0$, and an i.i.d. sequence of $\mathrm{Ber}(0.01)$, which has an entropy rate of $0.08$.

On these sequences, we perform the following experiment:
\begin{enumerate}
    \item We first generate a sequence to compress of length $k_\text{max}$, which is $140$ for the two zero-entropy-rate sequences, and $50$ for the Bernoulli sequence.\footnote{As the codebook is drawn with probability approximately proportional to $2^{-LZ(\hat{X}^n)}$, sequences with a longer codelength are slower to compress because more samples must be drawn before the sequence can be reconstructed.}
    \item Looping over $k$ from $1$ to $k_\text{max}$, inclusive:
    \begin{enumerate}
        \item We generate length-$k$ sequences from $Q^{\text{LZ78}, \Uppi}$, where $\Uppi$ is the Dirichlet prior with parameter $0.1$,\footnote{Empirically, we found that this prior works best for compressing such short, low-entropy sequences.} until we find one that matches the first $k$ symbols of the sequence from (1).
        The total number of sequences generated is denoted $n_k$.
        \item The compression ratio for this subsequence is estimated as $\frac{\log n_k}{k}$: it takes $\log n_k$ bits to represent the number of sequences to generate, assuming the encoder and decoder have a shared seed, and the original binary sequence is represented by $k$ bits.
    \end{enumerate}
\end{enumerate}
We repeat this experiment for $200$ trials, for each sequence being compressed.

\begin{figure}[htbp]
    \centering
\subfloat{
    \includegraphics[width=0.33\linewidth]{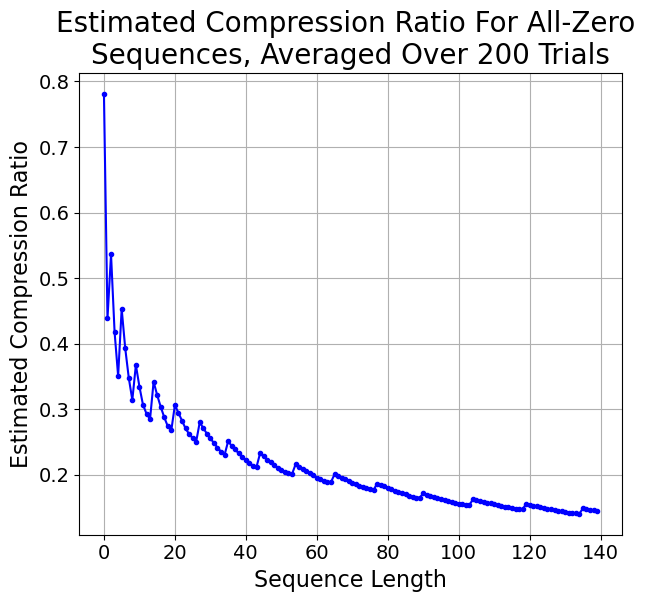}
}
\subfloat{
    \includegraphics[width=0.33\linewidth]{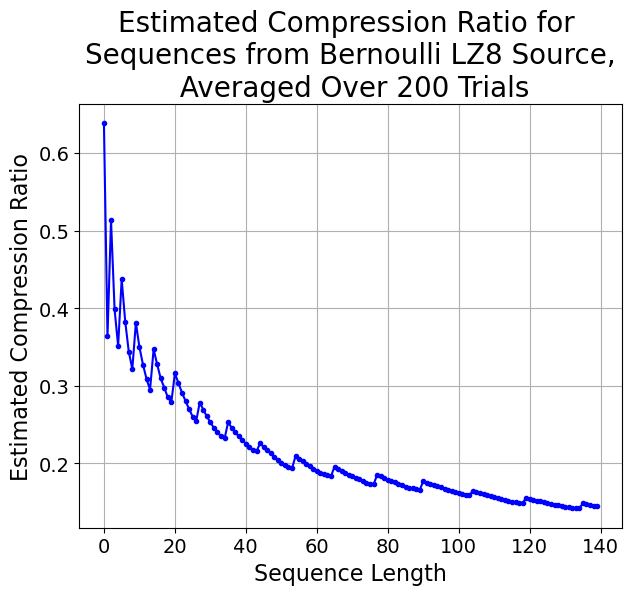}
}
\subfloat{\includegraphics[width=0.33\linewidth]{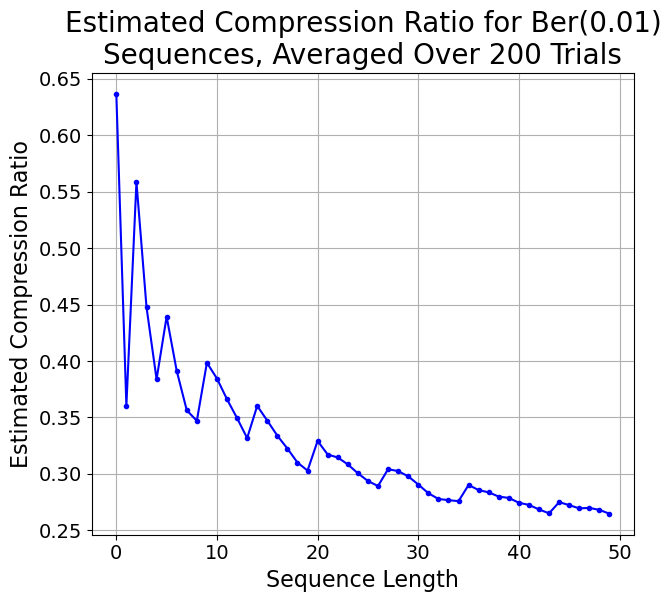}}
\caption{Compression ratios of $Q^{LZ}$ codebook compression for three simple sequences.}\label{fig:qlz-experiment}
\end{figure}

The compression ratios are plotted with respect to $k$ in \prettyref{fig:qlz-experiment}.
For comparison, LZ78 has a compression ratio of around $0.5$ at $k=140$ for the two zero-entropy-rate sequences, and compression ratios ranging from $0.7$ to $0.9$ for the length-$50$ Bernoulli sequences.

\subsubsection{Off-the-shelf Compressibility of the Bernoulli LZ78 Source}
As the entropy rate of the Bernoulli LZ78 Source from \prettyref{sec:bernoulli-lz78-source} is $0$, yet $\mu(\Xv)$ is almost surely $1$ for $\Xv$ generated from the source, it is of interest to explore how real-world compressors perform on a realization of this probability source.

The context length that is relevant for compressing this source grows indefinitely, so it is particularly well-suited to LZ78 compression, which naturally handles growing context lengths.
As we see, off-the-shelf LZ77-based compressors may or may not perform well, depending on the particular implementation.

\begin{wrapfigure}[12]{r}{0.4\textwidth}
    \centering
    \vspace{-2em}
    \includegraphics[width=0.99\linewidth,trim={0 1em 0 1em}, clip]{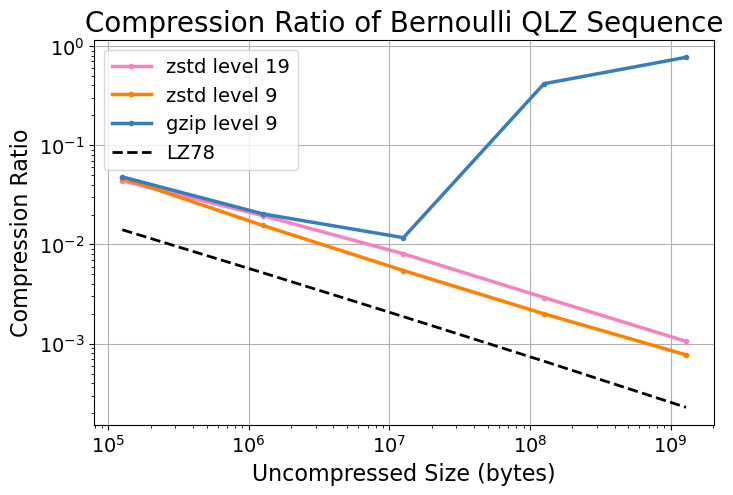}
    \vspace{-1.8em}
    \caption{Compression ratio of industry-standard for compressing different-length realizations of the Bernoulli LZ78 probability source, along with the compression ratio of LZ78.}
    \label{fig:ber-lz78-compressed}
\end{wrapfigure}

In \prettyref{fig:ber-lz78-compressed}, we compress different-length realizations of the Bernoulli LZ78 source using off-the-shelf compressors: ZSTD and GZIP.
For ZSTD, we consider level 19, which is the recommended setting for maximal compression at the cost of increased compute, and level 9, which is a moderate tradeoff between compression ratio and compute.
ZSTD does fairly well, scaling approximately with the LZ78 codelength.
Surprisingly, level 9 outperforms level 19.
The compression ratio of GZIP, however, suffers for longer sequences.
As the context length of GZIP is fixed, it can at best achieve the finite-state compressibility as the sequence length tends towards infinity.
By \prettyref{thm:lambda-eq-mu-eq-rho}, this is equal to $\mu(\Xv)$, which is almost surely $1$ for this probability source. 

\section{Conclusion}\label{sec:conclusion}
We defined a universal class of sequential probability assignments based on the LZ78 sequential parsing algorithm \cite{originalLZ78paper}.
The sequential probability assignment conditions on the LZ78 context associated with each symbol in the input sequence, and then applies a Bayesian mixture.
Under a Dirichlet prior, the sequential probability assignment becomes an additive perturbation of the empirical distribution (conditioned on LZ78 context), and, with a specific choice of Dirichlet prior, becomes the SPA from \cite{langdon1983LZNote,feder1991gambling,federMerhavGutman1992}.
We then proved that the log loss of any such LZ78 SPA converges to the normalized LZ78 codelength, uniformly over all individual sequences.
From there, we were well-situated to prove the universality of any SPA from this family, in the sense that, for any individual sequence, it achieves at most the log loss of the best finite-state SPA.  
As a prerequisite to this, we consolidated a group of results from throughout the literature: that the optimal log loss over Markovian SPAs, the optimal log loss over finite-state SPAs, and a scaled version of the finite-state compressibility are all equal.
Inspired by the LZ78 class of SPAs, we defined two equivalent formulations of a probability source that samples from an LZ78 SPA at each timestep.
Finally, we used the LZ78 SPA for text generation, image classification, and text classification, showing promise to compete with existing approaches in a compute-constrained environment.
We also performed preliminary experiments using the LZ78 probability source for text compression, as per \cite{merhav2022universalrandomcodingensemble}.

In ongoing work, we will explore the theoretical properties of the LZ78 probability source, including but not limited to its entropy rate; such results can be found in \cite{sagan2025lz78source}.
We will also thoroughly compare the capabilities of the LZ78 SPA to existing approaches in amount of data required, compute, and result quality.
Examples of this in genomics classification and music generation are presented in \cite{omri2025genomic,ding2025lzmidicompressionbasedsymbolicmusic}, and more extensive comparisons are in progress.
As an example, we will compare against the Ziv-Merhav cross-parsing \cite{zivMerhav1993CrossParsing} approach to classification.
We may improve the performance of the LZ78 SPA by incorporating ideas from Active-LeZi \cite{gopalratnam2007activeLeZi}, the LZ-based classifier from \cite{begleiter2004predictionVMM}, mixture of experts models, \etc\,
In addition, we will explore the capabilities of the LZ78 SPA and probability source for compression, via arithmetic coding \cite{rissanenLangdon1979} and the analysis in \cite{merhav2022universalrandomcodingensemble}, respectively.
All of threse directions are currently under investigation.

\section*{Acknowledgments}
Discussions with Amir Dembo, Divija Hasteer, and Andrea Montanari are acknowledged with thanks.

\newpage
{\appendices
\section{Notation Table}\label{app:notation-table}
The notation used throughout this paper is summarized in the following table:
\begin{center}
\renewcommand{\arraystretch}{1.15}
\small
    \begin{tabular}{ll}\toprule
        \multicolumn{2}{c}{\sc General Notation} \\
        \midrule
        $\Acal$ & Alphabet over which sequences take values. \\
        $A$ & Size of the alphabet, \ie, $|\Acal|$. \\
        $\xv \in \Acal^\infty$ & Infinite individual sequence $(x_1\,\, x_2\,\, \cdots)$, where $x_i \in \Acal$. \\
        $x^n \in \Acal^n$ & First $n$ symbols in infinite sequence $\xv$. \\
        $x_k^\ell$ & $(x_k\,\, x_{k+1}\,\, \cdots \,\, x_\ell)$ if $\ell \geq k$, otherwise the empty sequence. \\
        $\Acal^\ast$ & The set of all finite-length sequences over the alphabet $\Acal$ (including the empty sequence). \\
        $x^n\,^\frown y^m$ & The concatenation of $x^n$ and $y^m$. \\
        $N(a|x^n)$ & The number of times $a\in\Acal$ appears in $x^n \in \Acal^n$. \\
        $\Xv$ & Probability source $(X_1\,\,X_2\,\,\cdots)$, where $X_i$ is a random variable over $\Acal$. \\
        $\Mcal(\Acal)$ & The simplex of probability mass functions over alphabet $\Acal$. \\
        $\Theta[a]$ & For $\Theta \in \Mcal(\Acal)$ and $a \in \Acal$, this is the probability that PMF $\Theta$ assigns to $a$. \\
        $\indic{\text{Event}}$ & Indicator function for an event. \\
        % $\#\{E(i)\}_{i \in \text{Set}}$ & For set of events $E(i)$, indexed by $i \in \text{Set}$, this counts the number of times that $E(i) = 1$.\\
        \midrule
        \multicolumn{2}{c}{\sc Entropies} \\
        \midrule
        $H(X)$ & Shannon entropy of random variable $X$. \\
        $H(X|Y)$ & Conditional entropy of $X$ given $Y$, for jointly-distributed random variables $X$ and $Y$. \\
        $H_0(x^n)$ & Zero-order empirical entropy of sequence $x^n$. \\
        $\mathbb{H}(\Xv)$ & Entropy rate of stationary stochastic process $\Xv$. \\
        $D(P\lVert Q)$ & Relative entropy between distributions $P$ and $Q$. \\ 
        \midrule
        \multicolumn{2}{c}{\sc Sequential Probability Assignments} \\
        \midrule
        $q$ & Generic sequential probability assignment. \\
        $q_t(x_t|x^{t-1})$ & Sequential probability assignment for timestep $t$, also denoted $q(x_t|x^{t-1})$. \\
        $\gamma$ & Commonly, the parameter of a Dirichlet$(\gamma, \dots, \gamma)$ prior. \\
        \midrule
        \multicolumn{2}{c}{\sc Lempel-Ziv 78} \\
        \midrule
        $\Zcal(x^t)$ & Set of all LZ78 phrases in the parsing of $x^t$, \ie, nodes in the LZ78 prefix tree. \\
        $C(x^n)$ & Number of LZ78 phrases $C(x^n) = |\Zcal(x^n)|$. \\
        $Z_k = Z_k(x^n)$ & $k$\textsuperscript{th} phrase in the LZ78 parsing of $x^n$. \\ 
        $z \in \mathcal{Z}(x^n)$ & An LZ78 context, or node of the LZ78 prefix tree. \\
        $z_c(x^{t-1})$ & \parbox[t][][t]{\tablewidth}{The LZ78 context of $x_t$, \ie, the length-$(t-1)$ prefix of the phrase that $x_t$ belongs to, or the current node of the LZ78 tree when parsing $x_t$. } \\
         $\mathcal{Y}\{x^n, z\}$ & The ordered subsequence of $x^n$ that has LZ78 context $z$.\\
         $N_\text{LZ}(a|x^t, z)$ & The number of times that symbol $a$ appears in $\mathcal{Y}\{x^t, z\}$. \\
         $N_\text{LZ}(x^t, z)$ & The number of LZ78 phrases in $\Zcal(x^n)$ that start with $z$. \\
         \midrule
        \multicolumn{2}{c}{\sc LZ78 Sequential Probability Assignment} \\
        \midrule
        $q^{\Uppi}(x^n)$ & Probability assigned to sequence $x^n$ by a Bayesian mixture under prior $\Uppi$. \\
        $q^{\Uppi}(x_t|x^{t-1})$ & Bayesian mixture SPA under prior $\Uppi$. \\
        $q^{\text{LZ78},\Uppi}$ & LZ78 SPA with prior $\Uppi$. \\
        $q^{\text{LZ78}, \Uppi_0}$ & LZ78 SPA under a Dirichlet prior with parameter $\frac{1}{A-1}$. \\
        $\mathcal{L}(z)$ & Number of leaves below node $z$ of the modified tree formulation in \ref{con:tree-spa}. \\
        \midrule
        \multicolumn{2}{c}{\sc Fundamental Limits} \\
        \midrule
        $\Fcal_M$ & Set of all $M$-state SPAs. \\
        $\lambda_M(x^n)$ & Optimal log loss of any $M$-state SPA on sequence $x^n$. \\
        $\lambda_M(\xv)$ & Limit supremum of $\lambda_M(x^)$ as $\to\infty$. \\
        $\lambda(\xv)$ & Optimal finite-state SPA log loss on $\xv$; limit of $\lambda_M(\xv)$ as $M\to\infty$. \\
        $\Mcal_k$ & Set of all $k$-order Markovian SPAs. \\
        $\mu(\xv)$ & Optimal Markov SPA log loss. $\mu_k(x^n)$ and $\mu_k(\xv)$ are defined analogously to their finite-state counterparts. \\
    \end{tabular}
    
    (Table continued on next page)
\end{center}    
\begin{center}
    \small
    \begin{tabular}{ll}
        $\lambda_M^{g,s_1}(x^n)$ & Optimal log loss for $M$-state SPAs with fixed initial state $s_1$ and state transition function $g$. \\
            $\mathcal{P}_M$ & Set of all information lossless $M$-state encoders. \\
        $\rho(\xv)$ & Finite-state compressibility of $\xv$, with $\rho_M(\xv)$ and $\rho_M(x^n)$ defined analogously to the finite-state SPA case. \\
        \midrule
        \multicolumn{2}{c}{\sc LZ78 Probability Source} \\
        \midrule
        $Q^{\text{LZ78},\Uppi}$ & LZ78 probability source for prior $\Uppi$. \\
        \bottomrule
    \end{tabular}
\end{center}

\section{LZ778 Algorithmic Description}\label{app:lz78-algorithm}

The LZ78 compression algorithm is described in \prettyref{alg:lz78}.
\begin{algorithm}[h]
\caption{LZ78 Compression Algorithm}\label{alg:lz78}
\begin{algorithmic}[1]
\STATE $\Zcal \gets (())$ \COMMENT{List of phrases seen so far, represented as a prefix tree. It starts off with just the empty phrase.}
\STATE Output $\gets ()$ \COMMENT{Compression output}
\STATE $t' \gets 1$ \COMMENT{Start of the current phrase}
\STATE $k \gets 0$ \COMMENT{How many phrases we've seen so far} 
\WHILE{$t \leq n$}
    \STATE $t \gets$ smallest index such that $x_{t'}^t \not\in \Zcal$, or $n$ if no such index exists \COMMENT{End of the current phrase}
    \STATE \COMMENT{\textit{The phrase $x_{t'}^t$ now has the prefix $x_{t'}^{t-1}$, which is  $\in \Zcal$, followed by one new character, $x_t$.}}
    \STATE $i \gets$ the index of $\Zcal$ where you can find the prefix $x_{t'}^{t-1}$
    \STATE Output $\gets$ Output $^\frown (i)$ %\COMMENT{$\frown$ represents concatenation.}
    \STATE Output $\gets$ Output $^\frown$ $\left(x_t\right)$ \COMMENT{Add the last symbol of $x_{t'}^t$ to the compression output}
    \STATE $\Zcal(k+1) \gets x_{t'}^t$ \COMMENT{Add the new phrase to the list of phrases in the parsing}
    \STATE $t' \gets t + 1$
    \STATE $k \gets k + 1$
\ENDWHILE
\end{algorithmic}
\end{algorithm}

\section{LZ78 SPA: Parsing Example}\label{app:lz78-spa-examples}
We provide an example of building and evaluating the LZ78 SPA, for a sequence over the alphabet of DNA nucleotides ($\mathcal{A} = \{\texttt{A, C, G, T}\}$).
We will evaluate the LZ78 SPA under a Dirichlet prior with parameter $\frac{1}{2}$, also known as the Jeffreys prior.
Under this prior, for an alphabet size of $4$, the SPA is
\[q^{\text{LZ78}, \Uppi}(a|x^{t-1}) = \frac{2N_\text{LZ}(a|x^{t-1}, z_c(x^{t-1})) + 1}{2\sum_{b\in\Acal} N_\text{LZ}(b|x^{t-1}, z_c(x^{t-1})) + 4}.\]

To evaluate the SPA, we build an LZ78 prefix tree, and keep track of $N_\text{LZ}(a|z, x^n), \forall a \in \Acal$, for each node, $z$, of the tree.
Say we have already parsed $x^{16} = \texttt{A C A G T A C A C C A G A C A C}.$
This produces the following tree:
\begin{center}
    \begin{tikzpicture}[
    level distance=1.6cm,
      level 1/.style={sibling distance=4cm},
      level 2/.style={sibling distance=2cm},
      level 3/.style={sibling distance=1.8cm},
      level 4/.style={sibling distance=1.5cm}
      ]
      \node[rectangle, fill=yellow!25, rounded corners, draw]{\parbox[t]{4em}{\centering\small root \\[0.5em] \addtolength{\tabcolsep}{-0.5em} \begin{tabular}{cccc} A & C & G & T \\ 5 & 3 & 0 & 1 \end{tabular} }}
        child {
            node[rectangle, fill=gray!5, rounded corners, draw]{\parbox[t]{4em}{\centering\small \texttt{A} (phr. 1) \\[0.5em] \addtolength{\tabcolsep}{-0.5em} \begin{tabular}{cccc} A & C & G & T \\ 0 & 2 & 2 & 0 \end{tabular} }}
            child {
                node[rectangle, fill=gray!5, rounded corners, draw]{\parbox[t]{4.1em}{\centering\small \texttt{AC} (phr. 5) \\[0.5em] \addtolength{\tabcolsep}{-0.5em} \begin{tabular}{cccc} A & C & G & T \\ 0 & 1 & 0 & 0 \end{tabular} }}
                child {
                    node[rectangle, fill=gray!5, rounded corners, draw]{\parbox[t]{4.7em}{\centering\small \texttt{ACC} (phr. 6) }}
                }
            }
            child {
                node[rectangle, fill=gray!5, rounded corners, draw]{\parbox[t]{4.1em}{\centering\small \texttt{AG} (phr. 3) \\[0.5em] \addtolength{\tabcolsep}{-0.5em} \begin{tabular}{cccc} A & C & G & T \\ 1 & 0 & 0 & 0 \end{tabular} }}
                child {
                    node[rectangle, fill=gray!5, rounded corners, draw]{\parbox[t]{4.7em}{\centering\small \texttt{AGA} (phr. 7) }}
                }
            }
        }
        child {
            node[rectangle, fill=green!15, rounded corners, draw]{\parbox[t]{4em}{\centering\small \texttt{C} (phr. 2) \\[0.5em] \addtolength{\tabcolsep}{-0.5em} \begin{tabular}{cccc} A & C & G & T \\ 1 & 0 & 0 & 0 \end{tabular} }}
            child {
                node[rectangle, fill=gray!5, rounded corners, draw]{\parbox[t]{4.1em}{\centering\small \texttt{CA} (phr. 8) }}
            }
        }
        child {
            node[rectangle, fill=gray!5, rounded corners, draw]{\parbox[t]{4em}{\centering\small \texttt{T} (phr. 4) }}
        };
    \end{tikzpicture}
\end{center}

The table at each node represents $N_\text{LZ}(a|z, x^n), \forall a \in \Acal$, for the given node.
This table is omitted for the leaves, as all entries would be zero.
The node currently being traversed is highlighted in green.

Now, we evaluate the SPA for the next four symbols, $x_{17}^{20} = \texttt{A C A G}$.

\textbf{Timestep 17}: We are at node \texttt{C}.
While traversing this node previously, we have seen one \texttt{A} and no other symbols.
So,
\[q^{\text{LZ78}, \Uppi}(A|x^{t-1}) = \frac{2+1}{2 + 4} = \frac{1}{2};\quad q^{\text{LZ78}, \Uppi}(C|x^{t-1})=q^{\text{LZ78}, \Uppi}(G|x^{t-1})=q^{\text{LZ78}, \Uppi}(T|x^{t-1})=\frac{1}{6}.\]
The current symbol being parsed is $x_{17} = \texttt{A}$. So, we increment the count for \texttt{A} at the current node (from $1$ to $2$) and traverse from \texttt{C} to \texttt{CA}.

\textbf{Timestep 18}: We are at node \texttt{CA}.
This is a leaf, so the SPA is a uniform distribution
\[q^{\text{LZ78}, \Uppi}(a|x^{t-1}) = \frac{1}{4},\quad \forall a \in \Acal.\]
The current symbol being parsed is $x_{18} = \texttt{C}$.
We increment the count for \texttt{C} at the current node by 1, and add a new leaf, \texttt{CAC}.
Since we just added a new leaf, we return to the root.

\begin{center}
    \begin{tikzpicture}[
    level distance=1.6cm,
      level 1/.style={sibling distance=4cm},
      level 2/.style={sibling distance=2cm},
      level 3/.style={sibling distance=1.8cm},
      level 4/.style={sibling distance=1.5cm}
      ]
      \node[rectangle, fill=green!15, rounded corners, draw]{\parbox[t]{4em}{\centering\small root \\[0.5em] \addtolength{\tabcolsep}{-0.5em} \begin{tabular}{cccc} A & C & G & T \\ 5 & 3 & 0 & 1 \end{tabular} }}
        child {
            node[rectangle, fill=gray!5, rounded corners, draw]{\parbox[t]{4em}{\centering\small \texttt{A} (phr. 1) \\[0.5em] \addtolength{\tabcolsep}{-0.5em} \begin{tabular}{cccc} A & C & G & T \\ 0 & 2 & 2 & 0 \end{tabular} }}
            child {
                node[rectangle, fill=gray!5, rounded corners, draw]{\parbox[t]{4.1em}{\centering\small \texttt{AC} (phr. 5) \\[0.5em] \addtolength{\tabcolsep}{-0.5em} \begin{tabular}{cccc} A & C & G & T \\ 0 & 1 & 0 & 0 \end{tabular} }}
                child {
                    node[rectangle, fill=gray!5, rounded corners, draw]{\parbox[t]{4.7em}{\centering\small \texttt{ACC} (phr. 6) }}
                }
            }
            child {
                node[rectangle, fill=gray!5, rounded corners, draw]{\parbox[t]{4.1em}{\centering\small \texttt{AG} (phr. 3) \\[0.5em] \addtolength{\tabcolsep}{-0.5em} \begin{tabular}{cccc} A & C & G & T \\ 1 & 0 & 0 & 0 \end{tabular} }}
                child {
                    node[rectangle, fill=gray!5, rounded corners, draw]{\parbox[t]{4.7em}{\centering\small \texttt{AGA} (phr. 7) }}
                }
            }
        }
        child {
            node[rectangle, fill=gray!5, rounded corners, draw]{\parbox[t]{4em}{\centering\small \texttt{C} (phr. 2) \\[0.5em] \addtolength{\tabcolsep}{-0.5em} \begin{tabular}{cccc} A & C & G & T \\ 2 & 0 & 0 & 0 \end{tabular} }}
            child {
                node[rectangle, fill=gray!5, rounded corners, draw]{\parbox[t]{4.1em}{\centering\small \texttt{CA} (phr. 8) \\[0.5em] \addtolength{\tabcolsep}{-0.5em} \begin{tabular}{cccc} A & C & G & T \\ 0 & 1 & 0 & 0 \end{tabular} }}
                child {
                    node[rectangle, fill=gray!5, rounded corners, draw]{\parbox[t]{4.7em}{\centering\small \texttt{CAC} (phr. 9) }}
                }
            }
        }
        child {
            node[rectangle, fill=gray!5, rounded corners, draw]{\parbox[t]{4em}{\centering\small \texttt{T} (phr. 4) }}
        };
    \end{tikzpicture}
\end{center}

\textbf{Timestep 19}: We are at the root.
Based on the counts table displayed above, the SPA evaluates to
\[q^{\text{LZ78}, \Uppi}(A|x^{t-1}) = \frac{2\cdot 5+1}{2\cdot 9 + 4} = \frac{1}{2};\quad q^{\text{LZ78}, \Uppi}(C|x^{t-1})= \frac{7}{22};\quad q^{\text{LZ78}, \Uppi}(G|x^{t-1})= \frac{1}{22};\quad q^{\text{LZ78}, \Uppi}(T|x^{t-1})=\frac{3}{22}.\]
We are parsing $x_{19} = \texttt{A}$, so we increment the count for \texttt{A} at the root from $5$ to $6$ and traverse to $A$.

\textbf{Timestep 20}: We are at node \texttt{A}.
At this node, we have seen \texttt{C} and \texttt{G} twice, and not seen \texttt{A} or \texttt{T}.
So,
\[q^{\text{LZ78}, \Uppi}(C|x^{t-1}) = q^{\text{LZ78}, \Uppi}(G|x^{t-1}) = \frac{2\cdot 2+1}{2\cdot 4 + 4} = \frac{5}{12};\quad q^{\text{LZ78}, \Uppi}(A|x^{t-1})= q^{\text{LZ78}, \Uppi}(T|x^{t-1}) = \frac{1}{12}.\]
We are parsing $x_{20} = \texttt{G}$, so we increment the count for \texttt{G} at node \texttt{A} from $2$ to $3$ and traverse to \texttt{AG}.

After parsing $x_{20}$, we are at node \texttt{AG}, and the LZ78 prefix tree looks like:

\begin{center}
    \begin{tikzpicture}[
    level distance=1.6cm,
      level 1/.style={sibling distance=4cm},
      level 2/.style={sibling distance=2cm},
      level 3/.style={sibling distance=1.8cm},
      level 4/.style={sibling distance=1.5cm}
      ]
      \node[rectangle, fill=yellow!25, rounded corners, draw]{\parbox[t]{4em}{\centering\small root \\[0.5em] \addtolength{\tabcolsep}{-0.5em} \begin{tabular}{cccc} A & C & G & T \\ 6 & 3 & 0 & 1 \end{tabular} }}
        child {
            node[rectangle, fill=gray!5, rounded corners, draw]{\parbox[t]{4em}{\centering\small \texttt{A} (phr. 1) \\[0.5em] \addtolength{\tabcolsep}{-0.5em} \begin{tabular}{cccc} A & C & G & T \\ 0 & 2 & 3 & 0 \end{tabular} }}
            child {
                node[rectangle, fill=green!15, rounded corners, draw]{\parbox[t]{4.1em}{\centering\small \texttt{AC} (phr. 5) \\[0.5em] \addtolength{\tabcolsep}{-0.5em} \begin{tabular}{cccc} A & C & G & T \\ 0 & 1 & 0 & 0 \end{tabular} }}
                child {
                    node[rectangle, fill=gray!5, rounded corners, draw]{\parbox[t]{4.7em}{\centering\small \texttt{ACC} (phr. 6) }}
                }
            }
            child {
                node[rectangle, fill=gray!5, rounded corners, draw]{\parbox[t]{4.1em}{\centering\small \texttt{AG} (phr. 3) \\[0.5em] \addtolength{\tabcolsep}{-0.5em} \begin{tabular}{cccc} A & C & G & T \\ 1 & 0 & 0 & 0 \end{tabular} }}
                child {
                    node[rectangle, fill=gray!5, rounded corners, draw]{\parbox[t]{4.7em}{\centering\small \texttt{AGA} (phr. 7) }}
                }
            }
        }
        child {
            node[rectangle, fill=gray!5, rounded corners, draw]{\parbox[t]{4em}{\centering\small \texttt{C} (phr. 2) \\[0.5em] \addtolength{\tabcolsep}{-0.5em} \begin{tabular}{cccc} A & C & G & T \\ 2 & 0 & 0 & 0 \end{tabular} }}
            child {
                node[rectangle, fill=gray!5, rounded corners, draw]{\parbox[t]{4.1em}{\centering\small \texttt{CA} (phr. 8) \\[0.5em] \addtolength{\tabcolsep}{-0.5em} \begin{tabular}{cccc} A & C & G & T \\ 0 & 1 & 0 & 0 \end{tabular} }}
                child {
                    node[rectangle, fill=gray!5, rounded corners, draw]{\parbox[t]{4.7em}{\centering\small \texttt{CAC} (phr. 9) }}
                }
            }
        }
        child {
            node[rectangle, fill=gray!5, rounded corners, draw]{\parbox[t]{4em}{\centering\small \texttt{T} (phr. 4) }}
        };
    \end{tikzpicture}
\end{center}

For a more details on the implementation of the LZ78 SPA, see \prettyref{sec:implementation}.
Note that the implementation discussed in that section does not store $N_\text{LZ}(a|x^{t-1}, z)$, $\forall a\in\Acal, z \in \Zcal(x^t)$, as this is not the most efficient implementation.
\prettyref{sec:implementation} uses an equivalent but more memory-efficient data structure.

\section{Proofs: The LZ78 Family of Sequential Probability Assignments}\label{app:lz78-spa}
\subsection{LZ78 Compression Algorithm}\label{app:proofs-lz78-compression}
\begin{lemma} \label{lem:phrase-length-infty}
    For any individual sequence $\xv$, the length of the $k$\textsuperscript{th} phrase, denoted $\ell_k$, grows unbounded. \ie, $\ell_k \to \infty$.
    \begin{proof}
        For contradiction, assume that $\lim_{k\to\infty} \ell_k \neq \infty$.
        This means that $\exists M < \infty$ s.t., $\forall k' > 0$, $\exists k > k'$ where $\ell_k \leq M$.
        As a consequence, there are infinitely many phrases such that $\ell_k \leq M$ (otherwise, we could set $k'$ to be the last phrase with $\ell_k \leq M$).
        This is impossible because there are at most $\sum_{i=1}^M A^i \leq MA^M$ phrases with length $\leq M$.
    \end{proof}
\end{lemma}

\subsection{Special Cases of the LZ78 Sequential Probability Assignment}\label{app:proofs-special-case-spas}
The following theorems concern properties of the log loss of the particular LZ78-based SPA in \prettyref{con:tree-spa}.
\begin{lemma}\label{lem:tree-spa-phrase-loss}
For any full phrase in the LZ78 parsing of $x^n$, \ie, $\alpha \triangleq x_{t_0+1}^{t_1}$ s.t. $z(x^{t_1}) = \alpha$ and $z_c(x^{t_1}) = \emptyset$, the log loss incurred is
\[\log \frac{1}{q^{\text{LZ78}, \Uppi_0}(x_{t_0+1}^{t_1}|x^{t_0})} \leq \log((A-1)C(x^{t_0}) + A),\]
with equality for all phrases $\alpha$ in the LZ78 parsing, except perhaps the last one.

    \begin{proof}
        Consider one full phrase that starts at $t_0$, and denote the nodes of the prefix tree visited during that phrase $z_0, z_1, \dots, z_m$.
        The first node, $z_0$ is always the root, and $z_m$ is a leaf unless $\alpha$ is the last phrase in the parsing of $x^n$.
        
        The log loss incurred by this phrase is:
        \[\log \frac{1}{q^{\text{LZ78}, \Uppi_0}(x_{t_0+1}^{t_1}|x^{t_0})} = \log \left(\frac{\Lcal(z_0)}{\Lcal(z_1)} \cdot \frac{\Lcal(z_1)}{\Lcal(z_2)} \cdots \frac{\Lcal(z_{m-1})}{\Lcal(z_m)}\right) = \log \frac{\Lcal(z_0)}{\Lcal(z_m)}.\]
        $\Lcal(z_m) \geq 1$, with equality if $z_m$ is a leaf itself, and $\Lcal(z_0)$, the number of leaves in the whole tree, is $(A-1) C(x^{t_0}) + A$.
        This is because the tree starts with $A$ leaves, and each phrase removes one leaf (a node that was once a leaf now has children) and adds $A$ new leaves (by construction).

        Plugging these values in,
        \[\log \frac{1}{q^{\text{LZ78}, \Uppi_0}(x_{t_0+1}^{t_1}|x^{t_0})} \leq \log \left((A-1) C(x^{t_0}) + A\right),\]
        with equality if $z_m$ is a leaf, which must hold except for the last phrase of the LZ78 parsing.
    \end{proof}
\end{lemma}

\textbf{Lemma \ref{lem:tree-spa-log-loss}} (Log loss of \prettyref{con:tree-spa}).
    For any individual sequence and $q^{\text{LZ78}, \Uppi_0}$ from \prettyref{con:tree-spa},
    \[\max_{x^n} \left|\frac{1}{n}\log \frac{1}{q^{\text{LZ78}, \Uppi_0}(x^n)} - \frac{C(x^n)\log C(x^n)}{n}\right| = \epsilon(A, n),\]
    where $\epsilon(A, n) = \bigO\!\left(\frac{(\log A)^2}{\log n}\right)$.
    By Theorem 2 of \cite{originalLZ78paper}, this means that $-\frac{1}{n}\log q^{\text{LZ78}, \Uppi_0}(x^n)$ uniformly converges to the normalized LZ78 codelength.
    \begin{proof}
         We will first prove an upper bound on $\max_{x^n} \left(\frac{1}{n}\log \frac{1}{q^{\text{LZ78}, \Uppi_0}(x^n)} - \frac{C(x^n)\log C(x^n)}{n}\right)$, followed by a lower bound on $\min_{x^n} \left(\frac{1}{n}\log \frac{1}{q^{\text{LZ78}, \Uppi_0}(x^n)} - \frac{C(x^n)\log C(x^n)}{n} \right)$.
        
        \textbf{Upper Bound}: As a consequence of \prettyref{lem:tree-spa-phrase-loss}, the $\ell$\textsuperscript{th} phrase has scaled log loss $\leq \frac{1}{n}\log((A-1)\ell + A)$, with equality for all but potentially the last phrase.
        \begin{equation}
            \frac{1}{n} \sum_{\ell=1}^{C(x^n)-1}\log((A-1)\ell + A) \leq \frac{1}{n}\log \frac{1}{q^{\text{LZ78}, \Uppi_0}(x^n)} \leq \frac{1}{n} \sum_{\ell=1}^{C(x^n)}\log((A-1)\ell + A).\label{eqn:upper-lower-bound-tree-spa-log-loss}
        \end{equation}
        
        The upper bound of \prettyref{eqn:upper-lower-bound-tree-spa-log-loss} evaluates to
        \begin{align*}
            \frac{1}{n}\log \frac{1}{q^{\text{LZ78}, \Uppi_0}(x^n)} &\leq \frac{1}{n} \sum_{\ell=1}^{C(x^n)}\log((A-1)\ell + A) \leq \frac{1}{n} \sum_{\ell=1}^{C(x^n)} \log(2A\ell) \\
            &= C(x^n)\frac{\log(A) + 1}{n} + \frac{1}{n}\sum_{\ell=1}^{C(x^n)} \log(\ell) \leq C(x^n)\frac{\log A + 1}{n} + \frac{C(x^n)\log C(x^n)}{n}.
        \end{align*}
         By \cite{originalLZ78paper}, the number of LZ78 phrases satisfies $\max_{x^n} \frac{C(x^n)}{n} \leq C_1 \frac{\log A}{\log n}$, where $C_1$ is a universal constant.
        Thus,
        \begin{align*}
            \max_{x^n} \left(\frac{1}{n}\log \frac{1}{q^{\text{LZ78}, \Uppi_0}(x^n)} - \frac{C(x^n)\log C(x^n)}{n}\right) \leq \bigO\!\left(\frac{(\log A)^2}{\log n}\right).
        \end{align*}

        \textbf{Lower Bound}: The lower bound of \prettyref{eqn:upper-lower-bound-tree-spa-log-loss} simplifies to
        \begin{align*}
            \frac{1}{n}\log \frac{1}{q^{\text{LZ78}, \Uppi_0}(x^n)} &\geq \frac{1}{n} \sum_{\ell=1}^{C(x^n)-1}\log((A-1)\ell + A) \geq \frac{1}{n} \sum_{\ell=1}^{C(x^n)-1}\log(\ell) \\
            &= \frac{1}{n}\sum_{\ell=1}^{C(x^n)}\log(\ell) - \frac{1}{n}\log C(x^n) \geq \frac{1}{n}\sum_{\ell=1}^{C(x^n)}\log(\ell) - \frac{\log ( C_1 n\log A/\log n)}{n},
        \end{align*}
        where $C_1$ is a universal constant.
        As the last term decays faster than the desired $\epsilon(A, n)$, we focus on bounding 
        $\frac{1}{n}\sum_{\ell=1}^{C(x^n)}\log(\ell)$.      

        By Stirling's approximation,
        \begin{align*}
            \frac{1}{n}\sum_{\ell=1}^{C(x^n)}\log(\ell) &= \frac{1}{n}\log(C(x^n)!) \geq \frac{1}{n} C(x^n) \log C(x^n) - \frac{1}{n} \bigO(C(x^n))
            % &\geq \frac{1}{n}\sum_{\ell=C(x^n)/\log n}^{C(x^n)}\log(\ell) \geq \frac{1}{n} \left(1-\frac{1}{\log n}\right) C(x^n)(\log C(x^n) - \log \log n) \\
            % &\geq \frac{1}{n}C(x^n) \log C(x^n) - \frac{1}{n \log n} C(x^n)\log C(x^n) - \frac{1}{n}C(x^n) \log \log n.
        \end{align*}
        Using the bound $C(x^n) \leq C_1 \frac{n\log A}{\log n}$,
        \begin{align*}
            \frac{1}{n}\sum_{\ell=1}^{C(x^n)}\log(\ell) &\geq \frac{1}{n}C(x^n) \log C(x^n) - \bigO \left( \frac{\log A}{\log n}\right).
        \end{align*}
        Thus, 
        \begin{align*}
            \min_{x^n} \left(\frac{1}{n}\log \frac{1}{q^{\text{LZ78}, \Uppi_0}(x^n)} - \frac{C(x^n) \log C(x^n)}{n}\right) &\geq -\bigO \left( \frac{\log A}{\log n}\right) - \frac{\log ( C_1 n\log A/\log n)}{n}
            = \bigO\!\left(\frac{\log A}{\log n}\right).
        \end{align*}
        Taking the maximum of the lower and upper bounds produces the result of this lemma.        
    \end{proof}

\subsection{Correspondence of LZ78 Sequential Probability Assignment Log Loss and LZ78 Codelength}\label{app:lz78-log-loss-eq-codelength}
We wish to show that the sequential probability assignment log loss incurred by any SPA in the family \prettyref{con:lz78-spa}\footnote{with a prior bounded away from $0$} approaches the normalized (\ie, $\frac{1}{n}$-scaled) LZ78 codelength, uniformly over all individual sequences.
As \cite{originalLZ78paper} proves that the distance between the scaled LZ78 codelength and $\frac{C(x^n)\log C(x^n)}{n}$ uniformly converges to $0$, it suffices to show that the distance between the log loss and $\frac{C(x^n)\log C(x^n)}{n}$ uniformly approaches $0$ as well.

From \prettyref{lem:tree-spa-log-loss}, we know that this result holds for the specific instance of the LZ78 SPA described in \prettyref{con:tree-spa}.
So, we will show that the log loss achieved by \prettyref{con:lz78-spa} for any two priors with a density bounded away from zero is asymptotically equivalent, uniformly over all individual sequences.

To do so, we need the following result:
\begin{theorem}\label{thm:prior-optimality}
    Let $q^\Uppi(x^n)$ be a Bayesian mixture SPA (\prettyref{con:bayesian-mixture}) such that $\supp(\Uppi) = \Mcal(\Acal)$.
    Then
    \[ \lim_{n \rightarrow \infty }\max_{x^n \in \Acal^n} \left|\frac{1}{n} \log \frac{1}{q^\Uppi(x^n)} - H_0(x^n)\right| = 0,\]
    where $H_0(x^n)$ is the (zero-order) empirical entropy of $x^n$.

    \begin{proof}
        Fix some $0 < \epsilon < \frac{1}{2A}$, and consider the subset of $\Mcal(\Acal)$ defined by
        \[\mathcal{V}_\epsilon \triangleq \left\{ \theta\,:\, {\sum}_{a\in\Acal} \theta[a] = 1,\, \theta[a] > \epsilon, \forall a \in \Acal \right\}.\]
        For $\epsilon < \frac{1}{2A}$, this subset has nonzero measure under $\Uppi$ (given the condition $\supp(\Uppi) = \Mcal(\Acal)$), as the following is a subset of $\mathcal{V}$ and has nonzero measure:
        \[\left\{ \theta\,:\, \epsilon \leq \theta[a] \leq 2\epsilon,\,\forall a \neq a_0;\, \theta[a_0] = 1 - {\sum}_{a \neq a_0} \theta[a]\right\}.\]
        Define $\mathcal{U}_\epsilon$ as an arbitrary, fixed, finite set consisting of subsets of $\mathcal{V}_\epsilon$ such that: (1) $\bigcup_{u \in \mathcal{U}_\epsilon} u = \mathcal{V}_\epsilon$, (2) every $u \in \mathcal{U}_\epsilon$ has nonzero measure under $\Uppi$, and (3) no $u \in \mathcal{U}_\epsilon$ has a width larger than $\epsilon$ in any coordinate dimension.
        It is possible to construct such a $\mathcal{U}_\epsilon$ because $\Uppi(\mathcal{V}_\epsilon) > 0$ and $\supp(\Uppi) = \Mcal(\Acal)$.
        
        Consider the integral for $q^\Uppi(x^n)$, evaluated over any $u \in \mathcal{U}$:
        \begin{align*}
            q^\Uppi(x^n) &= \int_{\Theta\in\Mcal(\Acal)} {\prod}_{a\in\Acal} \Theta[a]^{N(a|x^n)} d\Uppi(\Theta) \geq \int_{\Theta\in u} {\prod}_{a\in\Acal} \Theta[a]^{N(a|x^n)}  d\Uppi(\Theta) \\
            &\geq \Uppi(u) \min_{\theta \in u} {\prod}_{a\in\Acal} \theta[a]^{N(a|x^n)}.
        \end{align*}
        Let $\alpha_\epsilon \triangleq \min_{u\in\mathcal{U}_\epsilon} \Uppi(u)$.
        By the definition of $\mathcal{U}_\epsilon$, $\alpha_\epsilon > 0$.

        Now, consider $-\frac{1}{n}\log q^\Uppi(x^n)$ for arbitrary $x^n \in \Acal^n$.
        For any $u\in\mathcal{U}_\epsilon$,
        \begin{align*}
            0 \leq \frac1n \log \frac{1}{q^\Uppi(x^n)} - H_0(x^n) &\leq \frac{1}{n}\log \frac{1}{\alpha_\epsilon} + \max_{\theta \in u}  {\sum}_{a\in\Acal} \left(\frac{N(a|x^n)}{n} \log \frac{1}{\theta[a]} - \frac{N(a|x^n)}{n} \log \frac{n}{N(a|x^n)} \right) \\
            &= \frac{1}{n}\log \frac{1}{\alpha_\epsilon} + \max_{\theta \in u} D(\Theta_{x^n} \,\lVert\, \theta),
        \end{align*}
        where $\Theta_{x^n}$ is the empirical distribution of $x^n$.
        Set $u_\ast$ to be an element of $\mathcal{U}_\epsilon$ that contains $\Theta_{x^n}$, if one exists, or otherwise the element of $\mathcal{U}_\epsilon$ that is closest to $\Theta_{x^n}$ in $\ell_\infty$ distance.
        \[\frac1n\log \frac{1}{q^\Uppi(x^n)} - H_0(x^n) \leq \frac{1}{n}\log \frac{1}{\alpha_\epsilon} + \max_{\theta \in u^\ast} D(\Theta_{x^n} \,\lVert\, \theta).\]
        
        As $\alpha_\epsilon$ is a positive constant, the first term decays as $n\to\infty$.
        We now wish to bound the second term by a quantity constant in $n$ but decaying as $\epsilon \to 0$.
        \begin{fact}
            As $\log x \leq x - 1$, relative entropy is upper-bounded by $\chi^2$ distance: for random variables $P, Q$ over alphabet $\Acal$,
            \[D(P\lVert Q) \leq D_{\chi^2}(P \lVert Q) \triangleq \sum_{a \in \Acal} \frac{(P[a]-Q[a])^2}{Q(x)} \leq A\frac{\max_{a\in\Acal} (P[a] - Q[a])^2}{\min_{a\in\Acal} Q[a]}.\]
        \end{fact}

        By the definition of $u_\ast$, $\left|\Theta_{x^n}[a] - \theta[a]\right| \leq 2\epsilon$, $\forall a \in \Acal$.
        Also, based on the definition of $\mathcal{V}_\epsilon$, $\min_{a\in\Acal}\theta[a] \geq \epsilon$, $\forall \theta \in \mathcal{V}_\epsilon$.
        So, using the above bound on relative entropy,
        \begin{align*}
            0 \leq \frac1n \log \frac{1}{q^\Uppi(x^n)} - H_0(x^n) \leq \frac{1}{n} \log \frac{1}{\alpha_\epsilon} + \frac{8A\epsilon^2}{\epsilon} = \frac{1}{n} \log \frac{1}{\alpha_\epsilon} + 8A\epsilon.
        \end{align*}
        This bound is independent of $x^n$, so it also applies to $\max_{x^n} \left|\frac1n\log \frac{1}{q^\Uppi(x^n)} - H_0(x^n)\right|$.
        Taking the limit supremum as $n\to\infty$,
        \[\limsup_{n\to\infty} \max_{x^n} \left|\frac1n\log \frac{1}{q^\Uppi(x^n)} - H_0(x^n)\right| \leq 8A\epsilon,\quad \forall \epsilon \leq \frac{1}{2A}.\]
        Since the above applies to $\epsilon$ arbitrarily small, it must hold that
        \[\lim_{n\to\infty} \max_{x^n} \left|\frac1n\log \frac{1}{q^\Uppi(x^n)} - H_0(x^n)\right| = 0.\]
    \end{proof}
\end{theorem}
\begin{corollary}\label{cor:prior-optimality-bounded-density}
    \prettyref{thm:prior-optimality} is a purely asymptotic result.
    If we assume that the prior $\Uppi$ admits a density that is bounded away from zero, we can obtain the following finite-sample result:
    \[\max_{x^n \in \Acal^n}  \left|\frac1n\log \frac{1}{q^\Uppi(x^n)} - H_0(x^n)\right| \leq \alpha(\Uppi)A\frac{\log n}{n},\]
    where $\alpha(\Uppi)$ is a constant depending on the minimum value that the density of $\Uppi$ attains.
    This is a combination of equation (50) of \cite{MR0914346}, which states
    \[\max_{x^n \in \Acal^n}  \frac1n\log \frac{1}{q^\Uppi(x^n)} - H_0(x^n) \leq \alpha(\Uppi)A\frac{\log n}{n},\]
    and the fact that the scaled log loss of $q^\Uppi(x^n)$ is lower-bounded by the empirical entropy, as the mixture distribution is upper-bounded by the maximum likelihood of $x^n$ over i.i.d. $\mathrm{Ber}(\Theta)$ laws, so
        \[\log \frac{1}{q^\Uppi(x^n)} \geq \log \frac{1}{\max_\Theta \prod_{i=1}^n \Theta(x_i)} = H_0(x^n).\]
\end{corollary}

Now, using \prettyref{thm:prior-optimality}, we can show that the asymptotic log loss achieved by any SPA in the LZ78 family is the same:
\begin{lemma}\label{lem:lz78-log-losses-approach-emp-entropies}
    For any prior such that $\supp(\Uppi) = \Mcal(\Acal)$,
    \[\max_{x^n} \left|\frac{1}{n} \log \frac{1}{q^{\text{LZ78}, \Uppi}(x^n)} - \frac{1}{n} \sum_{z \in \Zcal(x^n)} \left|\mathcal{Y}\{x^n, z\}\right| H_0\left( \mathcal{Y}\{x^n, z\}\right)\right| = o(1).\]
    \begin{proof}
        By construction, the log loss of $q^{\text{LZ78}, \Uppi}$ can be divided into the subsequences corresponding to each LZ78 context:
        \begin{equation}
            \frac{1}{n}\log \frac{1}{q^{\text{LZ78}, \Uppi}(x^n)} = \frac{1}{n}\sum_{z\in\Zcal(x^n)}  \log \frac{1}{q^{\text{LZ78}, \Uppi}(\mathcal{Y}\{x^n, z\})} \stackrel{(a)}{=} \frac{1}{n}\sum_{z\in\Zcal(x^n)} \log \frac{1}{q^\Uppi\left(\mathcal{Y}\{x^n, z\}\right)},\label{eqn:split-q-hat-into-phrases}
        \end{equation}
        where (a) directly applies \prettyref{con:lz78-spa}.

        By \prettyref{thm:prior-optimality}, $\forall y^m \in \Acal^*$,
        \begin{equation}
            \left|\frac{1}{m}\log \frac{1}{q^\Uppi(y^m)} - H_0(y^m)\right| \leq \xi(m) = o(1),\label{eqn:rewritten-q-assumption}
        \end{equation}
        where $\xi(m)$ is a function purely of $m$.
        For \prettyref{thm:prior-optimality} to hold, $q^\Uppi$ cannot incur unbounded loss, so $\xi(m) \leq B, \forall m \geq 1$.

        For simplicity of notation, define $y_z^{m_z} \triangleq \mathcal{Y}\{x^n, z\}$, where $m_z = \left| \mathcal{Y}\{x^n, z\} \right|$.
        By \eqref{eqn:split-q-hat-into-phrases} and \eqref{eqn:rewritten-q-assumption},
        \begin{align*}
            \max_{x^n \in \Acal^n} \left|\frac{1}{n}\log \frac{1}{q^{\text{LZ78}, \Uppi}(x^n)} - \frac{1}{n} \sum_{z \in \Zcal(x^n)} m_z H_0(y_z^{m_z})\right| &\leq \frac{1}{n} \max_{x^n \in \Acal^n} \sum_{z\in\Zcal(x^n)} \left|\log \frac{1}{q^\Uppi(y_z^{m_z})}- m_z H_0(y_z^{m_z})\right| \\
            &\leq \max_{x^n \in \Acal^n} \frac{1}{n}\sum_{z\in\Zcal(x^n)} m_z \xi\left(m_z\right) = \max_{x^n \in \Acal^n}\frac{1}{n}\sum_{t=1}^n \xi(m_{z_t}),
        \end{align*}
        where $z_t$ is shorthand for $z_c(x^{t-1})$.
        We now show that
        \[\epsilon > 0,\,\exists N > 0 \text{ s.t. } \forall n > N,\, \max_{x^n} \frac{1}{n}\sum_{t=1}^n \xi(m_{z_t}) \leq \epsilon.\]
        We divide the summation over timepoints into three parts, each of which we bound by $\frac{\epsilon}{3}$ (for large enough $n$).

        \textbf{Part 1: timesteps where $\xi(m)$ is small enough.}
        By assumption, $\exists M > 0$ such that, $\forall m > M$, $\xi(m) < \frac{\epsilon}{3}$.
        Let the timesteps with $m_{z_t} > M$ be denoted $\mathcal{T}_1$.
        \[\max_{x^n} \frac{1}{n}\sum_{t\in\mathcal{T}_1} \xi(m_{z_t}) \leq \frac{|\mathcal{T}_1|}{n} \frac{\epsilon}{3} \leq \frac{\epsilon}{3}.\]

        \textbf{Part 2: final timesteps of long phrases.}
        $m_z$ is always upper-bounded by the number of nodes that are descendants of $z$, as node $z$ must have been traversed before each such node was added to the tree.
        So, in each phrase, at most the final $M$ symbols have a corresponding $m_{z_t} < M$.
        For long enough phrases, only a small fraction of symbols have $m_{z_t} \leq M$: specifically, we consider phrases longer than $L \triangleq \frac{3MB}{\epsilon}$, where $m_{z_t} \leq M$ for at most a $\frac{\epsilon}{3B}$ fraction of symbols (where $B$ is the upper bound on $\xi$).
        Denote the symbols in such phrases with $m_{z_t} \leq M$ by $\mathcal{T}_2$.
        \[\max_{x^n} \frac{1}{n}\sum_{t\in\mathcal{T}_2} \xi(m_{z_t}) \leq \frac{|\mathcal{T}_2|}{n} B \leq \frac{\epsilon}{3}.\]

        \textbf{Part 3: timesteps in short phrases.}
        Each LZ78 phrase is unique (except perhaps the last), so at most $\sum_{\ell=1}^L A^\ell + 1 < LA^L + 1$ phrases have length $< L$.
        There are at most $L^2A^L + L$ symbols in those phrases.
        For $N = \frac{3(L^2A^L+L)B}{\epsilon}$, these phrases are at most an $\frac{\epsilon}{3B}$ fraction of the sequence.
        Define the corresponding timesteps as $\mathcal{T}_3$, $\forall n > N$,
        \[\max_{x^n} \frac{1}{n}\sum_{t \in \mathcal{T}_3} \xi(m_{z_t}) \leq \frac{|\mathcal{T}_3|}{n}B \leq \frac{\epsilon}{3B} B = \frac{\epsilon}{3}.\]

        $\mathcal{T}_1 \cup \mathcal{T}_2 \cup \mathcal{T}_3 = \{1, \dots, n\}$, so, $\forall n > N$,
        \[\max_{x^n} \frac{1}{n}\sum_{t=1}^n \xi(m_{z_t}) \leq \max_{x^n} \frac{1}{n}\sum_{t \in \mathcal{T}_1} \xi(m_{z_t}) + \frac{1}{n}\sum_{t \in \mathcal{T}_2} \xi(m_{z_t}) + \frac{1}{n}\sum_{t \in \mathcal{T}_3} \xi(m_{z_t}) \leq 3\frac{\epsilon}{3} = \epsilon,\]
        so, as $\epsilon$ is arbitrary, $\max_{x^n} \frac{1}{n}\sum_{t=1}^n \xi(m_{z_t}) \to 0$ as $n\to\infty$.
    \end{proof}
\end{lemma}
\begin{corollary}
    Using \prettyref{cor:prior-optimality-bounded-density}, for any prior with a density bounded away from zero, we can get version of \prettyref{lem:lz78-log-losses-approach-emp-entropies} that includes a rate of decay:
    \[\max_{x^n} \left|\frac{1}{n} \log \frac{1}{q^{\text{LZ78}, \Uppi}(x^n)} - \frac{1}{n} \sum_{z \in \Zcal(x^n)} \left|\mathcal{Y}\{x^n, z\}\right|H_0\left( \mathcal{Y}\{x^n, z\} \right)\right| = \bigO\!\left(\frac{A \log A \log\log n}{\log n}\right).\]

    \begin{proof}
        As in the proof of \prettyref{lem:lz78-log-losses-approach-emp-entropies}, let $y_z^{m_z}$ be shorthand for $ \mathcal{Y}\{x^n, z\}$, and divide $q^{\text{LZ78}, \Uppi}$ into the subsequences corresponding to each LZ78 context:
        \[\log \frac{1}{q^{\text{LZ78}, \Uppi}(x^n)} = \sum_{z \in \Zcal(x^n)} \log \frac{1}{q^\Uppi(y_z^{m_z})}.\]
        By \prettyref{cor:prior-optimality-bounded-density},
         \begin{align*}
             \max_{x^n} \frac{1}{n} \left|\log \frac{1}{q^{\text{LZ78}, \Uppi}(x^n)} - \sum_{z \in \Zcal(x^n)} m_z H_0(y_z^{m_z})\right| &\leq \max_{x^n} \frac{1}{n} \sum_{z \in \Zcal(x^n)} \alpha A \log m_z,
         \end{align*}
         where $\alpha$ is a constant that depends on the choice of prior.
         Multiplying and dividing by $C(x^n)$, we can apply Jensen's inequality to get
         \begin{align*}
             &\max_{x^n} \frac{1}{n} \left|\log \frac{1}{q^{\text{LZ78}, \Uppi}(x^n)} - \sum_{z \in \Zcal(x^n)} m_z H_0(y_z^{m_z})\right| \leq \max_{x^n}  \frac{\alpha A C(x^n)}{n} \sum_{z \in \Zcal(x^n)} \frac{1}{C(x^n)}\log m_z \\
             &\hspace{15em}\leq \max_{x^n}  \frac{\alpha A C(x^n)}{n} \log \left(\frac{\sum_{z\in\Zcal(x^n)} m_z}{C(x^n)}\right) = \alpha A \max_{x^n} \frac{C(x^n)}{n} \log \frac{n}{C(x^n)}.
         \end{align*}
         By \cite{originalLZ78paper}, $\frac{C(x^n)}{n} \leq C_1 \frac{\log A}{\log n}$, for universal constant $C_1$.
         The function $x\log \frac{1}{x}$ is increasing in the range $\left[0, 2^{-1/\ln 2}\right]$,
         \footnote{$\frac{d^2}{dx^2}x\log\frac{1}{x} = \frac{1}{x\ln 2}$, so it concave for $x > 0$.
         $\therefore$, a single global maximum is reached when $0 = \frac{d}{dx}x\log\frac{1}{x} = -\log x - \frac{1}{\ln 2}$, or when $x = 2^{-1/\ln 2}$.}
        so, if $C_1 \frac{\log A}{\log n} \leq 2^{-1/\ln 2}$, the upper bound of $\frac{C(x^n)}{n}\log \frac{n}{C(x^n)}$ can be found by plugging in the upper bound of $\frac{C(x^n)}{n}$.
        Let $C_2\triangleq C_1 2^{\frac{1}{\ln 2}}$.
        Then, for $n \geq A^{C_2}$, 
        \[\max_{x^n}\frac{C(x^n)}{n}\log \frac{n}{C(x^n)} = \frac{C_1 \log A}{\log n} \log\left(\frac{\log n}{C_1 \log A}\right) \leq \frac{C_1 \log A \log \log n}{\log n}.\]
        Plugging this in,
        \[\max_{x^n} \frac{1}{n} \left|\log \frac{1}{q^{\text{LZ78}, \Uppi}(x^n)} - \sum_{z \in \Zcal(x^n)} m_z H_0(y_z^{m_z})\right| = \bigO\!\left(\frac{A \log A \log \log n}{\log n}\right).\]
    \end{proof}
\end{corollary}
\begin{remark}
    The above result indicates slow convergence for large alphabet sizes.
    It is important to note that this is a worst-case result; as demonstrated in the text generation examples in \prettyref{sec:experiments}, the LZ78 SPA can still achieve reasonable performance on larger alphabets ($\approx\!50$ in the case of the text generation).
    However, the LZ78 SPA is most effective, both theoretically and empirically, on small alphabets.
\end{remark}

Putting \prettyref{lem:lz78-log-losses-approach-emp-entropies} together with \prettyref{lem:tree-spa-log-loss},

\textbf{Theorem \ref{thm:lz78-log-loss-codelength-correspondence}}.
    For any prior such that $\supp(\Uppi) = \Mcal(\Acal)$,
    \[ \lim_{n\to\infty} \max_{x^n} \left| \frac{1}{n} \log \frac{1}{q^{\text{LZ78}, \Uppi}(x^n)} - \frac{C(x^n) \log C(x^n)}{n}\right| = 0.\]

    \begin{proof}
        Denote the SPA corresponding to \prettyref{con:tree-spa} as $q^{\text{LZ78}, \Uppi_0}$, where $\Uppi_0$ is the Dirichlet$(\frac{1}{A-1}, \dots \frac{1}{A-1})$ prior.
        Also, for brevity, denote the asymptotic LZ78 codelength $C(x^n) \log C(x^n)$ by $\ell_\text{LZ78}(x^n)$, and $\sum_{z \in \Zcal(x^n)} m_z H_0(y_z^{m_z})$ from \prettyref{thm:lz78-log-loss-codelength-correspondence} by $H^\text{LZ78}(x^n)$.
        
        By the triangle inequality and \prettyref{lem:tree-spa-log-loss},
        {\small
        \begin{align*}
            \max_{x^n} \frac{1}{n}\left|\log \frac{1}{q^{\text{LZ78}, \Uppi}(x^n)} - \ell_\text{LZ78}(x^n)\right| &\leq \max_{x^n} \frac{1}{n}\left| \log \frac{1}{q^{\text{LZ78}, \Uppi_0}(x^n)} - \ell_\text{LZ78}(x^n)\right| + \max_{x^n} \frac{1}{n}\left|\log \frac{1}{q^{\text{LZ78}, \Uppi_0}(x^n)} - \log \frac{1}{q^{\text{LZ78}, \Uppi}(x^n)}\right| \\
            &\leq o(1) +\max_{x^n}  \frac{1}{n} \left|\log \frac{1}{q^{\text{LZ78}, \Uppi}(x^n)} - H^\text{LZ78}(x^n)\right| + \max_{x^n} \frac{1}{n} \left|\log \frac{1}{q^{\text{LZ78}, \Uppi_0}(x^n)} - H^\text{LZ78}(x^n)\right|.
        \end{align*}}
        By \prettyref{lem:lz78-log-losses-approach-emp-entropies}, the final two terms are $o(1)$ as well, so
        \[\max_{x^n} \frac{1}{n}\left|\log \frac{1}{q^{\text{LZ78}, \Uppi}(x^n)} - \ell_\text{LZ78}(x^n)\right| = o(1)\text{ as } n\to\infty.\]
    \end{proof}

\subsection{Optimal Markov and Finite-State Log Loss in terms of Empirical Entropies}\label{app:classes-of-spas}
\subsubsection{Empirical Distributions}
In order to precisely define the empirical entropies mentioned in \prettyref{sec:mu-lambda-entropies}, we define some empirical distributions over individual sequence $x^n$.
In addition, for finite-state SPAs, we define empirical distributions over the list of states, $s^n$.

\begin{definition}[Zero-order empirical distribution]\label{def:zero-order-emp}
    $X$, the random variable following the zero-order empirical distribution of $x^n$, has law
$\Pbb(X = a) = \frac{N(a|x^n)}{n}.$
\end{definition}

\begin{definition}[$k$-order empirical distribution]\label{def:k-order-emp}
    The $k$-order empirical distribution of $x^n$ is defined as the law
    \[\Pbb(X^k = a^k) = \frac{1}{n} \sum_{i=1}^n \indic{x_i^{i+k-1} = a^k},\]
    where indices greater than $n$ ``wrap around'' to the beginning of the sequence.
    This wrapping is known as the \textbf{circular convention}.
\end{definition}
\begin{remark}\label{rem:k-order-conditional-dist}
    From \prettyref{def:k-order-emp}, a conditional distribution can also be defined:
    \[\Pbb(X_k=a_k|X^{k-1}=a^{k-1}) = \frac{1}{|\{t \in [n]: x_t^{t+k-2} = a^{k-1}\}|} \sum_{i=1}^n \indic{x_i^{i+k-1} = a^k}.\]
    For a fixed $a^{k-1}$, this is equivalent to the zero-order empirical distribution of the subsequence $\{x_t : x_{t-k}^{t-1} = a^{k-1}\}$, where $x_{t-k}^{t-1}$ is evaluated using the circular convention.
\end{remark}

\begin{definition}[Finite-state empirical distribution]
    For finite-state SPAs, we can define the empirical distribution of the states $s^n$ exactly as we defined the distribution of $X$.
    The joint empirical distribution of $(x^n, s^n)$ is defined as
    \[\Pbb(X=a, S=s) = \frac{1}{n}\sum_{i=1}^n \indic{x_i=a, s_i=s}.\]
    The $k$-order joint empirical distribution of $(x^n, s^n)$ is defined analogously as in \prettyref{def:k-order-emp}, as is $\Pbb(X|S)$.
\end{definition}

For any individual sequence $x^n$ and an associated set of states $s^n$, the $k$-order empirical distribution of $(x^n, s^n)$ satisfies the following property:
\begin{lemma}\label{lem:stationary-k-empirical}
    Let $(X^k, S^k)$ follow the joint $k$-order empirical distribution of $(x^n, s^n)$.
    Then,
    $(X_\ell, S_\ell) \stackrel{d}{=} (X, S),\, \forall \ell \in [k],$
    where $(X, S)$ is the zero-order joint empirical distribution of $(x^n, s^n)$.

    \begin{proof}
        The distribution of $(X^{k}, S^{k})$ is
        \[\Pbb\left(X^{k}=a^{k}, S^{k}=s^{k}\right) = \frac{1}{n} \left|\left\{i \in [n] \,:\,x_{i}^{i+k-1}=a^{k}, s_i^{i+k-1} = s^{k}\right\}\right|,\]
        where we follow the circular convention.

        Then, the distribution of $(X_\ell, S_\ell)$ is defined by
        \begin{align*}
            \Pbb(X_\ell = a, S_\ell = s) &= \sum_{\substack{a^{k} \in \Acal^{k}: a_\ell = a\\s^{k} \in [M]^{k} : s_\ell = s}} \Pbb\left(X^{k}=a^{k}, S^{k}=s^{k}\right) \\
            &= \frac{1}{n} \sum_{\substack{a^{k} \in \Acal^{k}: a_\ell = a\\s^{k} \in [M]^{k} : s_\ell = s}} \left|\left\{i\in[n] \,:\, x_{i}^{i+k-1}=a^{k}, s_i^{i+k-1} = s^{k}\right\}\right| \\
            &= \frac{1}{n} \left|\left\{i\in[n] \,:\, x_{i+\ell-1} = a, s_{i+\ell - 1} = s \right\}\right| = \frac{1}{n} N\left((a, s) | (x^n, s^n)\right) = \Pbb(X=a, S=s),
        \end{align*}
        where, the second-to-last step is a result of the circular convention.
    \end{proof}
\end{lemma}
\begin{corollary}\label{cor:stationary-k-empirical}
    Via a similar  proof to that of \prettyref{lem:stationary-k-empirical}, we can show that 
    $(X_{\ell}^{\ell+r-1}, S_{\ell}^{\ell+r-1}) \stackrel{d}{=} (X^r, S^r),\,\forall \ell, r \in [k].$
\end{corollary}

\subsubsection{Equivalence of Log Losses and Empirical Entropies}
We can now show the equivalence of $\mu_k(\xv)$ and $\lambda_M(\xv)$ to conditional entropies of the corresponding empirical distributions, the specifics of which are discussed in the following lemmas.

\begin{lemma}[Equivalence of Zero-Order Markov Loss to Empirical Entropy]\label{lem:mu-zero-equals-H}
    For all individual sequences $x^n$, the optimal zero-order Markov log loss is equal to the empirical entropy.
    \ie, $\mu_0(x^n) = H(X)$, where $X$ follows the zero-order empirical distribution of $x^n$.

    \begin{proof}
        The zero-order optimal log loss is
        \begin{align*}
            \mu_0(x^n) &= \min_{q \in \Mcal(\Acal)} \frac{1}{n} \sum_{t = 1}^n \log \frac{1}{q(x_t)} \stackrel{(a)}{=} \min_{q \in \Mcal(\Acal)} \sum_{a \in \Acal} \frac{N(a|x^n)}{n} \log \frac{1}{q(a)} = \min_{q \in \Mcal(\Acal)} \sum_{a \in \Acal}  \Pbb(X=a) \log \frac{1}{q(a)} \\
            &= H(X) + \min_{q \in \Mcal(\Acal)}D(X||(Y\sim q)) \stackrel{(b)}{=} H(X),
        \end{align*}
        where $D(X||Y)$ represents relative entropy.
        $(a)$ follows from rearranging the sum, and $(b)$ is a result of the fact that relative entropy is a non-negative quantity, with a minimum of $0$ when the two distributions are identical.
    \end{proof}
\end{lemma}

In order to extrapolate this to $k$-order Markov SPA log loss, we make use of the following property: the optimal $k$-order Markov log loss can be achieved by dividing the input sequence into subsequences for every possible length-$k$ context and then applying a probability assignment that achieves the optimal zero-order Markov log loss to each subsequence.
\begin{lemma}[Relationship between zero-order and $k$-order Markov log loss]\label{lem:k-order-markov-loss-sum-zero-order-loss}
    For any sequence $\yv$,
    \[m \cdot \mu_k(y^m) = \sum_{z\in\Acal^k} \left|\{ k+1 \leq t \leq n : y_{t-k}^{t-1}=z\}\right| 
    \mu_0\left(\{y_t : k+1 \leq t \leq n,  y_{t-k}^{t-1}=z\}\right), \]
    where $\{y_t : k+1 \leq t \leq n,  y_{t-k}^{t-1}=z\}$ is taken to be the \textbf{ordered subsequence} of $y^m$ with a given $k$-order context.
    \begin{proof}
        As a $k$-order Markov SPA is allowed to attain a log loss of zero over the first $k$ symbols,
        \begin{align*}
            m\mu_k(y^m) &= \min_{q \in \Mcal_k} \sum_{t=k+1}^m \log \frac{1}{q(y_t|y_{t-k}^{t-1})} = \sum_{z\in\Acal^k} \min_{q_z \in \Mcal(\Acal)} \log \frac{1}{q_z\left(\{y_t :  k+1 \leq t \leq n, y_{t-k}^{t-1}=z\}\right)} \\
            &= \sum_{z\in\Acal^k} \left|\{y_t :  k+1 \leq t \leq n, y_{t-k}^{t-1}=z\}\right| \mu_0\left(\{y_t :  k+1 \leq t \leq n, y_{t-k}^{t-1}=z\}\right).
        \end{align*}
    \end{proof}
\end{lemma}

\begin{lemma}[Equivalence of $k$-Order Markov Loss to Conditional Entropy]\label{lem:mu-equals-H}
    $\forall k \geq 1$ and sequence $x^n$,
    \[H(X_{k+1}|X^k) - \epsilon(k, A, n) \leq \mu_k(x^n) \leq H(X_{k+1}|X^k),\]
    where $X^{k+1}$ follows the $(k+1)$-order empirical distribution of $x^n$ and $\lim_{n\to\infty} \epsilon(k, A, n) = 0$.

    \begin{proof}
        We first prove the upper bound, $\mu_k(x^n) \leq H(X_{k+1}|X^k)$, followed by the lower bound, $H(X_{k+1}|X^k) - \epsilon(k, A, n)$.
        
        \textbf{Upper bound}:
        By \prettyref{lem:k-order-markov-loss-sum-zero-order-loss},
        \[\mu_k(x^n) = \sum_{z\in\Acal^k} \frac{\left|\{k+1\leq t \leq n : x_{t-k}^{t-1}=z\}\right|}{n} \mu_0\left(\{x_t : k+1\leq t \leq n, x_{t-k}^{t-1}=z\}\right),\]
        where the circular convention is not used.
        As applying the circular convention can only increase each term of the summation (as it allows for $t \leq k$ to be included), $\mu_k(x^n)$ can be upper-bounded by applying the circular convention:
        \begin{align*}
            \mu_k(x^n) &\leq \sum_{z\in\Acal^k} \frac{\left|\{t \in [n] : x_{t-k}^{t-1}=z\}\right|}{n} \mu_0\left(\{x_t : t \in [n], x_{t-k}^{t-1}=z\}\right) \quad\text{(with circular convention)} \\
            &= \sum_{z\in\Acal^k} \Pbb(X^k=z) \mu_0\left(\{x_t : t \in [n], x_{t-k}^{t-1}=z\}\right) \\
            &= \sum_{z\in\Acal^k} \Pbb(X^k=z) H(X_{k+1}|X^k=z) = H(X_{k+1}|X^k).
        \end{align*}

        \textbf{Lower bound}: First, define $\tilde{X}^{k+1}$ according to the following empirical distribution:
        \[\Pbb\left(\tilde{X}^{k+1} = a^{k+1}\right) = \frac{1}{n-k} \sum_{i=1}^{n-k} \indic{x_{i}^{i+k} = a^{k+1}}.\]
        This is similar to the $(k+1)$-order empirical distribution of $x^n$, but we only consider the first $n-k$ length-$(k+1)$ sequences instead of using the circular convention.
        
        By \prettyref{lem:k-order-markov-loss-sum-zero-order-loss} and analysis identical to that in the proof of the upper bound,
        \begin{align*}
            \mu_k(x^n) &= \frac{n-k}{n}\sum_{z\in\Acal^k} \frac{\left|\{k+1 \leq t \leq n : x_{t-k}^{t-1}=z\}\right|}{n-k} \mu_0\left(\{x_t : k+1 \leq t \leq n, x_{t-k}^{t-1}=z\}\right) \quad \text{(without circular convention)} \\
            &= \frac{n-k}{n}\sum_{z\in\Acal^k} \Pbb\left(\tilde{X}^{k} = z\right) H(\tilde{X}_{k+1}|X^k=z) = \frac{n-k}{n} H(\tilde{X}_{k+1}|\tilde{X}^k).
        \end{align*}
        
        $H(\tilde{X}_{k+1}|\tilde{X}^k)$ is bounded and $\frac{k}{n} \to 0$ as $n \to \infty$, so the second term is $o(1)$ as $n \to \infty$ and $k$ is fixed.
        Therefore, it is sufficient to show that $H(\tilde{X}_{k+1}|\tilde{X}^k) = H(X_{k+1}|X^k) - \xi(k, n)$, where $\lim_{n\to\infty} \xi(k, n) = 0$.
        
        Equivalently, we show that $\lim_{n \to \infty} \left| H(X_{k+1}|X^k) - H(\tilde{X}_{k+1}|\tilde{X}^k) \right| = 0$.
        
        By the chain rule of entropy and the triangle inequality,
        \[\left| H(X_{k+1}|X^k) - H(\tilde{X}_{k+1}|\tilde{X}^k) \right| \leq \left| H(\tilde{X}^{k+1})- H(X^{k+1})\right| + \left| H(\tilde{X}^k) - H(X^k) \right|.\]
        Entropy is uniformly continuous with respect to the probability mass function and the $\ell_1$ metric,\footnote{The $\ell_1$ metric for probability mass functions $P$, $Q$ is taken to be $\norm{P-Q}_1 = \sum_{a \in \Acal} |P(a) - Q(a)|$.} as each term of the summation $H(X) = \sum_{a \in \Acal} p_a \log\frac{1}{p_a}$ is continuous and bounded over a compact domain.
        
        By the definition of uniform continuity,
        \[\forall \epsilon > 0,\, \exists \delta_\epsilon > 0 \,\st\, \forall P, Q \in \Mcal(\Acal),\, \norm{P-Q}_1 < \delta_\epsilon \implies |H_P(X) - H_Q(X)| < \epsilon.\]
        To apply uniform continuity to show that $\lim_{n\to\infty} \left| H(X_{k+1}|X^k) - H(\tilde{X}_{k+1}|\tilde{X}^k) \right| = 0$, we must verify that \\$\norm{\Pbb(\tilde{X}^{k+1}) - \Pbb(X^{k+1})}_1 \to 0$ and $\norm{\Pbb(\tilde{X}^{k}) - \Pbb(X^{k})}_1 \to 0$ as $n \to \infty$.
        
        By the definitions of the empirical distributions for $X^{k+1}$ and $\tilde{X}^{k+1}$, $\forall a^{k+1} \in \Acal^{k+1}$,
        \begin{align*}
            \Pbb(X^{k+1} = a^{k+1}) &= \frac{1}{n} \left|\left\{i \in [n] \,:\, x_i^{i+k} = a^{k+1}\right\}\right| \\
            &= \frac{1}{n} \left|\left\{i\in [n-k]\,:\,x_i^{i+k} = a^{k+1}\right\}\right| + \frac{1}{n} \left|\left\{n-k < i \leq n \,:\, x_i^{i+k} = a^{k+1}\right\}\right| \\
            &= \frac{n-k}{n} \Pbb(\tilde{X}^{k+1} = a^{k+1}) + \frac{1}{n} \left|\left\{n-k < i \leq n \,:\, x_i^{i+k} = a^{k+1}\right\}\right|.
        \end{align*}
        As $\left|\{i : n-k < i \leq n \,:\, x_{i}^{i+k} = a^{k+1}\}\right| \leq \left|\{i : n-k < i \leq n\}\right| = k$,
        \[\Pbb(\tilde{X}^{k+1} = a^{k+1}) - \frac{k}{n} \leq \Pbb(X^{k+1} = a^{k+1}) \leq  \Pbb(\tilde{X}^{k+1} = a^{k+1}) + \frac{k}{n}.\]
        Via an identical argument, the exact same relation holds for $\Pbb(X^{k} = a^{k})$.
        
        We can now directly show that $\lim_{n\to\infty} \left| H(X_{k+1}|X^k) - H(\tilde{X}_{k+1}|\tilde{X}^k) \right| = 0$.
        Choose arbitrarily-small $\epsilon > 0$.
        As entropy is uniformly continuous, $\exists \delta_1 > 0$ and $\delta_2 > 0$ such that 
        \begin{align*}
            \norm{\Pbb(X^{k+1}) - \Pbb(\tilde{X}^{k+1})}_1 < \delta_1 &\implies \left|H(\tilde{X}^{k+1})- H(X^{k+1})\right| < \epsilon/2,\\
            \norm{\Pbb(X^{k}) - \Pbb(\tilde{X}^{k})}_1 < \delta_2 &\implies \left|H(\tilde{X}^{k})- H(X^{k})\right| < \epsilon/2.
        \end{align*}
        Let $N = \left\lceil\frac{A^{k+1}k}{\min(\delta_1, \delta_2)}\right\rceil$.
        Then, $\forall n > N$,
        \[\norm{\Pbb(X^{k+1}) - \Pbb(\tilde{X}^{k+1})}_1 \leq \frac{k}{n}A^{k+1} < \delta_1,\text{ and } \norm{\Pbb(X^{k}) - \Pbb(\tilde{X}^{k})}_1 \leq \frac{k}{n}A^{k} < \delta_2.\]
        Therefore, $\forall n > N$,
        \begin{align*}
            \left| H(X_{k+1}|X^k) - H(\tilde{X}_{k+1}|\tilde{X}^k) \right| \leq \epsilon/2 + \epsilon/2 = \epsilon.
        \end{align*}
        As $\epsilon$ is arbitrary, $\lim_{n\to\infty} \left| H(X_{k+1}|X^k) - H(\tilde{X}_{k+1}|\tilde{X}^k) \right| = 0$.
    \end{proof}
\end{lemma}

To draw a similar connection between the optimal finite state-log loss and the empirical entropy of $x^n$ given the corresponding states, we analyze the behavior of SPAs in $\Fcal_M^{g, s_1}$, which denotes the set of $M$-state SPAs with fixed state transition function $g$ and initial state $s_1$.
Define the optimal loss over this class of SPAs as
\[\lambda^{g, s_1}_M(x^n) \triangleq \min_{q^{g, s_1, f} \in \Fcal_M^{g, s_1}} \frac{1}{n} \log \frac{1}{q^{g, s_1, f}(x^n)}.\]

\begin{lemma}[Equivalence of Finite-State Log Loss to Conditional Entropy]\label{lem:lambda-equals-H}
    $\forall$ sequences $x^n$, state transition functions $g$, and initial states $s_1$,
    \[\lambda^{g, s_1}_M(x^n) = H(X|S),\]
    where $(X, S)$ follow the joint empirical distribution of $(x^n, s^n)$, where $s^n$ is a function of $x^n$, $g$ and $s_1$.

    \begin{proof}
        By the definition of $\lambda^{g, s_1}_M(x^n)$,
        \begin{align*}
            \lambda^{g, s_1}_M(x^n) &= \min_{q^{g, s_1, f} \in \Fcal_M^{g, s_1}} \frac{1}{n} \log \frac{1}{q^{g, s_1}(x^n)} = \min_{f: [M] \to \Mcal(\Acal)} \frac{1}{n} \sum_{i = 1}^n \log \frac{1}{f(s_i)(x_i)},
        \end{align*}
        where $s_i$ is the $i$\textsuperscript{th} state, determined by $x^i$, $g$ and $s_1$.
        Rearranging the summation,
        \begin{align*}
            \lambda^{g, s_1}_M(x^n) &= \min_{f: [M] \to \Mcal(\Acal)} \frac{1}{n} \sum_{(a, s) \in \Acal \times [M]} N\left((a, s)|(x^n, s^n)\right) \log \frac{1}{f(s)(a)} \\
            &= \min_{f: [M] \to \Mcal(\Acal)} \sum_{(a, s) \in \Acal \times [M]} \Pbb(X=a, S=s) \log \frac{1}{f(s)(a)} \\
            &= \sum_{s \in [M]} \Pbb(S=s) \min_{f_s \in \Mcal(\Acal)} \sum_{a \in \Acal} \Pbb(X=a|S=s) \log \frac{1}{f_s(a)}.
        \end{align*}
        As in \prettyref{lem:mu-zero-equals-H}, we write this expression as the sum of a condition entropy and a sum of relative entropies.
        \begin{align*}
            \lambda^{g, s_1}_M(x^n) &= H(X|S) + \sum_{s \in [M]} \Pbb(S=s) \min_{f_s \in \Mcal(\Acal)} D\left(\Pbb_{X|S}(\cdot|S=s) || f_s\right).
        \end{align*}
        $D\left(\Pbb_{X|S}(\cdot|S=s) || f_s\right)$ achieves the minimum of $0$ when $f_s = \Pbb(X|S=s)$, so
        \[\lambda^{g, s_1}_M(x^n) = H(X|S).\]
    \end{proof}
\end{lemma}

\subsection{Equivalence of Optimal Finite-State Log Loss and Markov Log Loss, and Finite-State Compressibility} \label{app:proofs-lambda-eq-mu-eq-rho}
The proof that, for any individual sequence, $\lambda(\xv) = \mu(\xv) = \rho(\xv) \log A$, is divided into three parts.
First, we prove that $\lambda(\xv) = \mu(\xv)$ by proving that $\lambda(\xv)$ is both upper- and lower-bounded by $\mu(\xv)$.
Then, we directly prove that $\mu(\xv) = \rho(\xv) \log A$.

\begin{lemma}[Upper bound of $\lambda(\xv)$ by $\mu(\xv)$]\label{lem:lambda-leq-mu}
    For all individual sequences,
    \[\lambda(\xv) \leq \mu(\xv).\]

    \begin{proof}
        By \prettyref{fact:markov-finite-state-correspondence}, as any $k$-order Markov SPA is also an $A^k$-state SPA, $\Mcal_k \subseteq \Fcal_{A^k}$, so $\lambda_{A^k}(x^n) \leq \mu_k(x^n)$.
        Taking $\limsup_{n\to\infty}$ on both sides, $\lambda_{A^k}(\xv) \leq \mu_k(\xv)$.
        So, as $\lambda_M(\xv)$ is monotonically non-increasing in $M$, $\lambda(\xv) \leq \mu_k(\xv)$, $\forall k \geq 0$.
        Taking the limit as $k \to \infty$, we have $\lambda(\xv) \leq \mu(\xv)$.
    \end{proof}
\end{lemma}

The proof of the upper bound, $\lambda(\xv) \geq \mu(\xv)$, is primarily based on the following claim:
\begin{claim}\label{claim:mu-minus-lambda-bound}
    $\forall$ finite sequences $x^n$, number of states $M$, and Markov order $k$,
    \[\mu_k(x^n) - \lambda_M(x^n) \leq \frac{\log M}{k+1}.\]

    \begin{proof}
        By \prettyref{lem:mu-equals-H}, \prettyref{lem:lambda-equals-H}, and \prettyref{lem:stationary-k-empirical}, 
        \[\mu_k(x^n) - \lambda_M^{g,s_1}(x^n) \leq H(X_{k+1}|X^k) - H(X|S) = H(X_{k+1}|X^k) - H(X_{k+1}|S_{k+1}),\]
        where $(X^{k+1}, S^{k+1}$) is the $(k+1)$-order joint empirical distribution of the sequence and corresponding states.
        
        We will now apply rules of conditional entropy such that we can take advantage of the deterministic finite state dynamics and the chain rule of entropy.

        As conditioning reduces entropy,
        \[\mu_k(x^n) - \lambda_M^{g,s_1}(x^n) \leq \frac{1}{k+1} \sum_{j=-1}^{k-1} \left( H(X_{k+1}|X_{k-j}^k) - H(X_{k+1}| S_{k+1}, X_{k-j}^k, S_{k-j})\right),\]
        because the right-hand side is an average where each term is $\geq \mu_k(x^n) - \lambda_M^{g,s_1}(x^n)$ (we condition on fewer variables than in $H(X_{k+1}|X^k)$ for the first component of each term, and we condition on more variables than $H(X|S)$ for the second component).

        Given $S_{k-j}$ and $X_{k-j}^k$, we deterministically know $S_{k+1}$ via applying the state-transition function $g$.
        So, \\$H(X_{k+1}| S_{k+1}, X_{k-j}^k, S_{k-j})$ is equivalent to $H(X_{k+1}| X_{k-j}^k, S_{k-j})$.
        Expanding the summation and applying \prettyref{cor:stationary-k-empirical},
        \begin{align*}
            \mu_k(x^n) - \lambda_M^{g,s_1}(x^n) &\leq \frac{1}{k+1}\bigg[\left(H(X_1) + H(X_2|X_1) \cdots + H(X_{k+1}|X^k)\right) \\
            &\hspace{4em}- \left( H(X_1|S_1) + H(X_2 | X_1, S_1) + \cdots +H(X_{k+1} | X^k, S_1) \right)\bigg].
        \end{align*}
        By pattern-matching with  the chain rule of entropy,
        \begin{align*}
            \mu_k(x^n) - \lambda_M^{g,s_1}(x^n) \leq \frac{1}{k+1} \left(H(X^{k+1}) - H(X^{k+1}|S_1)\right) = \frac{1}{k+1} I(X^{k+1}; S_1) \leq \frac{1}{k+1} H(S_1) \leq \frac{\log M}{k+1},
        \end{align*}
        where the second and third steps follow directly from the definition of mutual information.
    \end{proof}
\end{claim}

\begin{lemma}[Lower bound of $\lambda(\xv)$ by $\mu(\xv)$]\label{lem:lambda-geq-mu}
    For all individual sequences,
    \[\lambda(\xv) \geq \mu(\xv).\]

    \begin{proof}
        By \prettyref{claim:mu-minus-lambda-bound},
        \begin{align*}
            \limsup_{n \to \infty} \mu_k(x^n) \leq \limsup_{n\to\infty} \lambda_M(x^n) + \frac{\log M}{k+1} \implies \mu_k(\xv) \leq \lambda_M(\xv) + \frac{\log M}{k+1},\, \forall \xv, M, k.
        \end{align*}
        First, fix $M$ and take the limit as $k \to \infty$ to get
        \[\mu(\xv) \leq \lambda_M(\xv),\, \forall M.\]
        Then, we obtain the desired result by taking $M \to \infty$.
        \[\mu(\xv) \leq \lambda(\xv).\]
    \end{proof}
\end{lemma}

The proof that $\rho(\xv) \log A = \mu(\xv)$ follows from Theorem 3 of \cite{originalLZ78paper}, which states that $\rho(\xv) = \hat{H}(\xv) \log A$,\footnote{The definition of $\hat{H}(\xv)$ here is a factor of $\log A$ off from the definition in \cite{originalLZ78paper} in order to be more consistent with the work in \prettyref{sec:mu-lambda-entropies}.} where $\hat{H}(\xv)$ is defined as follows:
\begin{definition}\label{def:H-hat-lz78-paper}
    As in the proof for the lower bound in \prettyref{lem:mu-equals-H}, let $\tilde{X}^{k+1}$ follow empirical distribution
    \[\Pbb\left(\tilde{X}^{k+1} = a^{k+1}\right) = \frac{1}{n-k} \sum_{i=1}^{n-k} \indic{x_{i}^{i+k} = a^{k+1}}.\]
    Here, we only consider the first $n-k$ length-$(k+1)$ sequences instead of using the circular convention.
    $\hat{H}(\xv)$ is defined as
    \[\hat{H}(\xv) \triangleq \lim_{k\to\infty} \hat{H}_k(\xv),\text{ where } \hat{H}_k(\xv) \triangleq \limsup_{n\to\infty} \hat{H}_k(x^n) \triangleq \limsup_{n\to\infty} \frac{1}{k} H(\tilde{X}^k)).\]
\end{definition}

\begin{lemma}[Equivalence of $\rho(\xv)$ and $\mu(\xv)$]\label{lem:rho-eq-mu}
    For all individual sequences,
    \[\mu(\xv) = \rho(\xv) \log A.\]

    \begin{proof}
        Let $X^k$ be the empirical distribution from \prettyref{def:k-order-emp}, \ie, using the circular convention.
        As part of the proof of the lower bound in \prettyref{lem:mu-equals-H}, we showed that $\lim_{n\to\infty} |H(X^k) - H(\tilde{X}^k)| = 0$, so $\hat{H}_k(\xv) = \limsup_{n\to\infty} \frac{1}{k} H(X^k)$.

        By the chain rule of entropy and \prettyref{lem:mu-equals-H},
        \begin{align*}
            \hat{H}_k(\xv) = \frac{1}{k} \varlimsup_{n\to\infty} \left(H(X_1) + H(X_2|X_1) + \cdots H(X_k|X^{k-1})\right) = \frac{1}{k} \sum_{\ell=0}^{k-1} \mu_\ell(\xv).
        \end{align*}
        We would like to show that $\lim_{k\to\infty} \hat{H}_k(\xv) = \mu(\xv)$, \ie, that, $\forall \epsilon > 0$, 
        $\exists K > 0$, s.t., $\forall k > K$, $|\hat{H}_k(\xv)-\mu(\xv)| < \epsilon$.
        
        As $\lim_{k\to\infty} \mu_k(\xv) = \mu(\xv)$, $\exists L$ s.t., $\forall \ell > L$, $|\mu_\ell(\xv) - \mu(\xv)| < \frac{\epsilon}{2}$.
        Choose $K > \frac{2L\log A}{\epsilon}$.
        Then, $\forall k > K$,
        \begin{align*}
            \left|\hat{H}_k(\xv) - \mu(\xv)\right| &= \left|\frac{1}{k} \sum_{\ell=0}^{k-1} \mu_\ell(\xv) - \mu(\xv) \right| \leq \frac{1}{k}\sum_{\ell=0}^L \left|\mu_\ell(\xv) - \mu(\xv)\right| + \frac{1}{k} \sum_{\ell=L}^{k-1} \left|\mu_\ell(\xv) - \mu(\xv)\right| \\
            &< \frac{1}{k}\sum_{\ell=0}^L \left|\mu_\ell(\xv) - \mu(\xv)\right| + \frac{\epsilon}{2} < \epsilon,
        \end{align*}
        where the final inequality follows from the fact that $0 \leq \mu_\ell(\xv) \leq \log A$ and the same holds for $\mu(\xv)$.

        $\therefore$, $\hat{H}_k(\xv) \to \mu(\xv)$ as $k \to\infty$.
        And, as $\rho(\xv)\log A = \hat{H}(\xv)$, $\rho(\xv)\log A = \mu(\xv)$.
    \end{proof}
\end{lemma}
        
\textbf{Theorem \ref{thm:lambda-eq-mu-eq-rho}}.
    For any infinite individual sequence, the optimal finite-state log loss, optimal Markov SPA log loss, and finite-state compressibility are equivalent:
    \[\lambda(\xv) = \mu(\xv) = \rho(\xv) \log A.\]

    \begin{proof}
        This is encompassed in \prettyref{lem:lambda-leq-mu}, \prettyref{lem:lambda-geq-mu}, and \prettyref{lem:rho-eq-mu}.
    \end{proof}

\section{Proofs: Optimality of the LZ78 Family of SPAs}
\subsection{Results for Stationary and Ergodic Probability Sources}\label{app:stationary-and-ergodic-sources}

Here, we detail how much of the work on sequential probability assignments for individual sequences extends to sequential probability assignments for stationary stochastic processes.

First, the correspondence between the optimal $k$-order Markov loss and the corresponding empirical entropy, in the individual sequence setting, translates over to the following result:
\begin{lemma}\label{lem:mu-leq-h-stationary}
  For any stationary stochastic process $\Xv$,
  \begin{enumerate}[(a)]
      \item $\Ebb \mu_0(X^n) \leq H(X_1)$, and
      \item $\Ebb \mu_k(X^n) \leq H(X_{k+1}|X_k)$.
  \end{enumerate}

  \begin{proof}
      \begin{enumerate}[(a)]
          \item The expectation of $\mu_0(X^n)$ is
            \begin{align*}
                \Ebb \mu_0(X^n) &= \Ebb \left[\frac{1}{n} \min_{q_t \in \mu_0} \sum_{t=1}^n \log \frac{1}{q(X^t)}\right] \stackrel{(i)}{=} \Ebb \left[\min_{q \in \Mcal(\Acal)} \sum_{a \in \Acal} \frac{N(a | X^n)}{n} \log \frac{1}{q(a)}\right] \\
                &\stackrel{(ii)}{\leq} \min_{q \in \Mcal(\Acal)} \sum_{a \in \Acal} \Ebb\left[\frac{N(a | X^n)}{n}\right]\log \frac{1}{q(a)},
            \end{align*}
            where $(i)$ is because $q$ is a zero-order Markov model and $(ii)$ follows from Jensen's inequality.
            We can apply stationarity of $\Xv$ to evaluate
            \[\Ebb\left[\frac{N(a | X^n)}{n}\right] = \sum_{t=1}^n \frac{\Pbb(X_t = a)}{n} = \sum_{t=1}^n \frac{\Pbb(X_1 = a)}{n} = \Pbb(X_1 = a).\]

            So, applying logic from the proof of \prettyref{lem:mu-zero-equals-H},
            \begin{align*}
                \Ebb \mu_0(X^n) &\leq \min_{q \in \Mcal(\Acal)} \sum_{a \in \Acal} \Pbb(X_1 = a) \log \frac{1}{q(a)} = H(X_1).
            \end{align*}

            \item  We have defined $\Mcal_k$ such that $q_t \in \Mcal_k$ can attain zero loss for the first $k$ samples (as its behavior for those samples is entirely unconstrained).
            So,
            \[\Ebb \mu_k(X^n) = \Ebb \left[\frac{1}{n} \min_{q_t \in \Mcal_k} \sum_{t=k+1}^n \log \frac{1}{q(X_t|X^{t-1})} \right] \leq \Ebb \left[\frac{1}{n-k} \min_{q_t \in \Mcal_k} \sum_{t=k+1}^n \log \frac{1}{q(X_t|X^{t-1}_{t-k})} \right].\]
            As in part (a), we apply Jensen's inequality and stationarity to say 
            \begin{equation}
            \begin{split}
                \Ebb \mu_k(X^n) &\stackrel{(i)}{\leq}  \min_{q \in \Mcal_k} \sum_{a^{k+1} \in \Acal^{k+1}} \log\frac{1}{q(a_{k+1}|a^k)} \Ebb \frac{\left|\{k+1 \leq i \leq n : X_i^{i+k}=a^{k+1}\}\right|}{n-k} \\
                &\stackrel{(ii)}{=} \min_{q \in \Mcal_k} \sum_{a^{k+1} \in \Acal^{k+1}} \log\frac{1}{q(a_{k+1}|a^k)} \Pbb \left(X^{k+1} = a^{k+1}\right) \\
                &= \min_{q \in \Mcal_k} \sum_{a^{k} \in \Acal^{k}} \Pbb(X^k = a^k) \sum_{a_{k+1} \in \Acal} \Pbb\left(X_{k+1}=a_{k+1}|X^k=a^k\right) \log\frac{1}{q(a_{k+1}|a_k)} \\
                &= \sum_{a^{k} \in \Acal^{k}} \Pbb\left(X^k = a^k\right) \min_{q_{a^k} \in \Mcal(\Acal)}\sum_{a_{k+1} \in \Acal} \Pbb(X_{k+1}=a_{k+1}|X^k=a^k) \log\frac{1}{q_{a^k}(a_{k+1})} \\
                &= \sum_{a^{k} \in \Acal^{k}} \Pbb\left(X^k = a^k\right) H(X_{k+1}|X^k=a^k) = H(X_{k+1}|X^k).
            \end{split}\label{eqn:mu-k-stationary-proof}
            \end{equation}
            where $(i)$ is by Jensen's inequality and $(ii)$ is by stationarity.
      \end{enumerate}
  \end{proof}
\end{lemma}
Using these results, we can now show that any SPA such that the limit supremum of the log loss is at most $\mu(\xv)$, for all individual sequences, will have an expected log loss equal to the entropy rate of the process.
In particular, this holds for the LZ78 family of SPAs by \prettyref{sec:lz78-spa-optimality}.

\begin{theorem}\label{thm:log-loss-stationary-source}
    Suppose $\Xv$ is a stationary process and the SPA $q$ satisfies
    \[\limsup_{n\to\infty} \frac{1}{n}\log \frac{1}{q(x^n)} \leq \mu(\xv),\]
    for all individual sequences $\xv$.
    Then, if $\Hbb(\Xv)$ is the entropy rate of the stochastic process,
    \[\lim_{n\to\infty} \Ebb\left[\frac{1}{n} \log \frac{1}{q(X^n)}\right] = \Hbb(\Xv).\]
    \begin{proof}
        We will split this proof into two parts:
        \[\liminf_{n\to\infty} \Ebb\left[\frac{1}{n} \log \frac{1}{q(X^n)}\right] \stackrel{(a)}{\geq} \Hbb(\Xv);\quad \limsup_{n\to\infty} \Ebb\left[\frac{1}{n} \log \frac{1}{q(X^n)}\right] \stackrel{(b)}{\leq} \Hbb(\Xv),\]
        which, together, imply the result via the squeeze theorem.
        \begin{enumerate}[(a)]
            \item $\forall n > 0$, let $p^n$ be the joint PDF of $X^n$.
            Then,
            \[\Ebb \left[\frac{1}{n} \log \frac{1}{q(X^n)}\right] = \Ebb \left[\frac{1}{n} \log \left(\frac{1}{q(X^n)} \cdot \frac{p^n(X^n)}{p^n(X^n)}\right)\right] = \frac{1}{n}H(X^n) + \frac{1}{n}D(p^n||q) \geq \frac{1}{n}H(X^n),\]
            as relative entropy is a non-negative quantity.
            Taking $\liminf_{n\to\infty}$ on both sides,
            \[\liminf_{n\to\infty}  \Ebb\left[\frac{1}{n} \log \frac{1}{q(X^n)}\right] \geq \liminf_{n\to\infty} \frac{1}{n}H(X^n) = \Hbb(\Xv).\]

            \item By the assumption
            \[\limsup_{n\to\infty} \frac{1}{n}\log \frac{1}{q(x^n)} \leq \mu(\xv),\, \forall \text{ individual sequences } \xv,\]
            $\forall \epsilon > 0$, $\exists N > 0$ such that, $\forall n > N$,
            \[\frac{1}{n}\log \frac{1}{q(x^n)} \leq \mu(\xv) + \epsilon,\, \forall \xv.\]
            As $\mu(\xv) \leq \mu_k(\xv),\, \forall k \geq 0$ and individual sequence $\xv$,
            \[\frac{1}{n}\log \frac{1}{q(x^n)} \leq \mu_k(\xv) + \epsilon,\, \forall n > N, k \geq 0, \xv.\]
            Plugging this fact into $\Ebb\left[\frac{1}{n} \log \frac{1}{q(X^n)}\right]$ and applying the second result of \prettyref{lem:mu-leq-h-stationary},
            \begin{align*}
                \Ebb\left[\frac{1}{n} \log \frac{1}{q(X^n)}\right] &\leq \Ebb\left[\mu_k(\Xv) + \epsilon\right] \leq H(X_{k+1}|X^k) + \epsilon,\, \forall k \geq 0, n > N.
            \end{align*}
            Taking the limit as $k\to\infty$ on both sides,
            \[ \Ebb\left[\frac{1}{n} \log \frac{1}{q(X^n)}\right] \leq \Hbb(\Xv) + \epsilon,\, \forall n > N,\]
            where we applied strong stationarity to say that $\lim_{k\to\infty}H(X_{k+1}|X^k) = \Hbb(\Xv)$.
            As $\epsilon$ is arbitrary, we can apply the definition of a limit supremum to say
            \[\limsup_{n\to\infty} \Ebb\left[\frac{1}{n} \log \frac{1}{q(X^n)}\right] \leq \Hbb(\Xv).\]
        \end{enumerate}
    \end{proof}
\end{theorem}

\begin{theorem}\label{thm:loss-ergodic-source}
    If, additionally, the process is ergodic, then the result holds almost surely rather than in expectation:
    \[ \frac{1}{n} \log \frac{1}{q(X^n)} \stackrel{a.s.}{\to} \Hbb(\Xv).\]
    \begin{proof}
        We will again split the proof into that of two inequalities:
        \begin{enumerate}[(a)]
            \item $\liminf_{n\to\infty} \frac{1}{n} \log \frac{1}{q(X^n)} \geq \Hbb(\Xv)$ (a.s.).
            
            $\forall n > 0$, let $p^n$ be the joint PDF of $X^n$.
            Then,
            \[\frac{1}{n} \log \frac{1}{q(X^n)} = \frac{1}{n} \log \left(\frac{1}{q(X^n)} \cdot \frac{p^n(X^n)}{p^n(X^n)}\right) = \frac{1}{n} \log \frac{1}{p^n(X^n)} + \frac{1}{n} \log \frac{p^n(X^n)}{q(X^n)}.\]
           
            By the Shannon-McMillan-Breiman theorem \cite{breiman1957Ergodic}
            \[\lim_{n\to\infty} \frac{1}{n} \log \frac{1}{p^n(X^n)} = \Hbb(\Xv)\quad(a.s.).\]

            In addition, by Lemma 2 of \cite{algoet1992universalGambling}, $\liminf_{n\to\infty} \frac{1}{n} \log \frac{p^n(X^n)}{q(X^n)}$ is almost surely non-negative, so
            \[\liminf_{n\to\infty} \frac{1}{n} \log \frac{1}{q(X^n)} \geq \Hbb(\Xv)\quad(a.s.).\]
            
            \item $\limsup_{n\to\infty} \frac{1}{n} \log \frac{1}{q(X^n)} \leq \Hbb(\Xv)$ (a.s.).
            
            We first show the following:
            \begin{claim}
                For any stationary and ergodic source, the following holds with probability 1:
                \[\lim_{n\to\infty}\mu_k(X^n) = H(X_{k+1}|X^k).\]
                \begin{proof}[Proof of claim]
                    By Birkhoff's Ergodic Theorem \cite{Petersen_1983} and the continuous mapping theorem,
                    \[\frac{\left|\{i \in [n]: X_i^{i+k} = a^{k+1}\}\right|}{n} \stackrel{a.s.}{\to} \Pbb(X^{k+1} = a^{k+1}).\]
                    Then, almost surely,
                    \begin{align*}
                        \lim_{n\to\infty} \mu_k(X^n) &= \lim_{n\to\infty} \min_{\hat{q} \in \Mcal_k} \sum_{a^{k+1} \in \Acal^{k+1}} \frac{\left|\{i \in [n]: X_i^{i+k} = a^{k+1}\}\right|}{n} \log \frac{1}{\hat{q}(a_{k+1}|a^k)} \\
                        &=  \min_{\hat{q} \in \Mcal_k} \sum_{a^{k+1} \in \Acal^{k+1}} \Pbb(X^{k+1} = a^{k+1}) \log \frac{1}{\hat{q}(a_{k+1}|a^k)}. 
                    \end{align*}
                    We are now in an identical position as line 2 of \eqref{eqn:mu-k-stationary-proof}, and can therefore follow the rest of \eqref{eqn:mu-k-stationary-proof} to conclude that
                    \[ \lim_{n\to\infty} \mu_k(X^n) = H(X_{k+1}|X^k)\quad (a.s.).\]
                \end{proof}
            \end{claim}
            By the same logic as part (b) of \prettyref{thm:log-loss-stationary-source}, $\forall \epsilon > 0$, $\exists N \in \Nbb$ s.t., $\forall n > N$ and $k > 0$,
            \[\frac{1}{n} \log \frac{1}{q(X^n)} \leq \mu_k(\Xv) + \epsilon.\]
            By the claim,
            \[\frac{1}{n} \log \frac{1}{q(X^n)} \leq \mu_k(\Xv) + \epsilon \leq H(X_{k+1}|X^k) + \epsilon\quad(a.s.),\quad \forall n > N.\]
            As $\epsilon$ is arbitrary,
            \[\limsup_{n\to\infty} \frac{1}{n} \log \frac{1}{q(X^n)} \leq H(X_{k+1}|X^k)\quad(a.s.).\]
            Taking the limit as $k\to\infty$ and applying strong stationarity,
            \[\limsup_{n\to\infty} \frac{1}{n} \log \frac{1}{q(X^n)} \leq \Hbb(\Xv)\quad(a.s.).\]
        \end{enumerate}
    \end{proof}
\end{theorem} 

\section{Proofs: The LZ78 Sequential Probability Assignment as a Probability Source}
\subsection{The Bernoulli LZ78 Probability Source}\label{app:bernoulli-lz78-source}
\begin{lemma}
    The LZ78 probability source with prior $\Uppi = \mathrm{Ber}(1/2)$ has entropy rate $0$.

    \begin{proof}
        Let $\Xv$ be generated from the LZ78 probability source with a $\mathrm{Ber}(1/2)$ prior.
        Then.
        \[H(\Xv) = \lim_{n\to\infty} \frac{1}{n} H(X^n) = \lim_{n\to\infty} \frac{1}{n} \Ebb \log \frac{1}{q^{\text{LZ78}, \Uppi}(X^n)} = \lim_{n\to\infty} \frac{1}{n} \Ebb \sum_{k = 1}^{C(X^n)} \log \frac{1}{q^{\text{LZ78}, \Uppi}(C_k)},\]
        where $C_k$ is the $k$\textsuperscript{th} phrase in the LZ78 parsing of $x^n$.
        As each phrase is deterministic except for the final symbol, which is equally likely to be $0$ or $1$, a log loss of $1$ is incurred on each phrase.
        In addition, $C(X^n) = \bigO(\sqrt{n})$, so
        $H(\Xv) =  \lim_{n\to\infty} \frac{C(X^n)}{n} = 0$.
    \end{proof}
\end{lemma}

\begin{lemma}\label{lem:lz78-bernoulli-spa-log-loss}
    The LZ78 SPA of \prettyref{con:lz78-spa}, for any prior $\Upomega$ such that $\supp(\Upomega) = [0, 1]$, will achieve an asymptotic log loss of $0$ \underline{deterministically} on any sequence from the LZ78 probability source with a $\mathrm{Ber}(1/2)$ prior.

    \begin{proof}
        By \prettyref{thm:lz78-log-loss-codelength-correspondence}, for any individual sequence $x^n$,
        \[\lim_{n\to\infty} \left| \frac{1}{n} \log \frac{1}{q^{\text{LZ78}, \Upomega}(x^n)} - \frac{C(x^n) \log C(x^n)}{n}\right| = 0.\]
        As $C(X^n) = \bigO(\sqrt{n})$ deterministically,
        \[\frac{1}{n} \log \frac{1}{q^{\text{LZ78}, \Upomega}(x^n)} = \frac{C(x^n) \log C(x^n)}{n} + o(1) = \bigO\!\left(\frac{\log n}{\sqrt{n}} \right) + o(1) = o(1).\]
    \end{proof}
\end{lemma}

\begin{lemma}\label{lem:markov-log-loss-lz78-bernoulli}
    Almost surely, $\Xv$ generated from the LZ78 probability source with a $\mathrm{Ber}(1/2)$ prior satisfies $\mu(\Xv) = 1$.

        \begin{proof}
        To show that $\mu(\Xv) = 1$ (a.s.), we show that, for any fixed context length $k \in \Nbb$, $\mu_k(\Xv) = 1$, almost surely.

        Consider a realization, $X^n$, of the LZ78 source, and denote the final complete phrase by $Y^m$ (by construction, the location of this phrase is deterministic).
        This phrase, considered in isolation, is $\stackrel{\text{i.i.d.}}{\sim} \mathrm{Ber}(1/2)$.
        By \prettyref{lem:mu-equals-H}, the strong law of large numbers, and the continuous mapping theorem $\mu_k(Y^m) \stackrel{a.s.}{\to} h_2(1/2) = 1$ as $m\to\infty$, where $h_2$ is the binary entropy function $h_2(p) = p \log \frac{1}{p} + (1-p) \log \frac{1}{1-p}$.

        This result can be extended to the full sequence, $X^n$, via the following result:
        \begin{claim}
            Suppose that $\mu_k(u^m) \to 1$ for individual binary sequence $\uv$.
            Then, for $v^n$ defined as
            \[u_1;\, u_1, u_2;\, u_1, u_2, u_3;\, \dots;\, u_1, u_2, \dots, u_m;\, u_1, \dots, u_\ell,\]
            where $n = \frac{m(m+1)}{2} + \ell$, it also holds that $\mu_k(v^n) \to 1$.

            \begin{proof}[Proof of claim]
                First, $\mu_k(v^n) \leq 1$, as $\mu_k(v^n)$ is equal to the emprical entropy associated with binary sequence $v^n$, which is upper-bounded by $1$.
                So, it suffices to show that $\lim_{n\to\infty} \mu_k(v^n) \geq 1$.
                Specifically, we must show that, $\forall \epsilon > 0$, $\exists N > 0$ s.t., $\forall n > N$, $\mu_k(v^n) \geq 1 - \epsilon$.

                $\mu_k(u^m) \to 1$, so by definition, $\exists M > 0$ s.t., $\forall m > M$, $\mu_k(u^m) > 1-\epsilon/2$.
                We then choose $N$ such that fewer than $\frac{\epsilon}{2}N$ symbols are in phrases of length $\leq M$.\footnote{The only phrases that have length $\leq M$ are the first $M$ phrases, which comprise $\frac{M(M+1)}{2}$ symbols, and potentially the last phrase.
                As a result, the condition is satisfied by $N \geq \left\lceil \frac{M(M+1) + 2M}{\epsilon}\right\rceil$.}
                Then, for $n > N$,
                \begin{align*}
                    \mu_k(v^n) &= \min_{q \in \Mcal_k} \frac{1}{n} \left(\sum_{z \in \Zcal(v^n): |z| \leq M} \log \frac{1}{q(z)} + \sum_{z \in \Zcal(v^n): |z| > M} \log \frac{1}{q(z)}\right) \\
                    &\stackrel{(i)}{>} \frac{1}{n} \left(\sum_{z \in \Zcal(v^n): |z| \leq M} |z| \mu_k(z) + \sum_{z \in \Zcal(v^n): |z| > M} |z| \mu_k(z)\right) 
                    \stackrel{(ii)}{>} \frac{1}{n} \left(1- \frac{\epsilon}{2}\right) \sum_{z \in \Zcal(v^n): |z| \leq N} |z| \\
                    &\stackrel{(iii)}{\geq} \left(1-\frac{\epsilon}{2}\right)^2 > 1 - \epsilon,
                \end{align*}
                where $(i)$ is by Jensen's inequality, $(ii)$ follows from the fact that all phrases in the second summation have length $> N$, and $(iii)$ follows from the definition of $M$.
                $\therefore$, $\lim_{n\to\infty} \mu_k(v^n) \geq 1$, meaning $\lim_{n\to\infty} \mu_k(v^n) = 1$.
            \end{proof}
        \end{claim}
        Therefore, as $\lim_{m\to\infty}\mu_k(Y^m) = 1$ with probability 1, $\lim_{n\to\infty}\mu_k(X^n) = 1$ almost surely as well.
        So, by definition, $\mu_k(\Xv) = 1$ (a.s.), and therefore $\mu(\Xv) = 1$ (a.s.).
    \end{proof}
\end{lemma}
}
\newpage
\bibliographystyle{IEEEtran}
\bibliography{main.bib}

\end{document}